\documentclass[acmsmall]{acmart}

\usepackage{tikz}
\usetikzlibrary{positioning}
\usepackage{subcaption}
\usepackage{caption}
\usepackage{graphicx}
\usepackage{float}
\usetikzlibrary{fit, positioning, backgrounds}
\usetikzlibrary{decorations.pathreplacing, decorations.markings}
\usetikzlibrary{arrows.meta}
\usepackage{pgfplots}
\usepgfplotslibrary{fillbetween}
\pgfplotsset{compat=1.18}
\usepackage{titlesec}

\usepackage{lipsum}
\usepackage{symbols}
\usepackage{thmtools}
\usepackage{thm-restate}
\usepackage{balance}
\usepackage{subcaption}
\usepackage[linesnumbered, ruled, vlined]{algorithm2e}[2023/02/08]

\usepackage{xcolor}
\usepackage{xspace}
\usepackage{listings}

\usepackage{url}
\usepackage{soul}
\sethlcolor{lightgray}
\usepackage[export]{adjustbox}
\usepackage{enumitem}
\usepackage{tabularx}
\usepackage{graphicx}
\usepackage{pdfpages}

\usepackage{multirow}  
\usepackage{array}     

\setlist[enumerate]{leftmargin=20pt}
\setlist[itemize]{leftmargin=20pt}

\graphicspath{{figures/}}

\definecolor{GreenColor}{HTML}{38A528}

\usepackage[inline]{aplcomments}

\newcommand{\edit}[1]{\textcolor{black}{#1}}

\newcommand{\reviseBegin}{\color{black}}
\newcommand{\reviseEnd}{\color{black}}

\let\svfigurename\figurename
\newcommand\figcolor[1]{
  \renewcommand\figurename{\color{black}\svfigurename}
}

\newif\ifshowcomments
\showcommentsfalse

\newcommand{\mcomment}[1]{\ifshowcomments #1\fi}

\setlist[enumerate,1]{%
  label=\arabic*.,
}

\newlist{inlinelist}{enumerate*}{1}
\setlist*[inlinelist,1]{%
  label=(\arabic*),
}

\newcommand{\mahesh}[1]{\textcolor{purple} {mahesh: #1}}

\newcommand{\wjh}[1]{\textcolor{GreenColor} {hwj: #1}}
\newcommand{\gzh}[1]{\textcolor{orange} {hgz: #1}}

\newcommand{\secref}[1]{\S\ref{#1}}
\newcommand{\figref}[1]{Figure~\ref{#1}}
\newcommand{\algref}[1]{Algorithm~\ref{#1}}
\newcommand{\tabref}[1]{Table~\ref{#1}}

\newcommand*\ttvar[1]{\texttt{\expandafter\dottvar\detokenize{#1}\relax}}
\newcommand*\dottvar[1]{\ifx\relax#1\else
  \expandafter\ifx\string-#1\string-\allowbreak\else#1\fi
  \expandafter\dottvar\fi}
\let\oldnl\nl
\definecolor{comment-color}{rgb}{0.5,0.1,0.1}
\newcommand{\nonl}{\renewcommand{\nl}{\let\nl\oldnl}}
\newcommand{\codeComment}[1]{\textnormal{\color{comment-color}{\textit{\#
#1}}}\unskip}
\newcommand{\ignore}[1]{}

\sethlcolor{lightgray!75}
\newcommand{\shl}[1]{\hl{\mbox{#1}}} 

\newcommand{\name}{\textsf{Marlin}\xspace}
\newcommand{\system}{\name}
\newcommand{\mcommit}{MarlinCommit\xspace}

\newcommand{\zk}{ZooKeeper\xspace}
\newcommand{\szk}{\textit{S-ZK}\xspace}
\newcommand{\lzk}{\textit{L-ZK}\xspace}
\newcommand{\fdb}{\textit{FDB}\xspace}

\newcommand{\converged}{converged\xspace}
\newcommand{\convergedL}{Converged\xspace}

\newcommand{\syslog}{SysLog\xspace}

\newcommand{\glog}{GLog\xspace}
\newcommand{\gtbl}{GTable\xspace}
\newcommand{\mtbl}{MTable\xspace}

\xspaceaddexceptions{\}}

\newcommand{\boldtext}[1]{\vspace{1mm} \noindent \textbf{#1}}

\DeclareRobustCommand{\circleb}[1]{%
  \tikz[baseline=(char.base)]{%
    \node[shape=circle, fill=black, inner sep=.5pt] (char)
         {\textcolor{white}{\small #1}};%
  }%
}

\newcommand{\vldbavailabilityurl}{https://gitfront.io/r/user-5630758/jcmGNtQTsw62/Bonsai-private}

\AtBeginDocument{%
  \providecommand\BibTeX{{%
    \normalfont B\kern-0.5em{\scshape i\kern-0.25em b}\kern-0.8em\TeX}}}

\usepackage{subfiles}

\acmPrice{15.00}
\acmISBN{978-1-4503-XXXX-X/18/06}

\titlespacing*{\section}{0pt}{2.0ex minus .1ex}{1.0ex minus .1ex}
\titlespacing*{\subsection}{0pt}{1.8ex minus .1ex}{0.8ex minus .1ex}
\titlespacing*{\subsubsection}{0pt}{1.5ex minus .1ex}{0.5ex minus .1ex}
\titlespacing*{\subsubsubsection}{0pt}{1.5ex minus .1ex}{0.5ex minus .1ex}

\begin{document}

\title{\system: Efficient Coordination for Autoscaling Cloud DBMS (Extended Version)}
\author{Wenjie Hu}
\affiliation{%
  \institution{University of Wisconsin-Madison}
  \city{Madison}
  \country{USA}
}
\email{wenjiehuhippo@cs.wisc.edu}

\author{Guanzhou Hu}
\affiliation{%
  \institution{University of Wisconsin-Madison}
  \city{Madison}
  \country{USA}
}
\email{guanzhou.hu@wisc.edu}

\author{Mahesh Balakrishnan}
\authornote{Work done while unaffiliated.}
\affiliation{%
  \institution{Meta, Inc.}
  \city{Menlo Park}
  \country{USA}
}
\email{mbalakrishnan@meta.com}

\author{Xiangyao Yu}
\affiliation{%
  \institution{University of Wisconsin-Madison}
  \city{Madison}
  \country{USA}
}
\email{yxy@cs.wisc.edu}

\begin{abstract}

\mcomment{\mahesh{minor, feel free to ignore: title could be "A Disaggregated Control Plane for Autoscaling Cloud Databases"}}



Modern cloud databases are shifting from converged architectures to storage disaggregation, enabling independent scaling and billing of compute and storage. However, cloud databases still rely on external, converged coordination services (e.g., ZooKeeper) for their control planes. These services are effectively lightweight databases optimized for low-volume metadata. As the control plane scales in the cloud, this approach faces similar limitations as converged databases did before storage disaggregation: scalability bottlenecks, low cost efficiency, and increased operational burden.

We propose to disaggregate the cluster coordination to achieve the same benefits that storage disaggregation brought to modern cloud DBMSs. We present Marlin, a cloud-native coordination mechanism that fully embraces storage disaggregation. Marlin eliminates the need for external coordination services by consolidating coordination functionality into the existing cloud-native database it manages. To achieve failover without an external coordination service, Marlin allows cross-node modifications on coordination states. To ensure data consistency, Marlin employs transactions to manage both coordination and application states and introduces MarlinCommit, an optimized commit protocol that ensures strong transactional guarantees even under cross-node modifications. Our evaluations demonstrate that Marlin improves cost efficiency by up to 4.4× and reduces reconfiguration duration by up to 4.9× compared to converged coordination solutions.



\end{abstract}


\maketitle

\everypar{\looseness=-1}


\section{Introduction}
\label{sec:intro}
Modern cloud databases are moving from monolithic, converged architectures---where compute and storage reside on the same physical servers~\cite{taft2020cockroachdb, yang2022oceanbase,postgresql,cubukcu2021citus}---to storage disaggregation, where compute and storage are separated and connected via network~\cite{verbitski2017amazon, socrates, alloydb,  corbett12spanner, zhou2021foundationdb, cosmosdb, brantner2008building}. This disaggregation enables many benefits over the converged design, such as independent and elastic scaling of compute and storage, improved resource utilization, reduced cost, and flexible deployment. 

Despite modernizing their data planes through storage disaggregation, cloud databases still rely on external, centralized, and converged coordination services (e.g., ZooKeeper~\cite{hunt2010zookeeper}, etcd~\cite{etcd}, Chubby~\cite{burrows2006chubby}) to maintain shared state 
for their control planes. In effect, these services are miniature databases optimized for storing small-scale coordination states. However, as control plane workloads scale in size and throughput in tandem with user data, these coordination services face similar limitations as converged databases did before storage disaggregation. In particular, conventional coordination suffers from three major limitations:
\circleb{1} \textbf{Scalability bottleneck.} Converged coordination services for low-scale metadata typically adopt single-master design~\cite{hunt2010zookeeper, etcd, burrows2006chubby} where all traffic is routed to a single node. 
The problem becomes more pronounced on the cloud as 
frequent reconfigurations (e.g., rescaling, rebalancing) amplify coordination workloads. Moreover, cloud databases often employ geo-distributed deployments~\cite{shen2016follow}. The cross-region latency further increases the performance overhead of a centralized coordination service. \circleb{2} \textbf{Low cost-efficiency.} Cloud workloads change rapidly and are often bursty, with peak system utilization often many times higher than the 99th percentile~\cite{gmach2007workload,verma2009server, cortez2017resource,gmach2007workload}. Static coordination services are commonly overprovisioned to handle peak coordination demand~\cite{FaaSKeeper}, resulting in underutilized resources and poor cost efficiency. Furthermore, the converged design prevents true “scale-to-zero” elasticity when workloads diminish (e.g., ZooKeeper requires at least three VMs for reliability), incurring significant upfront costs. \circleb{3} \textbf{Operational burden.} Relying on an external, converged coordination service complicates deployment and maintenance. Operators must provision and manage this additional subsystem before deploying the database. Unlike stateless compute, converged coordination services must remain stateful, fault-tolerant, and always on, requiring specialized expertise to configure and operate. Moreover, external dependencies can reduce overall availability and increase troubleshooting efforts. According to the “golden rule” of component reliability~\cite{sloss2017calculus}, any critical dependency must be 10 times as reliable as the overall system's target to ensure its contribution to overall unreliability is negligible. Anecdotally, coordination services are notoriously difficult to operate reliably~\cite{chandra2007paxos,ongaro2014search,balakrishnan2020virtual} because they implement complex consensus protocols directly.






Prior works partially address some of the limitations but often exacerbate others, and therefore cannot fundamentally resolve the problem. One approach leverages an external distributed transactional database system (DDBMS) as the coordination service to improve scalability via partitioning~\cite{snowflake,byconity}. However, it significantly complicates the architecture, inflating the cost and introducing greater external dependencies and heavier operational burdens. Moreover, it can increase latency as each coordination request requires more network round-trips to traverse DDBMS's internal hierarchy. Another approach~\cite{FaaSKeeper} reduces cost by running a coordination service on serverless functions. Nevertheless, it does not address the scalability bottleneck and serverless functions are inherently 
unsuitable for performance-critical tasks. Other works try to solve all three limitations by utilizing internal infrastructure to eliminate external coordination services~\cite{Cockroach,yang2022oceanbase,kraft}. However, they focus on 
converged architectures where compute nodes can be protected by the storage layer's replication protocol. They are therefore inapplicable to cloud-native databases with compute-storage disaggregation, where the replication protocol resides solely in the storage layer and is not accessible at the compute layer~\cite{cornus, pang2024understanding}. 


In this paper, we aim to answer the following key question: \textit{Can we disaggregate the cluster coordination to achieve the same benefits that storage disaggregation brought to the data planes of modern cloud DBMSs?} Our answer is \system, a cloud-native coordination mechanism that fully embraces storage disaggregation. \system eliminates the need for external coordination services by consolidating coordination functionality \edit{into the database} it manages: coordination states are stored in the database’s highly available, disaggregated storage layer, while consistent coordination is handled in the compute layer using standard transaction managers. This design enables linear scalability through partitioned access paths, reduces costs via independent scaling of compute and storage, and simplifies operations by removing external dependencies. The design of \system 
faces the following key challenges:

\noindent \textbf{Challenge \#1: Scalability.} An intuitive approach for scalable coordination is to partition coordination states across compute nodes. 
However, without an external coordination service, this exclusive ownership can cause a “failover deadlock”: if a node owning failover-critical metadata fails, the remaining nodes cannot update that metadata to promote a new owner, stalling the failover process. \system solves this by allowing multiple nodes to access failover-critical states which sit in the disaggregated storage. \system identifies two types of coordination states essential to failover: (1) \textit{Data ownership}, mapping data partitions to their owner nodes, and (2) \textit{Group membership}, listing active nodes in the cluster. Data ownership information, which grows with user data, is partitioned by the owner node ID. Each node manages its own partition under normal conditions, while permitting other nodes to modify it if the owner fails. Group membership, which is relatively small and updated infrequently, is stored in a single, non-partitioned log accessible by all nodes. Other states unrelated to failover (e.g., security keys, monitoring metrics) is partitioned like regular user data and exclusively managed by their respective owner nodes at all times.

\noindent \textbf{Challenge \#2: Data Consistency.} \system must ensure data consistency without relying on an external coordination service, even during concurrent reconfigurations or arbitrary compute node failures. \system reuses existing database components (e.g., transaction and cache managers) and manages all coordination state accesses through special \textit{reconfiguration transactions}. 

Conventional transaction protocols (e.g., 2PL, OCC, and 2PC) cannot fully resolve metadata inconsistencies because their assumption that a compute node has exclusive control over the data it manages no longer holds in \system. For example, group membership metadata may be modified by multiple compute nodes concurrently, and compute node failures or network partitions can create ambiguous data ownership. To solve this, \system introduces \mcommit, an optimized commit protocol that ensures strong transactional guarantees even when multiple nodes can access the same data.
The key insight is to leverage a conditional append API, \textit{Append(updates, LSN)}, which succeeds only if the current log version matches the expected version. Via this API, \mcommit can detect cross-node modifications and commit a transaction only if no cross-node modifications have been made. This approach is storage-agnostic and can be supported by any disaggregated storage services offering compare-and-swap functionality~\cite{Redis2009,AzureBlobStorage, azuretablestorage, chang2008bigtable,cosmosdb,elhemali2022dynamodb}.

In summary, the paper makes the following key contributions:
\begin{itemize}[leftmargin=12pt, rightmargin=0pt,noitemsep,topsep=0pt]
    \item We revisit coordination solutions in modern cloud OLTP databases and identify their limitations in the context of autoscaling.
    \item We propose \system, a cloud-native coordination mechanism for modern cloud OLTP databases that is self-contained, scalable, and cost-efficient. \system eliminates the need for an external coordination service by integrating the coordination into the database itself.
    \item We implement \system in a cloud OLTP DBMS testbed. Our evaluations show that \system improves cost-efficiency by 4.4$\times$ and reduces reconfiguration duration by 4.9$\times$ compared to a converged coordination service.
\end{itemize}

The paper is organized as follows: \secref{sec:motivation} discusses architectural trends in cloud databases and provides background on cluster coordination. \secref{sec:sysmodel} outlines the system model and 
\secref{sec:cluster_management} details Marlin's core design, including coordination state management, primitives, and their usage examples. \edit{\secref{sec:impl} describes implementation details of the database testbed and \system.} \secref{sec:eval} evaluates \system's performance, \secref{sec:related} discusses related work, and \secref{sec:conclusion} concludes the paper.


\section{Background and Motivation}
\label{sec:motivation}

This section first describes the architecture of storage disaggregation (\secref{ssec:disagg}). Then, it explores the taxonomy of scalable cloud OLTP archetypes based on data access paths (\secref{ssec:arch}). Finally, it discusses the limitations of cluster coordination solutions in existing cloud DBMSs (\secref{ssec:metamanagement}).

\subsection{Storage Disaggregation in Cloud DBMS}
\label{ssec:disagg}
Modern cloud-native DBMSs~\cite{verbitski2017amazon, socrates, corbett12spanner, alloydb, cosmosdb, microsoft_sql_server} increasingly adopt a storage disaggregation architecture, where storage and compute services operate independently, connected via the data center network. This design contrasts with conventional shared-nothing systems that rely on direct-attached storage. 
Storage disaggregation enables independent scaling of compute and storage resources, enhancing elasticity and cost efficiency. The disaggregation model also supports limited computation within the storage layer, which can be performed on the storage nodes~\cite{verbitski2017amazon} or in a layer close to the storage nodes~\cite{redshift_spectrum, aqua_redshift}. Prior research has leveraged near-storage computation to enhance performance and reduce costs~\cite{cornus, fpdb}. Building on this concept, \system employs near-storage computation capabilities (i.e., Compare-And-Swap) for coordination purposes.

\subsection{Cloud OLTP DBMS}
\label{ssec:arch}

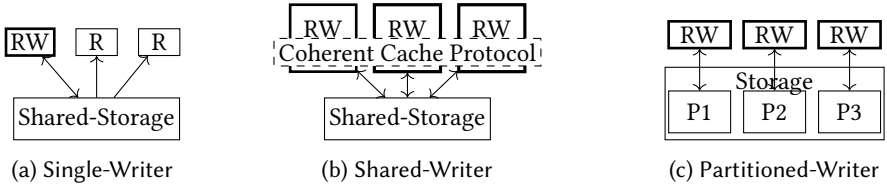
\begin{figure}[htbp]  
    \centering
    \begin{subfigure}[b]{0.26\linewidth}  
        \centering
        \begin{tikzpicture}[scale=0.55, transform shape, font=\huge]
            \node (storage) [draw, rectangle, minimum width=3cm, minimum height=1cm] {Shared-Storage};
            \node (rNode1) [draw, rectangle, above=1cm of storage, minimum width=1cm] {R};
            \node (rwNode) [draw, rectangle, left=0.5cm of rNode1, minimum width=1cm, line width=1pt] {RW};
            \node (rNode2) [draw, rectangle, right=0.5cm of rNode1, minimum width=1cm] {R};
            \draw[<->] (rwNode) -- (storage);
            \draw[<-] (rNode1) -- (storage);
            \draw[<-] (rNode2) -- (storage);
        \end{tikzpicture}
        \caption{Single-Writer}
        \label{fig:moti_f1_a}
    \end{subfigure}
    \begin{subfigure}[b]{0.33\linewidth}  
        \centering
        \begin{tikzpicture}[scale=0.55, transform shape, font=\huge]
            \node [draw, rectangle, minimum width=1.6cm, minimum height=1.6cm, text height=0.05cm, text depth=0.1cm, line width=1pt] (rw1) {RW};
            \node [draw, rectangle, right=0.4cm of rw1, minimum width=1.6cm, minimum height=1.6cm, text height=0.05cm, text depth=0.1cm, line width=1.2pt] (rw2) {RW};
            \node [draw, rectangle, right=0.4cm of rw2, minimum width=1.6cm, minimum height=1.6cm, text height=0.05cm, text depth=0.1cm, line width=1.2pt] (rw3) {RW};
            \node [draw, rectangle, below=0.6cm of rw2, minimum width=3cm, minimum height=1cm] (storage) {Shared-Storage};
            \node [draw, dashed, rectangle, fill=white, fill opacity=1, above=0.75cm of storage, minimum width=5cm, minimum height=0.5cm] (protocol) {Coherent Cache Protocol};
            
            \draw[<->] (storage) -- (rw1);
            \draw[<->] (storage) -- (rw2);
            \draw[<->] (storage) -- (rw3);
        \end{tikzpicture}
        \caption{Shared-Writer}
        \label{fig:moti_f1_b}
    \end{subfigure}
    \begin{subfigure}[b]{0.36\linewidth}  
        \centering
        \begin{tikzpicture}[scale=0.55, transform shape, font=\huge]
            \node (partition2) [draw, rectangle, minimum width=1.5cm, minimum height=1cm] {P2};
            \node (partition1) [draw, rectangle, left=0.3cm of partition2, minimum width=1.5cm, minimum height=1cm] {P1};
            \node (partition3) [draw, rectangle, right=0.3cm of partition2, minimum width=1.5cm, minimum height=1cm] {P3};
            \node (storage) [draw, rectangle, minimum width=5.3cm, minimum height=1.7cm, anchor=center, yshift=0.2cm] at (partition2) {\raisebox{1.1cm}{Storage}};
            \node (rwNode1) [draw, rectangle, above=1cm of partition1, minimum width=1.5cm, line width=1pt] {RW};
            \node (rwNode2) [draw, rectangle, above=1cm of partition2, minimum width=1.5cm, line width=1pt] {RW};
            \node (rwNode3) [draw, rectangle, above=1cm of partition3, minimum width=1.5cm, line width=1pt] {RW};
            \draw[<->] (rwNode1) -- (partition1);
            \draw[<->] (rwNode2) -- (partition2);
            \draw[<->] (rwNode3) -- (partition3);
        \end{tikzpicture}
        \caption{Partitioned-Writer}
        \label{fig:moti_f1_c}
    \end{subfigure}
    \vspace{-.1in}
    \caption{Three Archetypes of Cloud OLTP Databases}
    \label{fig:moti_f1}
\end{figure}

According to the taxonomy recently proposed by Tobias et al.~\cite{ziegler2023scalable}, contemporary scalable cloud OLTP DBMSs fall into three archetypes, categorized by the distinct data access paths in data planes, as shown in \figref{fig:moti_f1}.

\textbf{Single-Writer} archetype employs a single read-write node (RW-node) with multiple read-only nodes on top of the shared storage and is prevalent in commercial cloud databases~\cite{verbitski2017amazon, cao2020polardb, ScyllaDB, azure_sql_hyperscale,alloydb, neondb,cao2020polardb}. While being simple to manage, this archetype suffers from write-heavy workloads, as the single RW-node prohibits asymptotic scalability~\cite{ziegler2023scalable}---a concept that measures how well a system scales w.r.t the number of nodes for basic data access operations when increasing the load in the system. \textbf{Shared-Writer} archetype allows multiple RW-nodes to modify the same set of data items on the shared-storage, hence improving write scalability. Because data must be kept consistent across multiple nodes, these systems need to address the buffer-cache coherence problem~\cite{mohan1991recovery, rahm1989recovery}. However, due to the complexity of maintaining consistency and cache coherence, this approach is limited to a few commercial shared-disk systems~\cite{OracleRAC2022, IBMDB2DataSharing} and is primarily explored in research prototypes~\cite{zamanian2016end, ziegler2022scalestore, bernstein2011hyder}. \textbf{Partitioned-Writer} archetype partitions the database across multiple RW-nodes, each responsible for writes on an exclusive subset of data. This approach is widely used in modern commercial DBMSs~\cite{cosmosdb,zhou2021foundationdb,huang2020tidb,yugabytedb,Cockroach,elhemali2022dynamodb,corbett12spanner} as it offers asymptotic scalability of both reads and writes and reduces the complexity of Shared-Writer architectures. These databases typically use the two-phase commit protocol to ensure the atomicity of distributed transactions. Partitioned-Writer can be achieved either on local storage~\cite{Cockroach,yang2022oceanbase} or disaggregated storage~\cite{cosmosdb,zhou2021foundationdb}, with the latter being more prevalent in cloud-native databases. The storage choice drives two key differences: \circleb{1} Disaggregated storage allows independent scaling and billing of compute and storage resources, while local storage tightly binds compute and storage to the same node. \circleb{2} With disaggregated storage, RW-nodes need no replication because they are stateless and the storage layer is highly available. In contrast, local storage systems must replicate nodes for fault tolerance since compute and storage are coupled. We use Partitioned-Writer archetype with storage disaggregation as the target cloud OLTP DBMS for \system because it fully achieves asymptotic scalability and is widely used in commercial cloud databases.

\subsection{Cluster Coordination in Cloud DBMS}
\label{ssec:metamanagement}

\begin{figure}[htbp]

\begin{center}
\begin{tikzpicture}[scale=0.11]    
    \draw[-stealth, thick] (-20, 0) -- (20, 0);
    \draw[-stealth, thick] (0, 11) -- (0, -9);
    
    \node[below left] at (43, 2.5) {Disaggregated};
    \node[below right] at (-38, 2.5) {\convergedL};
    \node[right] at (-7, 14) {External};
    \node[right] at (-8, -12) {Integrated};

  \node[align=center] at (-13, 6) {\textit{Zookeeper}\\\textit{Etcd / Chubby}\\\textit{DDBMS}};
    \node[align=center] at (-13, -6) { \textit{KRaft}\\\textit{CockroachDB}};
    \node[align=center] at (13, 6) {\textit{FoundationDB}\\\textit{FaaSKeeper}};
  \node[align=center] at (13, -6) {\textit{Marlin}};

    \draw[dashed, gray] (-4.5, 0) -- (4.5, 0);
    \draw[dashed, gray] (0, -4.5) -- (0, 4.5);
\end{tikzpicture}
\figcolor{}
\setlength{\abovecaptionskip}{-0.01in}
\caption{\edit{Solution Quadrants for Cluster Coordination}}

\label{fig:moti_quadrant}
\end{center}

\end{figure}
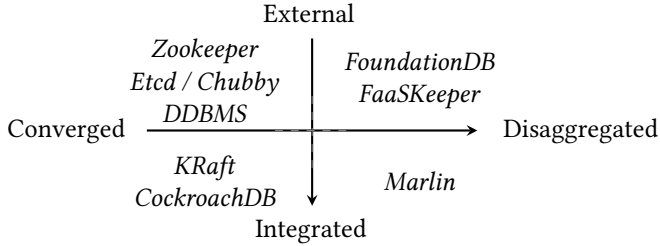

\edit{We categorize coordination solutions in existing cloud DBMSs into a quadrant chart in ~\figref{fig:moti_quadrant} by two dimensions: \textcircled{1} whether the coordination is integrated into the base system, and \textcircled{2} whether the coordination service embraces storage disaggregation.}

\reviseBegin

\reviseEnd

\edit{As shown in the top left quadrant, although many cloud databases implement their data planes following the storage disaggregation architecture, many of these  
databases~\cite{huang2020tidb, clickhouse,corbett12spanner,chang2008bigtable, kafka,lakshman2010cassandra,yang2014druid, neo4j} continue to rely on \textbf{external} and \textbf{\converged} coordination services such as ZooKeeper~\cite{hunt2010zookeeper}, Etcd~\cite{etcd}, or Chubby~\cite{burrows2006chubby} for their control planes.} However, in cloud environments, as cluster sizes and data volumes grow, metadata size expands rapidly\footnote{In this paper, "coordination states" and "metadata" are used interchangeably.}. This rising demand for elasticity necessitates frequent rescaling, rebalancing, and dynamic repartitioning, further increasing the amount of accesses to coordination states. For instance, a database with 1 PB of user data with 32 MB partitions would require approximately 32 GB metadata to map partitions to owner nodes, assuming each map entry consumes 1 KB. The size of metadata grows even more with the adoption of advanced data partitioning strategies, such as fine-grained or dynamic partitioning~\cite{curino2010schism, serafini2016clay,taft2014store, prasaad2020handling, le2019dynastar, liroz2013dynamic}. 

\begin{figure}[htbp]

        \begin{subfigure}[b]{0.7\linewidth}%
            \center
            \includegraphics[scale=0.8, height=3cm,keepaspectratio]{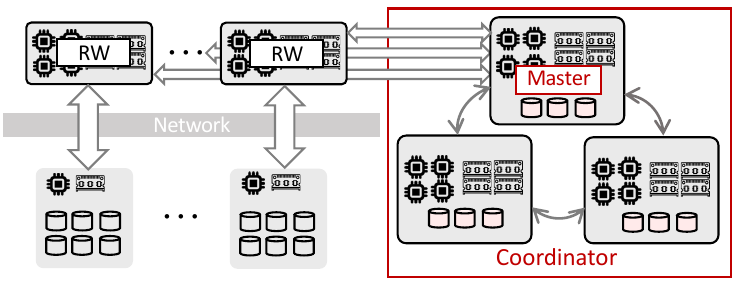}     
            \caption{Cloud DBMS with An External and Converged Coordination Service}
            \label{fig:moti_f2_a}
        \end{subfigure}%
    \begin{subfigure}[b]{0.3\linewidth}%
    	\center
        \includegraphics[scale=0.8, height=3cm,keepaspectratio]{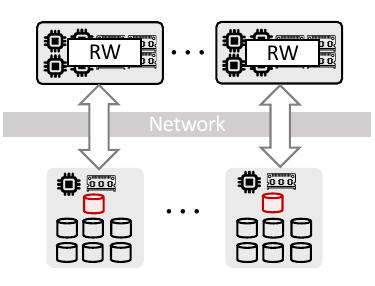}         
        \caption{Cloud DBMS with \system}
        \label{fig:moti_f2_b}
    \end{subfigure}%

    \vspace{-.12in}
    \caption{\textbf{Cluster Coordination in Partitioned-Writer DBMSs.} --- \normalfont{Using external, centralized, and converged coordination services present limitations: (1) scalability bottleneck (2) low cost-efficiency, and (3) increased operational burdens.}}

    \label{fig:moti_f2}
\end{figure}

As exemplified in \figref{fig:moti_f2_a}, existing external and \converged coordination services typically take a centralized (single-master) architecture. This design can cause scalability issues as all coordination requests must pass through the master node. Moreover, maintaining an external coordination service for bursty coordination requests is not cost-efficient and adds operational burdens.

\edit{
Previous works have explored ways to alleviate scalability and cost issues in external coordination solutions. To improve scalability, Snowflake \cite{snowflake} and ByConity \cite{byconity} replace the single-master coordinator with FoundationDB \cite{zhou2021foundationdb}, thereby achieving scalability through data partitioning. To reduce cost, systems like FaaSKeeper \cite{FaaSKeeper} run the coordination services on serverless functions and disaggregate the compute and storage resoruces. However, these solutions still  still rely on external coordination services and fail to fully resolve the three key limitations outlined in
~\secref{sec:intro}. Specifically, FoundationDB cannot automatically scale up and down as coordination loads fluctuate. It also incurs additional costs for maintaining an external coordination service and introduces external dependencies and operational burdens. FaaSKeeper does not address scalability and suffers from limitations inherent to serverless platforms, such as cold-start latency, unpredictable performance, and the lack of persistent storage, which are unsuitable for coordination tasks.}

\edit{As shown in the bottom left quadrant of ~\figref{fig:moti_quadrant}}, recent works like CockroachDB and KRaft~\cite{Cockroach,yang2022oceanbase,kraft} try to solve the three limitations by \textbf{integrating} the coordination into the database's internal infrastructure. For instance, CockroachDB~\cite{Cockroach} partitions metadata into ranges and employs a two-level structure for efficient key lookups. The first level stores metadata ranges and acts as the "metadata for metadata" and the second level addresses user data. Each metadata range is managed by a Raft group. 
\edit{The fundamental difference between our work and CockroachDB-like solutions is the \emph{architectural assumption} of the database system: whether it employs \textbf{storage disaggregation}. Systems like CockroachDB assume a \converged architecture where compute and storage are coupled on the same node, which is replicated through consensus protocols. In a storage disaggregation architecture, however, only the storage layer has built-in replication but the compute nodes are stateless and not replicated. 
Therefore, when a compute node fails, the coordination mechanism must carefully handle the coordination states through interactions between the remaining compute nodes and the storage layer---a challenge that does not exist in converged systems.}

In this paper, we aim to address prior limitations of conventional coordination services by developing a cloud-native coordination mechanism suitable for storage-disaggregated databases.

\section{System Model}
\label{sec:sysmodel}




\begin{figure}[t]

    \centering
    \includegraphics[width=0.6\linewidth]{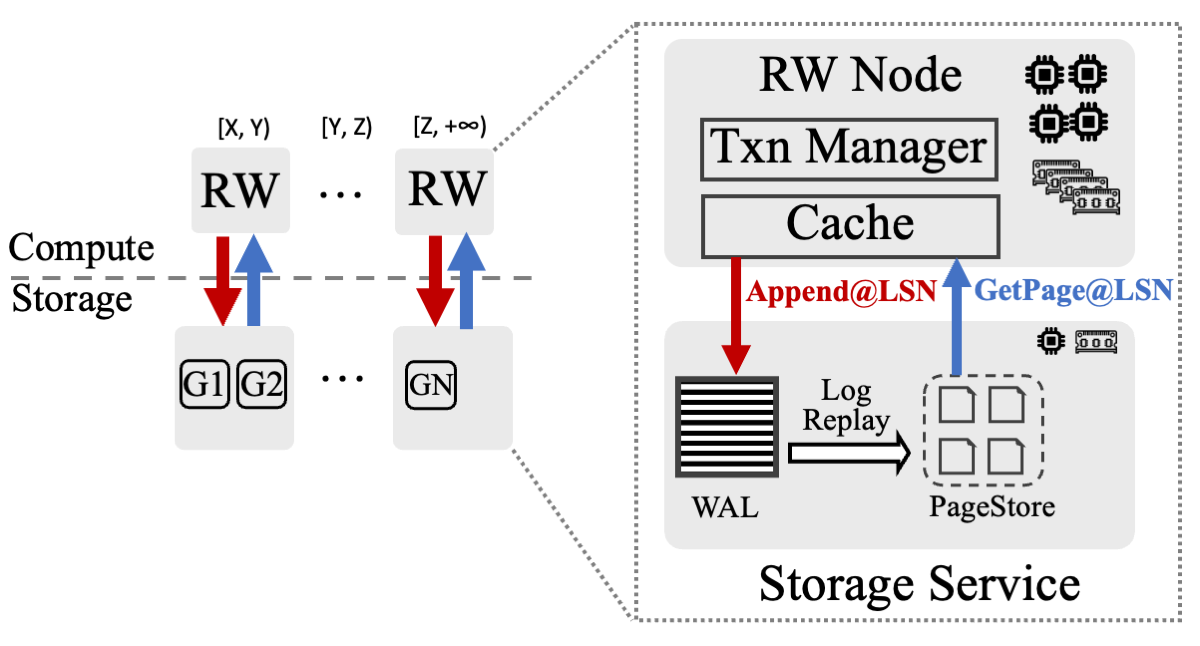}
    \vspace{-.1in}
    \caption{Reference Cloud Database. \normalfont{It follows key design principles~\cite{pang2024understanding, ziegler2023scalable} of modern scalable cloud-native databases: \normalfont{\protect\circleb{1} Storage disaggregation, \protect\circleb{2} Partitioned-Writer archetype, and \protect\circleb{3} Log-as-the-database paradigm.}}}

    \label{fig:arch_f1}
\end{figure}

Our target database architecture, illustrated in \figref{fig:arch_f1}, adopts the Partitioned-Writer archetype with storage disaggregation. The storage layer provides reliable data storage with limited near-storage compute capability, and the compute layer performs transaction processing and data access. Application data is partitioned across the cluster, with each partition assigned to a single compute node. For clarity, this section focuses on the scalable, cloud-native OLTP architecture without discussing the coordination mechanism.


\vspace{-.05 in}
\subsection{Storage Layer}
The disaggregated storage layer provides highly available and fault-tolerant storage for the persistent image of the database. As shown in \figref{fig:arch_f1} (right), the system follows the log-as-the-database (LogDB) paradigm \cite{pang2024understanding}, a design principle prevalent in storage-disaggregated databases~\cite{verbitski2017amazon,socrates,alloydb,depoutovitch2020taurus}. The compute nodes append updates to write-ahead logs (WAL) on the storage layer upon transaction commit, establishing these logs as the ground truth of the database. 
The storage materializes WAL into the data pages asynchronously through the log replay service, eliminating the need to write back dirty pages from compute nodes. In prevalent LogDBs~\cite{socrates,verbitski2017amazon,pang2024understanding, depoutovitch2020taurus}, the storage service provides two standard APIs: (1) \textit{Append(updates)} to append updates to the tail of WAL, and (2) \textit{GetPage(pageId, LSN)}, denoted \textit{GetPage@LSN}~\cite{socrates}, to retrieve a page from page stores that has applied all updates up to the expected log sequential number (LSN). Leveraging the near-storage computation capabilities (i.e., Compare-And-Swap), \system enhances the standard append API into \textit{Append(updates, LSN)}, a conditional append operation that can succeed only if the log end has that particular LSN at the time of append. 
We mark this operation as \textit{Append@LSN} and use it to enable coordination, which will be further discussed in ~\secref{subsec:cc}.



\vspace{-.05 in}
\subsection{Compute Layer}
Each compute node can service both reads and writes. 
The compute nodes are stateless and are not replicated. On the write path, an update is sent to the WAL using \textit{Append@LSN} upon transaction commit, and then updates its local cache. If the cache needs to evict a dirty page, the node simply deletes it without sending it to the storage layer. The compute node also tracks the highest LSN committed to the WAL, denoted \textit{H-LSN}. On the read path, the compute node first attempts to retrieve the page from its local cache. If there is a cache miss, it fetches the evicted page from the page store using \textit{GetPage@LSN} for the newest version \textit{H-LSN}. 
\section{\system}
\label{sec:cluster_management}
In contrast to relying on an external, converged coordination service for orchestrating the cloud DBMS (\figref{fig:moti_f2_a}), \system integrates coordination functionalities into the database itself to remove the external dependency.
This section presents the core design of \system. \secref{subsec:meta} \edit{discusses some candidate design options to achieve this integration idea} and proposes \system's solution to store and maintain coordination states. \secref{ssec:reconfigrationtxn} and \secref{subsec:cc} explain how \system accesses and modifies these states to ensure consistent coordination. \secref{subsec:app_examples} provides representative examples, including scale-out and node failover, to illustrate how \system achieves cluster management for cloud OTLP databases. \edit{\secref{subsec:invariants} summarizes the correctness invariants}.

\subsection{Coordination State Management}
\label{subsec:meta}

\reviseBegin
One way to integrate coordination into the cloud database is to store coordination states in a single-partition user table, managed exclusively by one compute node. While easy to implement, 
the owner node of the coordination states becomes a scalability bottleneck. Moreover, this exclusive access model introduces a "failover deadlock": if the metadata owner node fails, other nodes cannot safely update the critical metadata to promote a new owner. This stalls the failover process and leads to system-wide unavailability. To ensure safe failover, either an external coordination service or a specialized coordination protocol is required.

An alternative design is to manage coordination states as a partitioned user table: states are partitioned and stored in the disaggregated storage layer, with each partition managed exclusively by an owner compute node. This design requires a two-level metadata structure, where a top-level metadata partition is used to locate lower-level metadata partitions. While mitigating the scalability issue, it still suffers from the same failover problem.
\reviseEnd

\begin{figure}[t]

    \centering
    \includegraphics[width=0.7\linewidth]{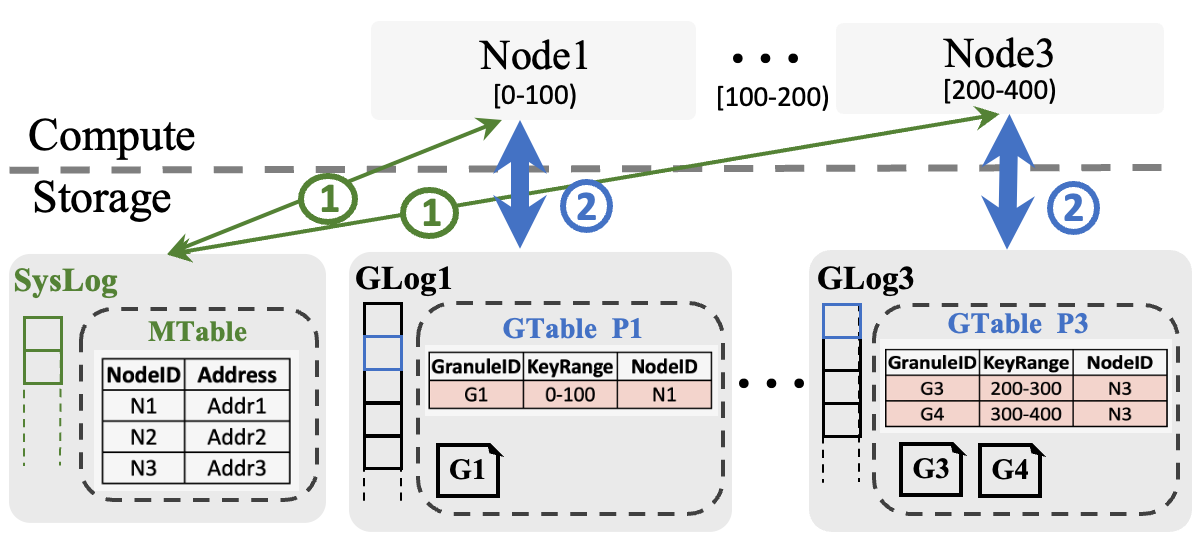}
    \vspace{-.1in}
    \caption{System Tables (\mtbl and \gtbl)}

    \label{fig:systbl_f1}
\end{figure}

\edit{To address this failover issue without relying on external coordination services, \system manages critical coordination states using special system tables that allow multiple compute nodes to read and update when necessary.} Specifically, \system identifies and organizes two types of coordination states 
into special system tables: (1) \textit{Group membership}, which lists the operational compute nodes in the cluster, is stored in the \textit{membership table} (\mtbl), and (2) \textit{Data ownership}, the mapping between data partitions and their owner compute nodes, is stored in the \textit{granule ownership table} (\gtbl). Granules are essentially fine-grained partitions with a small size (64 KB in the current implementation). \figref{fig:systbl_f1} illustrates schemas of these system tables. Each row in \mtbl contains a node ID and the corresponding server address, while each row in \gtbl specifies a granule ID, its key range, and the ID of the compute node that owns it. Because these states are critical for coordinating node failover, they must remain available for both read and write operations under arbitrary compute node failures.

\mcomment{\gzh{change \syslog to MLog to make it less confusing?}}

\mtbl is typically small in size and remains unpartitioned. All modifications to it are recorded in a single log, \syslog, as illustrated in \figref{fig:systbl_f1}. To prevent the failover problem, \syslog has no exclusive owner, allowing all compute nodes to access and modify it (\textcircled{1}). In contrast, \gtbl grows linearly with application data volume and may experience bursty access during large-scale cluster reconfigurations. Therefore, \system partitions \gtbl by the granule owner node ID to improve scalability. 
Each \gtbl partition is managed by the same node as the application data granules it describes, 
\edit{resulting in a \textbf{flat} metadata design and} obviating the need for a two-level “metadata-for-metadata” structure. 
\gtbl keeps that locality property even after reconfigurations since it is automatically and spontaneously repartitioned through migrations. Under normal circumstances, each \gtbl partition is accessed exclusively by its owner node (\textcircled{2}), and modifications are recorded in a per-node WAL (\glog). If the owner node fails, \system allows other compute nodes to access the partition, allowing the failover to progress smoothly. Other coordination states unrelated to node failover—such as security keys, ACLs, permissions, monitoring metrics, and user profiles—are managed by \system similarly to application data. \system organizes these states as regular partitioned tables, with each partition exclusively accessed by its owner compute node at all times. 



\subsection{Coordination Via Transactions}
\label{ssec:reconfigrationtxn}

\system must ensure data consistency under autoscaling context where system reconfigurations (e.g., scaling and load balancing), user transactions, and compute node failures can occur concurrently. Each of these activities involves multi-step operations that access or modify application and coordination states and should be performed together as a single logical unit. Therefore, \system manages all state access, including coordination and application states, through transactions. ACID properties are guaranteed for all transactions, which will be discussed in ~\secref{subsec:cc}. This section focuses on how to compose transactions on system tables and user tables to support consistent coordination.

\system provides two types of transactions. \textit{User transactions} are regular user-defined transactions. They can run interactively or as stored procedures. \textit{Reconfiguration transactions} handle operations on coordination states (i.e., system tables) and are predefined stored procedures that take parameters.

\begin{table}[h]
        \small
        \begin{tabular}{|p{2.4cm}|p{0.6\linewidth}|}
        \hline
    
        \textbf{Transaction} & \textbf{Description} \\
        \hline
        \textbf{AddNodeTxn} & Add a new node into the cluster \\
        \hline
        \textbf{DeleteNodeTxn} & Remove an existing node from the cluster \\
        \hline
        \textbf{MigrationTxn} & Migrate one or a list of granules from a source node to a destination node \\
        \hline
        \textbf{RecoveryMigrTxn} & Migrate one or a list of granules from an unresponsive source node to a destination node \\ 
        \hline
        \textbf{ScanGTableTxn} & Query the data ownership of the full cluster\\
        \hline
        \end{tabular}
        \vspace{0.05in}
        \caption{Reconfiguration Transactions}
        \label{tab:systbl}
\end{table}


As shown in~\tabref{tab:systbl}, five types of reconfiguration transactions serve as the core building blocks for common autoscaling scenarios, including scale-out/in, load balancing, and failover. The first two modify \mtbl, while the remaining three operate on \gtbl. All transactions in \system are detailed in~\algref{alg:cluster_c2}. \edit{RPC requests to peer nodes are expressed using the notation: {\(\textit{result} \leftarrow \mathit{RPC}_{\mathit{sync/async}}^{\mathit{node}}{::}\mathit{Request()}\)}, where the superscript \(\mathit{node}\) indicates the target node and the subscript \(\mathit{sync/async}\) denotes whether the RPC is synchronous or asynchronous.}



\mcomment{\gzh{clarify the meaning of inputs to MarlinCommit: node name means the node's DB WAL log, node.\glog means the node's \glog, \syslog means \syslog}}

\begin{algorithm}[htbp]
	\small 
    \SetKwProg{myfun}{Function}{}{}

    \nonl\texttt{\\}
    
    \myfun{\textbf{UserTxnRequest(key)}}{ 
      \nonl \codeComment{Acquire a read lock on GTable for the granule if using 2PL} \\
      \shl{\textit{nid  $\leftarrow$ GTable[getGranuleID(key)].NodeID}}\\
      \uIf{\shl{cur\_node is nid}} {
             \textit{executeUserRequest() } \\
      }
      \Else{\textit{\shl{\textbf{Abort} and} \Return \shl{WrongNodeError(nid)}} }
    }
    
    \nonl\texttt{\\}
    {\color{gray} \hrule}
    \nonl\texttt{\\}
    
    \myfun{\textbf{AddNodeTxn(new\_node)}}{
      \uIf{MTable.exist(new\_node.NodeID)} {
          \Return{NodeAlreadyExistError} \\
      }
      \Else{
        \textit{MTable.add(new\_node.NodeID, new\_node.Address)}\\ 
        \Return \textit{\mcommit(\{\syslog\})} \\
      }
    }

    \nonl\texttt{\\}

    \myfun{\textbf{DeleteNodeTxn(node\_to\_delete)}}{  
    \uIf{MTable.exist(node\_to\_delete.NodeID)} {
      \textit{MTable.delete(node\_to\_delete.NodeID)}\\ 
      \Return \textit{\mcommit(\{\syslog\})} \\ 
    }
    \Else{
      \Return{NodeNotExistError} \\
    }  
    }

    \nonl\texttt{\\}

    \myfun{\textbf{MigrationTxn(granule, src, dst)}}{ 
      \nonl \codeComment{Assume the transaction is triggered on the destination node} \\
      \textit{{nid $\leftarrow$ RPC$^{\textit{src}}_{\textit{sync}}$::GTable[granule].NodeID}}\\ 
      \uIf{nid is src} {
        \textit{{RPC$_{\textit{async}}^{\textit{src}}$::GTable[granule].NodeID = dst}}\\ 
        \textit{GTable[granule].NodeID = dst}\\
        \Return \textit{\mcommit(\{src, dst\}) } \\
      }
      \Else{\textit{\textbf{Abort} and \Return WrongNodeError(nid)} }
    }
    
    \nonl\texttt{\\}
    
    \myfun{\textbf{RecoveryMigrTxn(granule, src, dst)}}{ 
      \nonl \codeComment{The transaction is triggered on the destination node} \\
      \textit{{nid $\leftarrow$ GTable[granule].NodeID}}\\ 
      \If{nid is src} {
       \textit{GTable[granule].NodeID = dst}\\
       }
       \Return \textit{\mcommit(\{src.GLog, dst\}) } \\
    }
    
    \nonl\texttt{\\}

    \myfun{\textbf{ScanGTableTxn()}}{ 
      \textit{nodes $\leftarrow$ MTable.scan()} \\
      \textit{scan\_res $\leftarrow$ GTable.scan() asynchronously} \\
      \For{n in nodes} {
        \If{n is not cur\_node} {
          \textit{scan\_res.merge(RPC$_{\textit{async}}^{\textit{n}}$::GTable.scan())} \\
        }
      }
      \Return \textit{\mcommit(\{\syslog\} $\cup$ nodes) } \\
    }

    \nonl\texttt{\\}

    \caption{\textbf{User and Reconfiguration Transactions} --- Difference between \system \textit{UserTxn} and standard \textit{UserTxn} is highlighted in \shl{gray}. \mcomment{\gzh{(reordered to make it read better)}}}
     

    \label{alg:cluster_c2}
\end{algorithm}


\edit{The first four reconfiguration transactions are designed to modify system tables.} \textit{AddNodeTxn} is executed on the new node to be added and committed to \syslog. \textit{DeleteNodeTxn} is executed on the node to be deleted or on the node detecting a failure. \textit{MigrationTxn} is a cross-node transaction that is executed on both the source (\textit{src}) and destination (\textit{dst}) nodes involved in a granule migration. \textit{RecoveryMigrTxn} is a single-node transaction that is executed only on \textit{dst} node for the migration. However, it commits on \glog of both \textit{src} and \textit{dst} nodes. It is the key component to cope with failover. These reconfiguration transactions are composed of three main steps:

\begin{enumerate}[label=(\arabic*), leftmargin=12pt, rightmargin=0pt, noitemsep, topsep=0pt]
    \item \textbf{Check the data effectiveness} (lines 8, 14, 20-21, 28-29): Verify the system tables to ensure the database is in a valid state for the current reconfiguration. This step prevents data corruption during concurrent reconfigurations. For instance, \textit{DeleteNodeTxn} validates that the target node remains a member of the cluster before proceeding with deletion. Note that user transactions also incorporates this verification to ensure that data ownership is not altered by concurrent reconfiguration transactions. Before accessing application data, each \textit{UserTxnRequest} confirms that the granule to be accessed is owned by the current node by checking the \gtbl (lines 2-3).
    \item \textbf{Modify configuration states} (lines 11, 15, 22-23, 30): Update the corresponding system tables to reflect the desired changes.
    \item \textbf{Commit the transaction} (lines 12, 16, 24, 31): Commit the updates to the logs associated with the system tables to ensure durability. \system utilizes \mcommit to ensure atomic commit which will be discussed shortly.
\end{enumerate}




\reviseBegin
The last reconfiguraion transaction is a read-only \textit{ScanGTableTxn}, which serves routing purposes. It is triggered by routers to locate partition owners. It is a distributed transaction executed across all compute nodes to perform a full scan of the \gtbl. It will not cause much performance overhead as it does not need to be invoked per request, and routers can cache the scan results. Cache staleness in routers does not compromise system correctness, as \system ensures each compute node maintains the ground truth for its owned \gtbl partition. Consequently, if a request is misrouted due to stale routing information, the receiving node can detect that it no longer owns the granule (lines 2-3 and 6) and redirect the request to the correct owner. Moreover, compute nodes can periodically broadcast updates of their owned \gtbl partitions to routers, thereby reducing redirections.

Note that \system's correctness is independent of the isolation level used for user transactions. 
While user transactions must hold read locks on the corresponding \gtbl entries until they commit or abort, they can access user tables and operate at weaker isolation levels. For instance, if a user transaction employs 
Read Committed, the transaction could release a read lock on a user table right after the read. The long read lock on the \gtbl is orthogonal to the user table locks. 
\reviseEnd



\subsection{Consistency via \mcommit}
\label{subsec:cc}
Given that application data is strictly partitioned with each partition exclusively accessed by its owner compute node, conventional transaction protocols—such as concurrency control (e.g., 2PL~\cite{eswaran1976notions}, OCC~\cite{kung1981optimistic}) and atomic commit (e.g., 2PC)—are sufficient to ensure ACID. \system leverages existing protocols in reference databases to maintain data consistency within each compute node.

However, relying solely on traditional protocols is insufficient in \system, where their core assumption—exclusive control of data by a single node—does not always hold. While \system minimizes cross-node consistency overhead by partitioning most states, some critical types of coordination states (i.e., \mtbl and \gtbl) must be accessible by multiple compute nodes to address the failover problem (recall~\secref{subsec:meta}). For instance, transactions modifying \mtbl\xspace (e.g., \textit{AddNodeTxn} and \textit{DeleteNodeTxn}) can execute concurrently on different nodes. Additionally, during compute node failures or network partitions, \gtbl partitions owned by failed or unhealthy nodes may be accessed simultaneously by \textit{RecoveryMigrTxn} on multiple recovering nodes and by user transactions on the previously unhealthy node when it returns to normal.

This problem calls for additional mechanisms to ensure data consistency when the same data can be accessed by multiple nodes. \system addresses this challenge by \textit{MarlinCommit}, an optimized atomic commit protocol that commits a transaction only if no cross-node modification has occurred since each node’s last observed commit. Our key insight is to leverage a conditional append API provided by the log to detect whether any other node has appended to the log since the node’s last append. We first introduce the required Log API, then explain \textit{MarlinCommit} protocol in detail.



\begin{algorithm}[htbp]
	\small 
    \SetKwProg{myfun}{Function}{}{}

    \myfun{\textbf{MarlinCommit(participants)}} {  
      \edit{\textit{updates $\leftarrow$ writes made in the txn to each participant}} \\
      \uIf{len(participants) $=$ 1}{
        \nonl \codeComment{One-phase Commit if only one participant is involved} \\
        \Return{\shl{TryLog(participants[0], updates[participants[0]])}} \\
      }
      \Else{
        \nonl \codeComment{Two-phase Commit if multiple participants are involved} \\
        \For{p in participants} {
          \uIf{\shl{p is a log instance}} {
            \shl{\textit{TryLog(p, \{\texttt{VOTE-YES}\} $\cup$ updates[p])} asynchronously}\\       
          } 
          \Else {

            \nonl \codeComment{ \shl{Node p votes and runs \textit{TryLog}}} \\
            \textbf{send} \texttt{VOTE-REQ} to $p$ asynchronously \\
          }
        }
        \Return \textit{final decision according to responses of participants} \\
        \textit{while broadcasting decision to all participant asynchronously}   
      } 
    }

   {\color{gray} \hrule}
    \nonl\texttt{\\}

    \myfun{\textbf{TryLog(log, updates)}} {  
    \textit{(status, new\_lsn) $\leftarrow$ RPC$_{\textit{sync}}^{\textit{log}}$::Append(updates, lsn\_tracker[log])} \\
      \uIf{ status is \texttt{FAILURE}}{
        \textit{ClearMetaCache(log)} \\
        \textit{lsn\_tracker[log] $\leftarrow$ new\_lsn} \\
        \Return{ABORT} \\     
      } 
      \Else{
        \textit{lsn\_tracker[log] $\leftarrow$ new\_lsn} \\
        \Return{COMMIT} \\
      }
    }





    \caption{\textbf{MarlinCommit} --- Difference between \textit{\mcommit} and standard commit is highlighted in \shl{gray}.}
     

    \label{alg:c2}

\end{algorithm}

\subsubsection{Log APIs.}
\label{sssec:logapi}

Conventional commit protocols use a basic Log API, \textit{Append(updates)}, which simply appends transaction updates to the log’s tail. \system takes advantage of an enhanced API of \textit{Append(updates, LSN)}, a conditional append operation that succeeds only if the current log version matches an expected version. For convenience, we call it \textit{Append@LSN}, which is implemented as a remote procedure call (RPC) to the disaggregated storage service by leveraging near-storage computation capabilities. Formally, \textit{Append@LSN} is denoted as:
\vspace{-0.1in}

\[
    \textbf{(status, new\_lsn)  $\leftarrow$  \textit{RPC}$_{\textit{sync/async}}^\textit{log}$::\textit{Append(updates, target\_lsn)}}
\]
\vspace{-0.1in}

\mcomment{\gzh{the RPC notation appeared also in Algo 1 but is first introduced only at here -- may need to define it earlier}
\gzh{RPC targets can (and often will) be compute nodes, right?}}


\edit{The superscript \(\textit{log}\) identifies the target log instance; choices include per-node \glog (denoted by node.\glog) or the global \syslog.} \textit{Append@LSN} succeeds only if the log’s current LSN matches \(\textit{target\_lsn}\). Each compute node maintains a local \(\textit{H-LSN}\), which is the highest LSN it has successfully appended, and
uses \(\textit{H-LSN}\) as the \(\textit{target\_lsn}\) when issuing \textit{Append@LSN}. In cases where the target log is exclusively accessed by the node, \textit{H-LSN} will align with the newest LSN of the log, allowing the operation to succeed. However, if another node has updated the log, the operation will fail because the current LSN of the log will exceed the \( \textit{target\_lsn} \). In such scenarios, the newest LSN is returned to the caller, enabling it to retry the operation with an updated \( \textit{target\_lsn} \) if desired.

\textit{Append@LSN} can be implemented on any storage system supporting compare-and-swap (CAS), a feature commonly supported by modern cloud-native storage services~\cite{AzureBlobStorage, azuretablestorage, chang2008bigtable, cosmosdb, elhemali2022dynamodb, Redis2009}. For instance, many cloud storage services can provide CAS using entity tags (ETags) and conditional headers such as \textit{If-Match}~\cite{azure_etag, s3_etag, gcs_etag}. When a client issues an \textit{Append@LSN} request, the storage service checks whether the ETag on the log matches the ETag provided by the client. If they match, the updates are appended to the log and a new ETag is returned; otherwise, the operation fails and returns the current ETag instead. \edit{Further details can refer to ~\secref{sec:impl}.}




\subsubsection{MarlinCommit.}
\label{sssec:marlincommit}


Conventional commit protocols (e.g., 1PC, 2PC) focus on providing atomic commit guarantees for transactions. \mcommit modifies conventional protocols to expand their scope: it can ensure strong transactional guarantees even when multiple nodes can access the same data. \mcommit commits a transaction only if no cross-node modification has occurred since each node’s last observed commit.



To achieve this, each node maintains a \textit{lsn\_tracker} array to track the last committed LSN \textit{H-LSN} of each log in the cluster. \textit{\mcommit} leverages the conditional append API \textit{Append@LSN} to commit the transaction only when \textit{H-LSN} tracked by the current node matches the newest LSN of the log. This ensures that no other node has modified the log during the transaction, thereby preventing cross-node inconsistency. The pseudocode in ~\algref{alg:c2} outlines this protocol. 

\textbf{\textit{MarlinCommit(participants).}} After a transaction finishes the execution phase, the coordinator, normally the node that launches the transaction, calls \textit{MarlinCommit(participants)} to start the atomic commit protocol. Depending on the number of participants, the protocol chooses between one-phase commit and two-phase commit (line 3). \textit{\mcommit} generally follows the conventional commit protocol but with two key differences. First, it replaces the conventional \textit{Log()}, which directly appends the updates to the log, with \textit{TryLog()}, which conditionally appends the updates to the log if LSN matches or otherwise invalidates the caches of corresponding coordination states (lines 4, 8). Second, \mcommit does not limit participants to compute nodes; instead, a participant can be either a compute node or a log instance in the disaggregated storage. The rationale is twofold: (1) the log is the ground truth of the system, and (2) voting through a node is semantically identical to appending the vote directly to the log. This design allows a transaction to progress in the commit phase even if compute nodes fail, laying the foundation for failover transactions like \textit{RecoveryMigrTxn} as presented in ~\secref{ssec:reconfigrationtxn}.

\textbf{\textit{TryLog(log, updates).}} This method is the core mechanism for managing the log. It appends the updates via \textit{Append@LSN} (line 15) to the log, which succeeds only if the log’s current LSN matches the node’s last known LSN (\textit{H-LSN}). If successful, it updates \textit{H-LSN} and returns a \texttt{COMMIT} decision (lines 21-22). Otherwise, a cross‐node modification on the same log may have occurred, so the transaction is aborted to prevent inconsistencies. Because the log’s cache may now be outdated, it should be invalidated via \textit{ClearMetaCache(log)} (line 17). Specifically, if the log is \syslog, the \mtbl cache is evicted, whereas for a node.GLog, the corresponding node's GTable cache is evicted. Since only coordination states can encounter cross‐node modification, there is no need to evict the cache for user data. The compute node then updates \textit{lsn\_tracker} based on the latest LSN returned by \textit{Append@LSN} (line 18) and returns an \texttt{ABORT} decision, indicating a transaction retry. The next transaction that encounters a cache miss in system tables will fetch the latest data from the storage layer via \textit{GetPage@LSN}, guided by the updated \textit{H-LSN}.




\reviseBegin
We note that \system does not have the blocking issue~\cite{babaoglu1993nonblocking, bernstein1987concurrency, skeen1981nonblocking} that traditional 2PC protocols suffer from. The state-of-the-art 2PC protocol, Cornus~\cite{cornus}, has solved this blocking issue by leveraging disaggregated storage: it allows an active compute node to commit to the log of an unresponsive participant. This was possible because the disaggregated storage is highly-available and accessible by all compute nodes. We use the similar idea in \textit{MarlinCommit}.  

\reviseEnd


\subsection{Coordination At Work}
\label{subsec:app_examples}
This section demonstrates how system tables and reconfiguration transactions in \system can be composed to achieve cluster management for cloud OTLP databases under typical autoscaling scenarios. We use \textit{scale-out} and \textit{failover} as two examples. Other scenarios such as scale-in and rebalancing can be implemented similarly.

\subsubsection{Scale-Out}
\label{subsec:live-migration}



\begin{figure*}[htbp]
\centering
\vspace{-.1in}

\begin{minipage}{.48\textwidth}

    \includegraphics[width=1.08\linewidth]{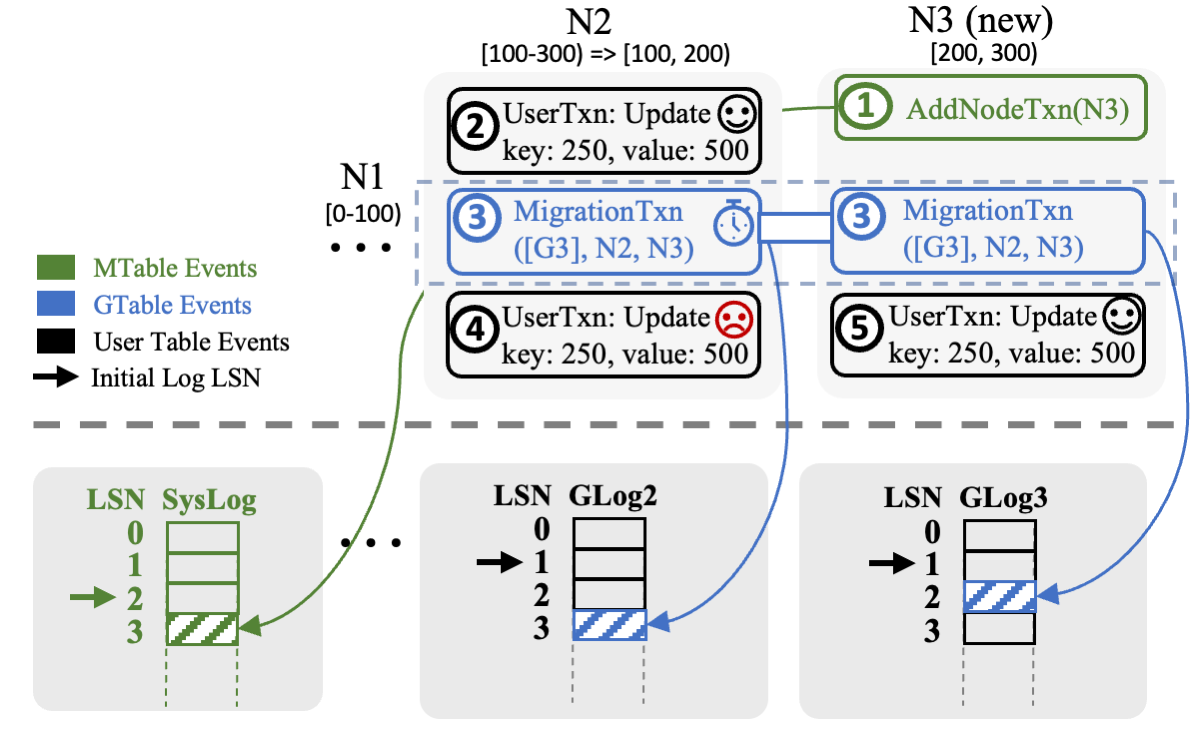}
            \vspace{-0.3in}
    \caption{Scale-Out Example}

    \label{fig:scaleout}
\end{minipage}
\hfill
\begin{minipage}{.48\textwidth}

    \includegraphics[width=1.08\linewidth]{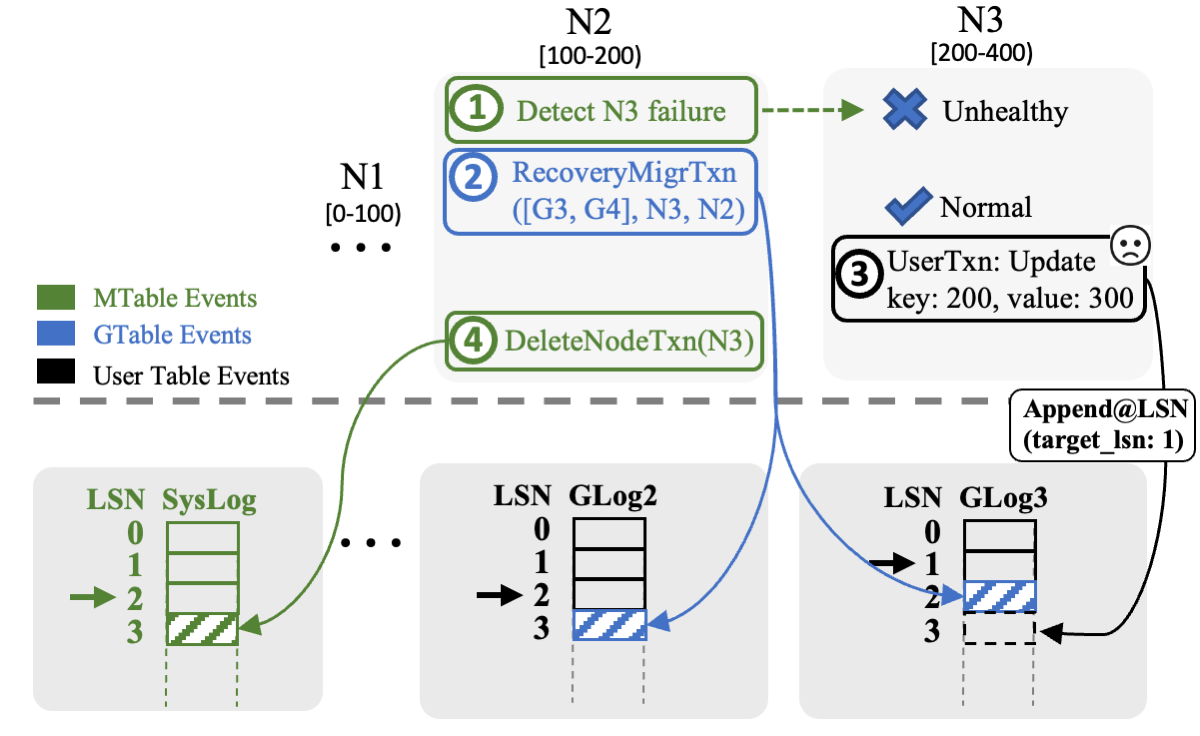}
    \vspace{-.3 in}
    \caption{Failover Example}

    \label{fig:failover}
\end{minipage}
\end{figure*}
When the user workload spikes, cloud databases will add more compute nodes to the cluster and spread the load evenly among the cluster. ~\figref{fig:scaleout} illustrates the scale-out process with a concrete example. \edit{Initially, node N2 manages the range [100, 300), including granules G2 ([100, 200)) and G3 ([200, 300)). After scale-out, a new node N3 takes over G3, reducing N2’s range to [100, 200).} Scale-out can be achieved in two steps: (1) adding the new node to the cluster and (2) rebalancing data from existing nodes to the new node. These steps are demonstrated in \textit{Membership Updates} and \textit{Live Migration}, respectively.

\boldtext{Membership Update.}
The new node triggers an \textit{AddNodeTxn} to add itself to \mtbl (\textcircled{1}). \textit{AddNodeTxn} is a single node transaction that commits on \syslog at LSN 3. If conflicting \textit{AddNodeTxn} and \textit{DeleteNodeTxn} run in parallel, \mcommit ensures that only one transaction is committed. The new node may broadcast this membership change to all nodes after the transaction commits to reduce cache staleness. However, this is not required for correctness, as other nodes will discover this change when issuing their next transaction on \mtbl. \mcommit of the next transaction will discover that \syslog has changed because \textit{Append@LSN} can no longer append to the log according to an outdated \textit{H-LSN} saying 2. The node then invalidates its local cache for \mtbl and retries the transaction by fetching the newest data from the page store in the storage layer.
\mcomment{\gzh{should we briefly talk about how we choose which granules to migrate from where to where?}}

\boldtext{Live Migration.} To redistribute load onto newly added nodes, the database transfers ownership of selected granules via a \textit{MigrationTxn}. As shown in \figref{fig:scaleout}, this transaction updates the relevant \gtbl partitions (P2 and P3) to move ownership of granule G3 from N2 to N3. It is a cross‐node transaction (\textcircled{3}) that commits on both the source (N2) and destination (N3): N2 logs its updates to \glog{}2 for \gtbl P2, while N3 logs its updates to \glog{}3 for \gtbl P3. Although a \textit{MigrationTxn} can originate on either node, it usually starts on the underutilized destination node.

If the source node is running a user transaction, the transactional concurrency control protocol prevents conflicts. For instance, under 2PL, an ongoing user transaction on N2 (\textcircled{2}) holds a write lock on G3, blocking the \textit{MigrationTxn} from acquiring its required write lock until the user transaction commits. After ownership changes, any new user transaction arriving at N2 (\textcircled{4}) aborts because it discovers that N2 no longer owns G3 during data-effectiveness check. The aborted transaction can inform the issuing clients of ownership change, prompting them to redirect subsequent transactions on G3 to N3 (\textcircled{5}). Because \gtbl is partitioned, \textit{MigrationTxn} can run in parallel for different granules on different nodes, thus avoiding the scalability bottleneck of centralized coordination.

After migration, user transactions (\textcircled{5}) on the newly added node may experience performance degradation due to a cold cache. Various live migration protocols have addressed this issue~\cite{elmore2015squall,das2011albatross}. Since \system focuses on coordination mechanisms, it can integrate with compatible protocols targeting similar reference cloud databases described in ~\secref{sec:sysmodel} by replacing their internal cluster coordination service. In our implementation, we mitigates the cold-cache issue by proactively warming up the cache after \textit{MigrationTxn} updates ownership, following an approach similar to Squall~\cite{elmore2015squall}. Specifically, the destination node issues a scan query to the source node and populates its local cache with the scan results for uncached data.

\subsubsection{Failure and Recovery}
\label{subsubsec:group_member}

This section shows how \system supports failover without relying on an external coordination service. We begin by briefly outlining failure detection, and then explain how to achieve consistent failover through a concrete example.

\boldtext{Failure Detection.}
Without a centralized coordination service, \system can use standard decentralized failure detection approaches such as ring-based heartbeating~\cite{bernstein2014orleans,stoica2003chord}, SWIM~\cite{das2002swim}, or gossip-based heartbeating~\cite{ganesh2003peer}. \system adopts a ring-based heartbeating approach akin to Orleans~\cite{bernstein2014orleans, orleans_cluster_management} to monitor node health. Compute nodes in \mtbl form a ring (sorted by node ID) and each node periodically sends heartbeat messages to its \textit{k} successors in the ring. If a successor fails to respond after a configurable number of attempts, the monitoring node assumes the successor has failed and initiates a \textit{Failover} procedure. This protocol can be further optimized to reduce false positives by letting compute nodes record "suspicious" votes for unresponsive nodes in \mtbl. A node is considered dead only when such votes exceed a threshold over a defined interval. Since these optimizations are orthogonal to our coordination mechanism, we leave them for future work.


\boldtext{Failover.}
~\figref{fig:failover} demonstrates how \system achieves failover. Suppose node N3, which initially owns granules G3 and G4, becomes unhealthy due to a temporary slowdown but eventually recovers. The failover process progresses through several key steps:

First, when N2 detects a heartbeat timeout from N3 (\textcircled{1}), it initiates a \textit{RecoveryMigrTxn} to transfer granules G3 and G4 from N3 to N2 by updating \gtbl P2 and P3 (\textcircled{2}). Although executed solely on N2, this transaction commits on both \glog{}2 at LSN 3 and \glog{}3 at LSN 2 through \mcommit. \textit{RecoveryMigrTxn}'s ability to modify coordination states owned by N3 (\gtbl P3) while N3 is unresponsive is key to solving the failover problem - the failover can progress even when the owner of coordination states has failed. However, if without care, data races might happen as the same data could be modified simultaneously by multiple nodes (N2 and N3) if the failed node recovers. \mcommit can ensure correctness even in such race conditions by leveraging \textit{Append@LSN}. 

For example, when N3 recovers and attempts a user transaction to update G3 (\textcircled{3}), the transaction aborts during \mcommit. Specifically, the conditional append operation \textit{Append(updates, H-LSN)} on \glog{}3 fails because the current LSN of \glog{}3 has advanced to 2, while the \textit{H-LSN} tracked by N3 remains at 1, causing a version mismatch. In response, N3 invokes \textit{ClearMetaCache} to invalidate the cache for \gtbl P3. Upon refreshing its local cache of \gtbl P3, triggered by the next cache miss, N3 detects that it no longer owns G3 and G4. Any ongoing or incoming transactions on N3 targeting these granules are thus aborted, preserving data integrity.

After migration, N2 runs \textit{DeleteNodeTxn} (\textcircled{4}) to remove N3, committing the change to \syslog. N2 may broadcast this membership change for quicker synchronization across nodes (akin to “Watch Notifications” in ZooKeeper~\cite{hunt2010zookeeper}), though such a broadcast is not mandatory for correctness.

\reviseBegin
\subsection{\edit{Correctness Invariants}}
\label{subsec:invariants}
We present the invariants maintained by \system's transactions to show their correctness. 
Central to \system's protocol is the Exclusive Granule Ownership invariant.


\noindent\textbf{I0: Exclusive Granule Ownership:} for any granule $G$ at any time, there is exactly one owner node $N$. This core invariant is a composition of the following sub-invariants:


\begin{itemize}[topsep=2pt,itemsep=1pt,leftmargin=10pt]
    \item \edit{\textbf{I1: Reconfiguration Transactions are Serializable:} for any log, transactions on the log are serialized via the \textit{TryLog} operation of \textit{\mcommit}. Given this, we only need to consider each reconfiguration transaction individually.}
    \item \edit{\textbf{I2: Node and GTable are 1-1 Mapped:} \textit{AddNodeTxn}, \textit{DeleteNodeTxn}, and \textit{ScanGTableTxn} never update GLogs and are only involved during membership changes; their serialization makes \syslog the only source of membership, where each node is associated with exactly one GLog (hence one GTable).}
    \item \edit{\textbf{I3: Owner Exists:} for each granule, there is always an owner, because a GTable can only be updated by a \textit{MigrationTxn} or \textit{RecoveryMigrTxn}, both swap a granule entry and never delete.}
    \item \edit{\textbf{I4: Owner is Unique:} each granule has at most one owner, because both \textit{MigrationTxn} and \textit{RecoveryMigrTxn} swap the ownership of a granule entry and never end with dual owners.}
\end{itemize}

\noindent These invariants jointly establish exclusive ownership. Due to space constraints, we defer the complete proof and the TLA$^+$ specification of \system's migration to the appendix. 

\reviseEnd

\reviseBegin
\section{Implementation}
\label{sec:impl}

We implement \system as a coordination mechanism on an OLTP database testbed\footnote{https://github.com/CloudOLTP-UWM/Marlin/tree/sigmod26}. 
Our testbed extends Sundial~\cite{yu2018sundial}, an open-source distributed OLTP DBMS, into a storage-disaggregated DBMS following the system model as decribed in ~\secref{sec:sysmodel}.
The testbed architecture includes three layers: the clients, the database servers, and the storage service. The client executes user transactions in interactive mode (in contrast to the stored procedure mode), where new request is issued only after previous response is received. The client communicates with the database server via gRPC~\cite{gRPC}, using either synchronous communication (e.g., data access requests) or asynchronous communication (e.g., two-phase commit messages). Exponential back-off is employed for aborted transactions. 


Each database server contains a transaction manager and a cache manager. The transaction manager implements the standard protocols such as two-phase locking (2PL) for concurrency control and two-phase commit (2PC) for atomic commit. We leverage group commit to reduce the storage access overhead by batching log records from multiple transactions and committing them through a single log operation. By default, all transactions follow serializable isolation through the \textit{NO\_WAIT} protocol which avoids deadlocks. The cache manager uses the clock replacement algorithm. On a read miss, the page is fetched from the disaggregated storage. If the requested data has a stale LSN, the storage node waits for log replay before replying. 
The storage layer contains a write-ahead log (WAL) for each compute node and a page store service. Following the log-as-database paradigm, committed transactions send only updates to the WAL, and data pages are asynchronously reconstructed via log replay. The page store uses Azure Table Storage~\cite{azure_table_storage}, with all services hosted under a single \textit{standard general-purpose v2} Azure storage account~\cite{AzureStorageGPv2} in West US 2. WALs are stored in Azure Append Blobs~\cite{AzureAppendBlobs}, which support atomic and concurrent appends. 

A key complication in our implementation is that the \textit{MarlinCommit} protocol requires atomic conditional append (i.e., \textit{Append@LSN}) from the storage service. We describe how we implemented it in Azure Blob Storage and how it can be realized in other cloud storage services.


\boldtext{Microsoft Azure Blob Storage:} Azure's \textit{append blobs} support atomic conditional appends via the \textit{AppendBlock} operation, using precondition headers like \textit{If-Match (ETag)} or \textit{x-ms-blob-condition-appendpos-equal}. An LSN can be implemented using either ETag or blob length (the latter is currently used in \system). \textit{AppendBlock} is called with one of these headers; the operation succeeds only if the condition holds. Otherwise, an error is returned and the compute node can read the new ETag or length and retry. This check-and-append is guaranteed to be atomic within Azure Storage.

\boldtext{Amazon S3:} S3 Express One Zone~\cite{aws_s3_express_one_zone} can achieve atomic conditional appends using a single PUT with headers like \texttt{If-Match} or \textit{x‑amz‑write‑offset‑bytes} as compare-and-swap primitives.

\boldtext{Google Cloud Storage (GCS):} GCS assigns each object a monotonically increasing \textit{generation number}, which can be used as LSN, and supports conditional writes via precondition headers like \textit{ifGenerationMatch}. The client uploads data to a temporary object, then issues a \textit{compose} operation that merges \textit{log@<target\_lsn>} with the temp object, using \textit{ifGenerationMatch:<target\_lsn>}.

Note that although \system is primarily designed for the Partitioned-Writer archetype, it is generalizable to other architectures discussed in~\secref{ssec:arch}. 
For both Single-Writer and Shared-Writer archetypes, the \gtbl is not needed since the data is not partitioned across multiple nodes---in Single-Writer only a single node can write to any data and in Shared-Writer any node can write to any data. In these two archetypes, membership management can still follow Marlin's design via \mtbl and its associated reconfiguration transactions. Since most of the design complexity of Marlin is in the {\gtbl}s, Marlin can be substantially simplified for these other two archetypes.




\reviseEnd

\section{Evaluation}
\label{sec:eval}

We empirically evaluate the performance and cost of \system in comparison to external converged coordination services. Our key findings are as follows:
\begin{itemize}[leftmargin=12pt, rightmargin=0pt]
\item \system outperforms \zk in both performance and cost efficiency in scale-out scenarios (\secref{ssec:exp_scaleout} \edit{and ~\secref{ssec:exp_scaleout_tpcc}}).
\item As coordination workloads scale, \system maintains superior performance and cost efficiency, while \zk and \edit{and FoundationDB} must trade off performance and cost efficiency (~\secref{ssec:exp_perf_tradeoff}).
\item The performance and cost benefits of \system are further amplified in geo-distributed environments (\secref{ssec:exp_perf_geo}) and preserve in more complicated and dynamic workloads  (\secref{ssec:exp_perf_sens}).
\reviseBegin
\item \system's membership management performance is comparable to other baselines under moderate contention but may degrade under high contention (~\secref{ssec:exp_mem_perf}).
\reviseEnd
\end{itemize}
\vspace{-0.1in}


\subsection{Experimental Setup}\label{ssec:exp_setup}
\subsubsection{\textbf{Configurations}}
We use the DBMS testbed described in \secref{sec:impl}. For a fair comparison, we implement \system and all baselines on this testbed. All experiments are conducted on Microsoft Azure Cloud Service platform~\cite{AzureCloudService}. Compute nodes are instances of Standard D4s v3 in the West US 2 region. Each runs on the 3rd Generation Intel(R) Xeon(R) Platinum 8370C processor with 4 vCPU, 16 GB memory, and a 2 Gbps network. 

\vspace{-0.05in}
\subsubsection{\textbf{Baselines}}\label{ssec:exp_baselines} 
We compare \system against \edit{two} external coordination services, \zk \edit{and FoundationDB}. \zk represents an external and \converged coordination solution. 
To better capture its scalability, we evaluate two \zk setups with different hardware configurations. \edit{FoundationDB is a distributed OLTP database offering high performance and scalability, and is used as a coordination service in several commercial cloud databases (e.g., Snowflake).} \system and all baselines adopt the same cache-warming strategy during reconfiguration (\secref{subsec:live-migration}).

\noindent \textbf{ZooKeeper-Small} (\szk): The baseline uses ZooKeeper 3.8.4 and JDK 12.0.2. The Zookeeper cluster consisted of one leader node and two follower nodes. Each node is a Standard D4s v3 instance with 4 vCPU, 16 GB memory, and 2 Gbps network.We set maxClientCnxns as 0 to disable client connection limits and set maximum heap memory of JVM as 15GB.

\noindent \textbf{ZooKeeper-Large} (\lzk): This baseline uses a similar deployment as \szk except that each node has a larger configuration: Standard D8s v3 with 8 vCPU, 32 GB memory, and 4 Gbps network bandwidth. The maximum heap memory of JVM is 30GB.

\reviseBegin
\noindent \textbf{FoundationDB} (\fdb): We deploy FoundationDB (7.3.63) on hardware comparable to \szk. Each node has one transaction process, one storage process, and one stateless process. The replication factor is three. Data is partitioned via a dynamic sharding mechanism based on key prefixes.

\reviseEnd



\vspace{-0.05in}

\subsubsection{\textbf{Workloads}}\label{ssec:exp_wl}

\edit{We use the YCSB and TPC-C OLTP workloads for performance evaluation.} 

\noindent \textbf{YCSB:} The Yahoo! Cloud Serving Benchmark (YCSB)~\cite{cooper2010benchmarking} is a synthetic benchmark designed to evaluate large-scale Internet applications. We use tables with different sizes (ranging from 3 GB to 20 GB) that are partitioned into granules across servers by range on the primary key. Unless otherwise specified, each tuple is around 1KB and each granule is 64 KB. Each transaction is single-site and has 16 requests with 50\% reads and 50\% updates accessing 16 tuples. We generate requests following a uniform distribution. \edit{We use the YCSB workload as the default unless otherwise stated.}

\reviseBegin
\noindent \textbf{TPC-C:} TPC-C~\cite{tpcc} models a warehouse-centric order processing application with nine tables and five transaction types. All tables except ITEM are partitioned by the warehouse ID. The ITEM table is replicated at each server. 10\% of NEW-ORDER and 15\% of PAYMENT transactions access multiple warehouses; other transactions access data on a single server. We use a warehouse as the unit of migration, and each granule contains one warehouse. To evaluate performance under heavy migration with a large number of warehouses, we tune down the size of each warehouse to $\sim$1MB by reducing the number of customers per district, and deploy 1600 warehouses per server.
\reviseEnd


We design \edit{four} testing scenarios to simulate real-world workloads. The \textbf{scale-out} scenario evaluates the system's ability to handle overloads by scaling out the cluster (\secref{ssec:exp_scaleout}--\secref{ssec:exp_perf_tradeoff}). The \textbf{geo-distributed} scenario models deployments where clients and compute nodes span multiple regions (\secref{ssec:exp_perf_geo}). The \textbf{dynamic} scenario simulates bursty workloads with fluctuating client counts (\secref{ssec:exp_perf_sens}). \edit{Finally, the \textbf{membership-update} scenario evaluates 
group membership management under intensive membership changes, such as cluster rescale or node failures and recoveries (\secref{ssec:exp_mem_perf}).}

\vspace{-0.05in}
\subsubsection{\textbf{Methodology}}
We fully warm up the cache so that performance is stable before running experiments. We run the experiments for a fixed amount of time and report throughput and latency for committed transactions. 
If a transaction aborts, the client will keep retrying with exponential backoff (bounded by 100 ms) until it succeeds. In \system, each reconfiguration transaction migrates one granule, and multiple reconfiguration transactions are executed concurrently. The number of concurrent migration transactions is increased as the number of compute nodes increases. 

\vspace{-0.05in}
\subsubsection{\textbf{Cost Calculation}}
\label{sec:eval_setup_cost} 
The total system cost includes data-plane and control-plane costs. \textit{DB Cost} accounts for computing servers and cloud storage, while \textit{Meta Cost} reflects coordination expenses. Since \system eliminates the external coordination service, its \textit{Meta Cost} is zero. 
Computing server costs are calculated based on the machine's hourly rate. Storage costs are excluded from comparisons due to their negligible impact. For example, a single Standard D4s v3 instance costs \$0.192 per hour, 384$\times$ higher than the cost of storing 20GB in the hot-tier blob storage.

\subsection{Scale-Out Performance on YCSB}
\label{ssec:exp_scaleout}
We evaluate the scale-out scenario \edit{using YCSB workload}. The experiment runs a static workload of 800 concurrent clients accessing a 24 GB table partitioned into $\sim$200K granules. Starting with 8 nodes, the cluster scales out to 16 nodes at the 10th second. This process executes $\sim$100K \textit{MigrationTxn}s, with each transaction migrating a single granule from an existing node to a newly added node.

\begin{figure}[t]
        \vspace{-.10in}

    	\begin{subfigure}{0.7\columnwidth}%
            \centering
            \includegraphics[width=\linewidth]{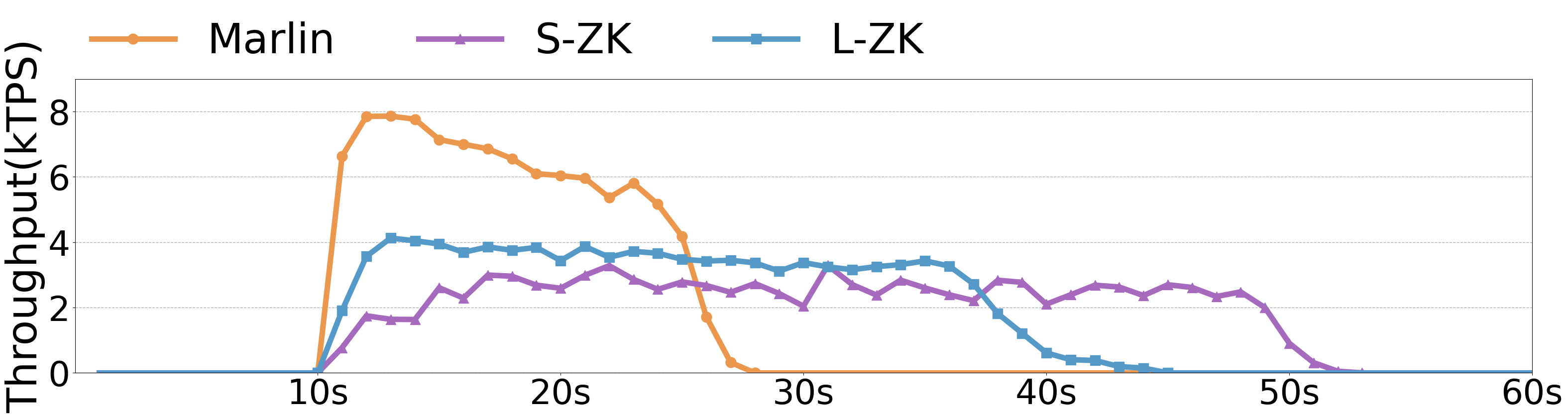}  
            \label{fig:ev1-f1}
        \end{subfigure}%
    \setlength{\abovecaptionskip}{-.15in}

    \caption{MigrationTxn Throughput Over Time.}
    \label{fig:ev1-f1}
\end{figure}


~\figref{fig:ev1-f1} shows that \system achieves 2.3$\times$ and 1.9$\times$ higher throughput for migration transactions than \szk and \lzk, respectively. This gain stems from \system's distributed design, which partitions \gtbl across multiple nodes, spreading the metadata update load. In contrast, \zk relies on a centralized single-writer node, which becomes a bottleneck. The throughput advantage of \lzk over \szk is due to \lzk's superior hardware, including higher network and disk bandwidth.


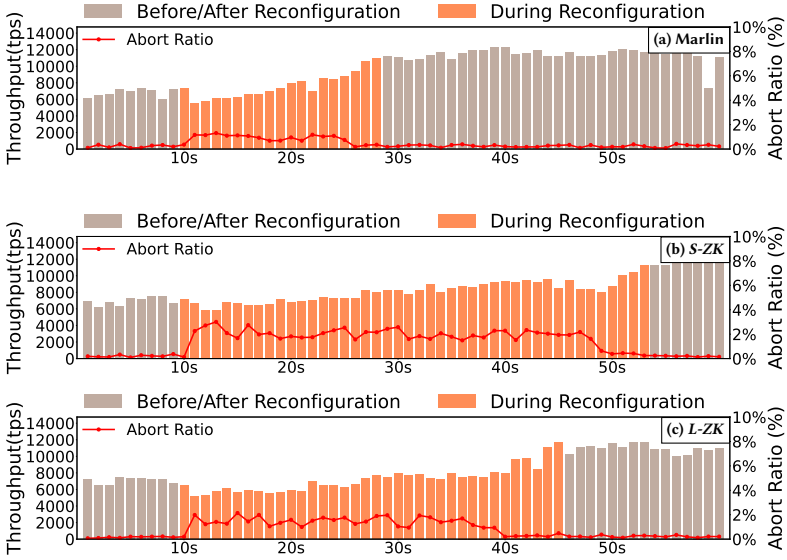
\begin{figure}[h]
        \vspace{-.10in}

    \centering
    \newcommand{\usrmarlin}{figures/plots/experiment/ev1-postheat-usr-timeline-Marlin.pdf}
    \newcommand{\usrszk}{figures/plots/experiment/ev1-postheat-usr-timeline-Small-ZK.pdf}
    \newcommand{\usrlzk}{figures/plots/experiment/ev1-postheat-usr-timeline-Large-ZK.pdf}
    
    \newlength{\figEvTwoSubfigspace}
    \newlength{\figEvTwoTitlexshift}
    \newlength{\figEvTwoTitleyshift}
    \newlength{\figEvTwoTitlefontsize}
    \setlength{\figEvTwoSubfigspace}{-.15in}
    \setlength{\figEvTwoTitlexshift}{-17pt}
    \setlength{\figEvTwoTitleyshift}{-9.8pt}
    \setlength{\figEvTwoTitlefontsize}{6pt}
    
    \begin{subfigure}[t]{0.75\columnwidth}
        \centering
        \begin{tikzpicture}
            \node[inner sep=0pt] (image) {\includegraphics[width=\linewidth]{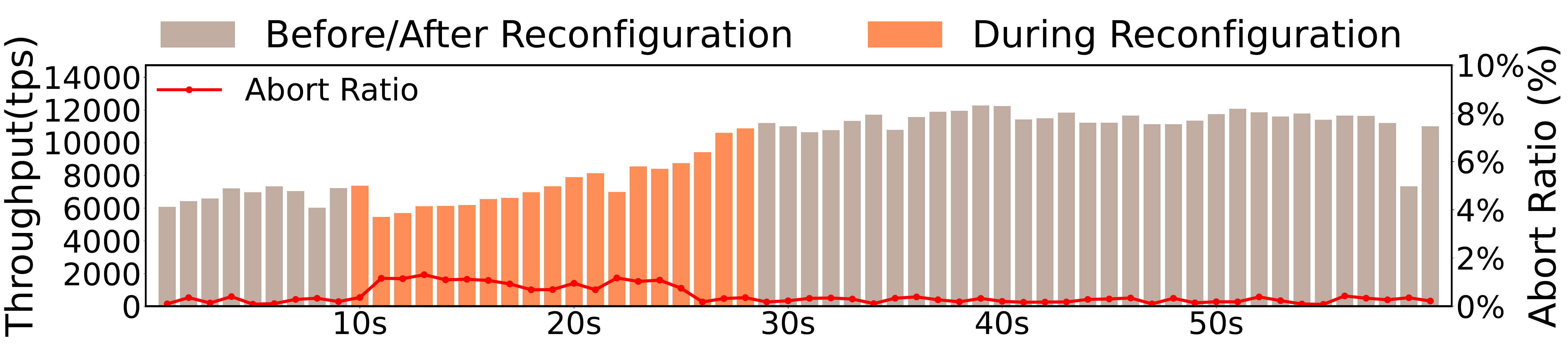}};
            \node[anchor=north east, font={\fontsize{\figEvTwoTitlefontsize}{1.2\figEvTwoTitlefontsize}\selectfont\bfseries}, 
                  draw, fill=white, inner sep=2pt,  xshift=1.3\figEvTwoTitlexshift, yshift=1.28\figEvTwoTitleyshift] 
                at (image.north east) {(a) \system};
        \end{tikzpicture}
        \phantomcaption
        \label{fig:ev1-f2-a}
    \end{subfigure}
    
    \begin{subfigure}[t]{0.75\columnwidth}
        \centering
        \begin{tikzpicture}
            \node[inner sep=0pt] (image) {\includegraphics[width=\linewidth]{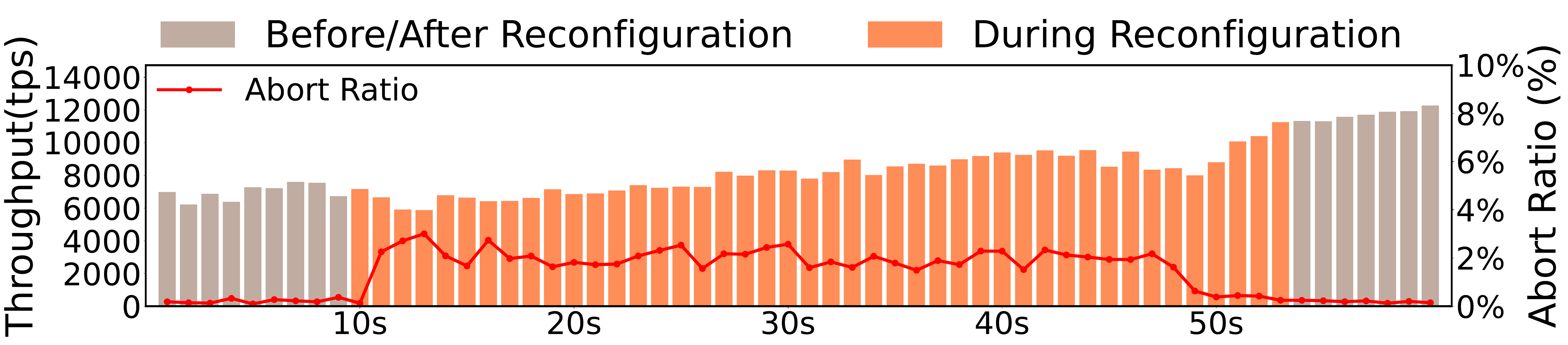}};
            \node[anchor=north east, font={\fontsize{\figEvTwoTitlefontsize}{1.2\figEvTwoTitlefontsize}\selectfont\bfseries}, 
                  draw, fill=white, inner sep=2pt,  xshift=1.3\figEvTwoTitlexshift, yshift=1.28\figEvTwoTitleyshift] 
                at (image.north east) {(b) \szk};
        \end{tikzpicture}
        \phantomcaption
        \label{fig:ev1-f2-b}
    \end{subfigure}
    \par\vspace{\figEvTwoSubfigspace}
    
    \begin{subfigure}[t]{0.75\columnwidth}
        \centering
        \begin{tikzpicture}
            \node[inner sep=0pt] (image) {\includegraphics[width=\linewidth]{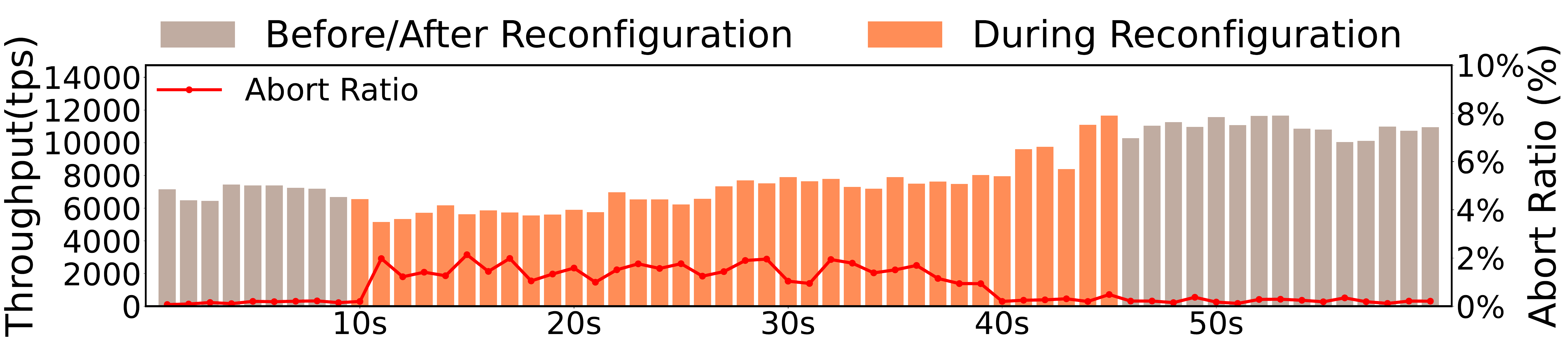}};
            \node[anchor=north east, font={\fontsize{\figEvTwoTitlefontsize}{1.2\figEvTwoTitlefontsize}\selectfont\bfseries}, 
                  draw, fill=white, inner sep=2pt,  xshift=1.3\figEvTwoTitlexshift, yshift=1.28\figEvTwoTitleyshift] 
                at (image.north east) {(c) \lzk};
        \end{tikzpicture}
        \phantomcaption
        \label{fig:ev1-f2-c}
    \end{subfigure}
    
    \setlength{\abovecaptionskip}{-.1in}
    \caption{Realtime Throughput of User Transactions {\small (YCSB)}}
    \label{fig:ev1-f2}
\end{figure}



\begin{figure}[h]
        \vspace{-.10in}

    \begin{subfigure}{0.44\columnwidth}%
    	\centering
        \includegraphics[width=\linewidth]
        {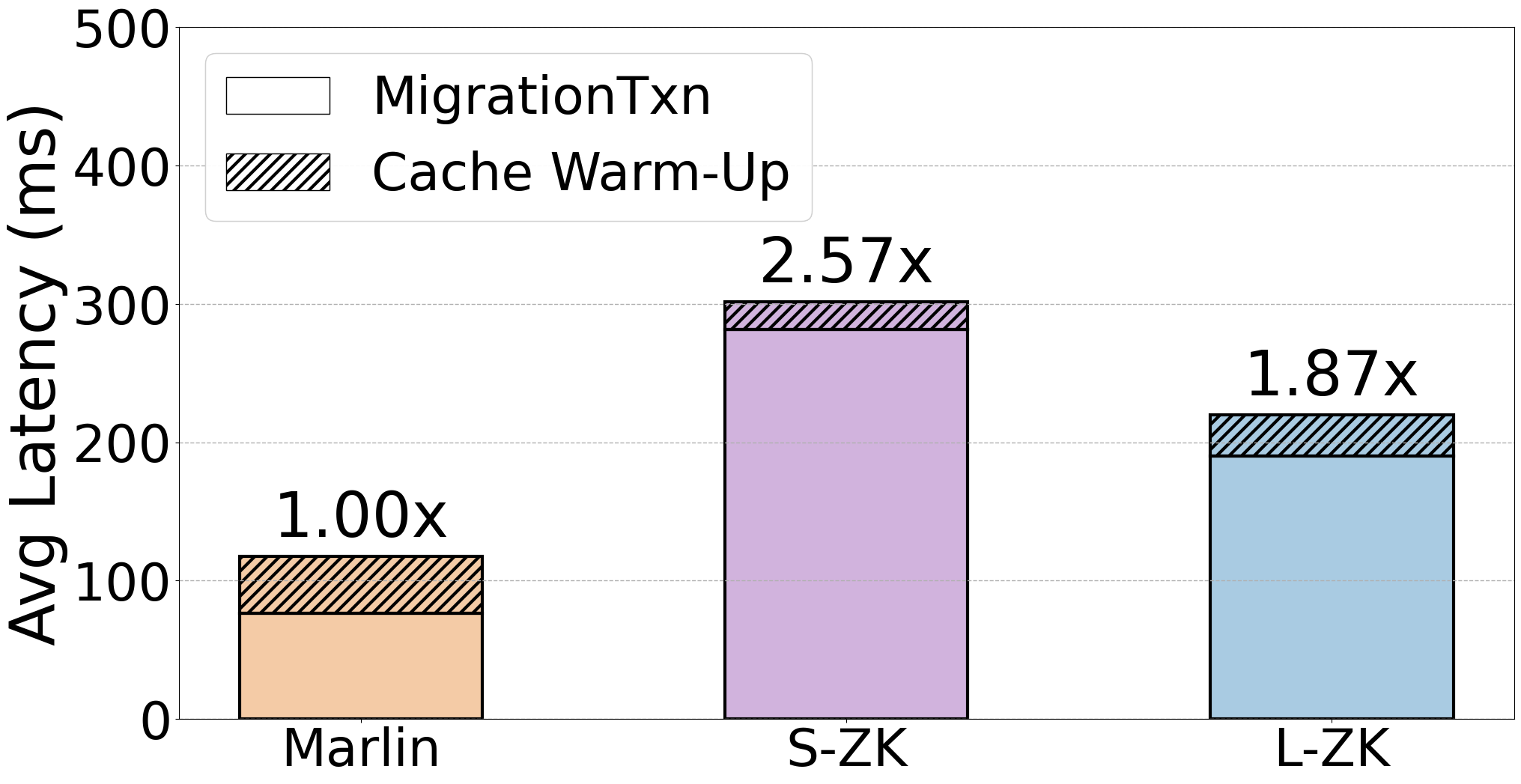}  
        \setlength{\abovecaptionskip}{-.15in}
        \caption{Migration Latency}
        \label{fig:ev1_f3_a}
    \end{subfigure}%
        \hspace{0.02\columnwidth}%
    \begin{subfigure}{0.44\columnwidth}%
    	\centering
        \includegraphics[width=\linewidth]
        {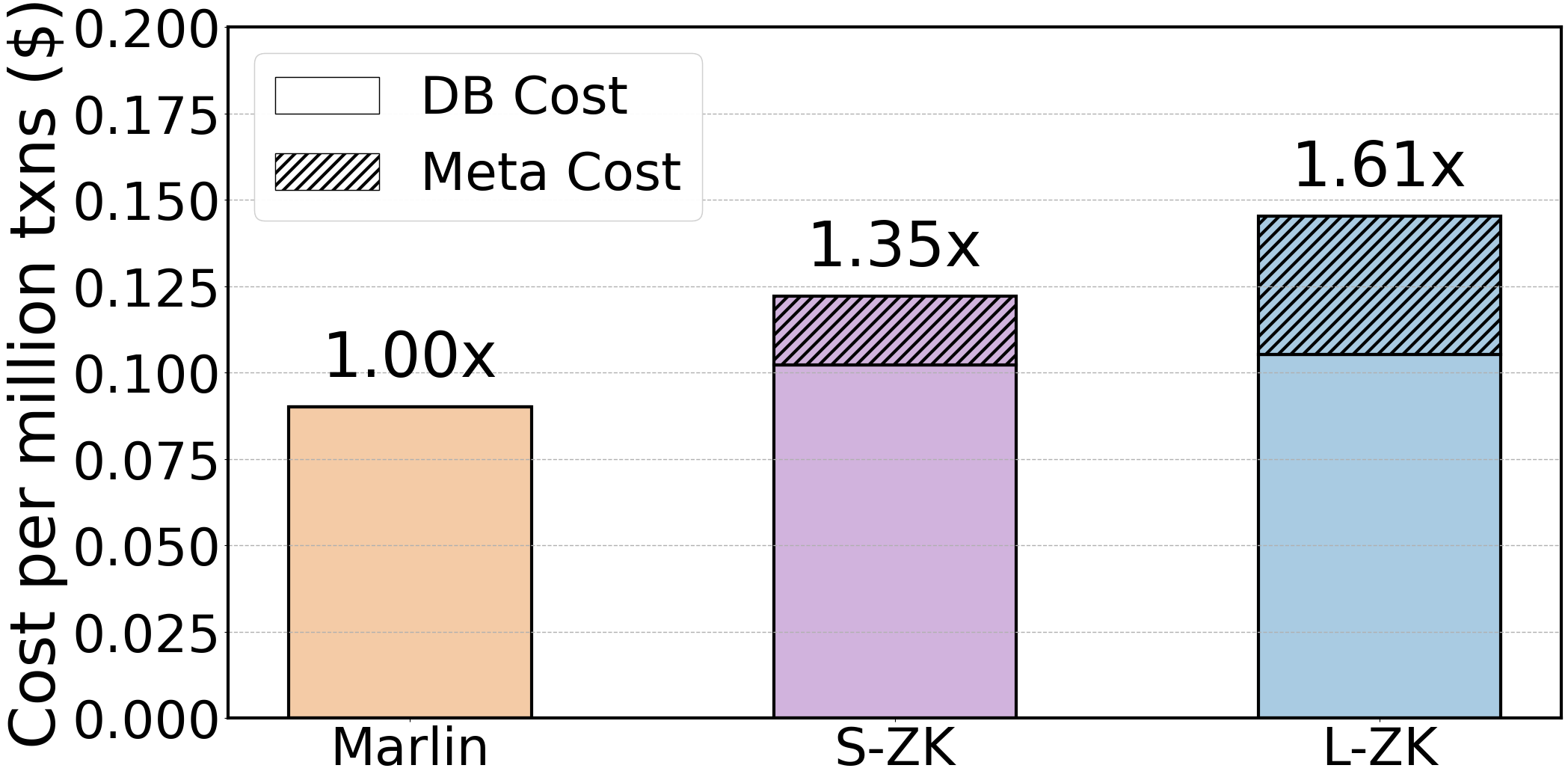}  
        \setlength{\abovecaptionskip}{-.15in}
        \caption{Cost of UserTxn}
        \label{fig:ev1_f3_b}
    \end{subfigure}%
            \vspace{-.15in}

    \caption{Cost and Latency Breakdown}

    \label{fig:ev1_f2}
\end{figure}

\system completes the scale-out process 2.6$\times$ and 1.9$\times$ faster than \szk and \lzk, due to its higher migration throughput. As shown in ~\figref{fig:ev1-f2}, the shorter migration duration promotes elasticity: the throughput of user transactions reaches a higher level of approximately 12k tps more rapidly than \zk-based approaches. Furthermore, \system has a lower abort rate for user transactions. 
This is because migration transactions in \system have lower latency and thus a lower chance of conflicting with user transactions. ~\figref{fig:ev1_f3_a} shows that \system reduces the migration latency by 2.57$\times$ and 1.87$\times$ compared to \szk and \lzk. While each migration includes a cache warm-up phase, 
the latency of \textit{MigrationTxn} remains the dominant factor of the performance difference.

Besides its performance benefits, \system also has the best cost efficiency. 
As shown in \figref{fig:ev1_f3_b}, \system reduces cost by 1.35$\times$ and 1.61$\times$ 
compared to \szk and \lzk, respectively. This is primarily because \system eliminates the upfront cost of a static, converged coordination service, 
which requires a cluster of at least 3 nodes in \zk. The hourly rate for \lzk and \lzk cluster is \$0.597 and \$1.173, respectively. Furthermore, \system slightly decreases \textit{DB Cost} per transaction by allowing the database to capitalize on scale-out benefits more quickly than \zk, thereby achieving higher throughput using the same \textit{DB Cost}.




\reviseBegin
\subsection{Scale-Out Performance on TPC-C}
\label{ssec:exp_scaleout_tpcc}

This section evaluates Marlin on the TPC-C benchmark. 
Similar to ~\secref{ssec:exp_scaleout}, the cluster initially consists of 8 nodes and scales out to 16 nodes at the 10th second. 
This process executes 6.4K \textit{MigrationTxn}s, each migrating a single warehouse from an existing node to a newly added node. Each new node launches 80 migration threads to execute \textit{MigrationTxn} concurrently.

\begin{figure}[h]
        \vspace{-.10in}

    \centering
    \newcommand{\usrmarlin}{figures/plots/experiment/ev1-postheat-usr-timeline-tpcc-Marlin.pdf}
    \newcommand{\usrszk}{figures/plots/experiment/ev1-postheat-usr-timeline-tpcc-Small-ZK.pdf}
    \newcommand{\usrlzk}{figures/plots/experiment/ev1-postheat-usr-timeline-tpcc-Large-ZK.pdf}
    
    \newlength{\figEvOneSubfigspace}
    \newlength{\figEvOneTitlexshift}
    \newlength{\figEvOneTitleyshift}
    \newlength{\figEvOneTitlefontsize}
    \setlength{\figEvOneSubfigspace}{-.15in}
    \setlength{\figEvOneTitlexshift}{-16pt}
    \setlength{\figEvOneTitleyshift}{-9pt}
    \setlength{\figEvOneTitlefontsize}{6pt}
    
    \begin{subfigure}[t]{0.7\columnwidth}
        \centering
        \begin{tikzpicture}
            \node[inner sep=0pt] (image) {\includegraphics[width=\linewidth]{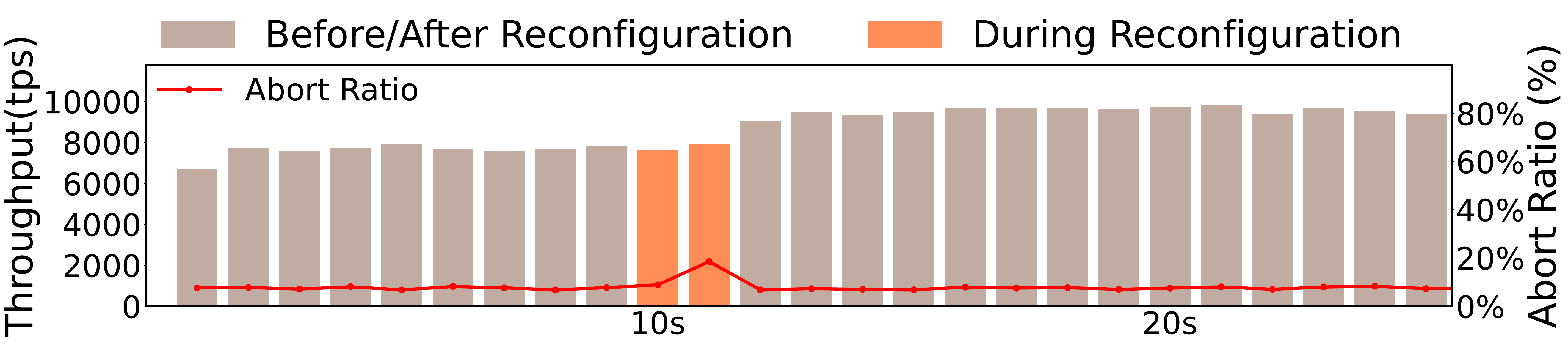}};
            \node[anchor=north east, font={\fontsize{\figEvOneTitlefontsize}{1.2\figEvOneTitlefontsize}\selectfont\bfseries}, 
                  draw, fill=white, inner sep=2pt, xshift=1.29\figEvOneTitlexshift, yshift=1.3\figEvOneTitleyshift] 
                at (image.north east) {(a) \system};
        \end{tikzpicture}
        \phantomcaption
        \label{fig:ev1-f4-a}
    \end{subfigure}
    \par\vspace{\figEvOneSubfigspace}
    
    \begin{subfigure}[t]{0.7\columnwidth}
        \centering
        \begin{tikzpicture}
            \node[inner sep=0pt] (image) {\includegraphics[width=\linewidth]{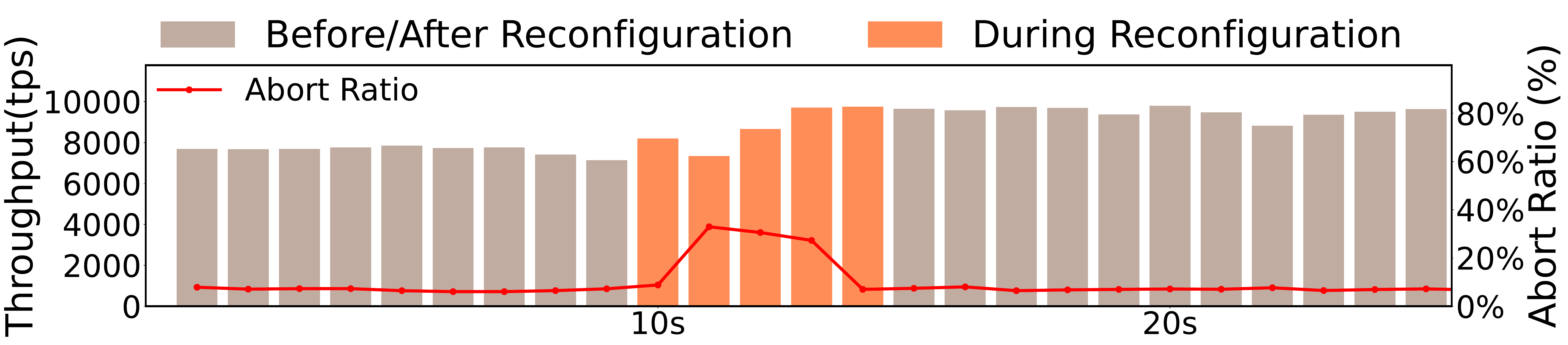}};
            \node[anchor=north east, font={\fontsize{\figEvOneTitlefontsize}{1.2\figEvOneTitlefontsize}\selectfont\bfseries}, 
                  draw, fill=white, inner sep=2pt, xshift=1.29\figEvOneTitlexshift, yshift=1.3\figEvOneTitleyshift] 
                at (image.north east) {(b) \szk};
        \end{tikzpicture}
        \phantomcaption
        \label{fig:ev1-f4-b}
    \end{subfigure}
    \par\vspace{\figEvOneSubfigspace}
    
    \begin{subfigure}[t]{0.7\columnwidth}
        \centering
        \begin{tikzpicture}
            \node[inner sep=0pt] (image) {\includegraphics[width=\linewidth]{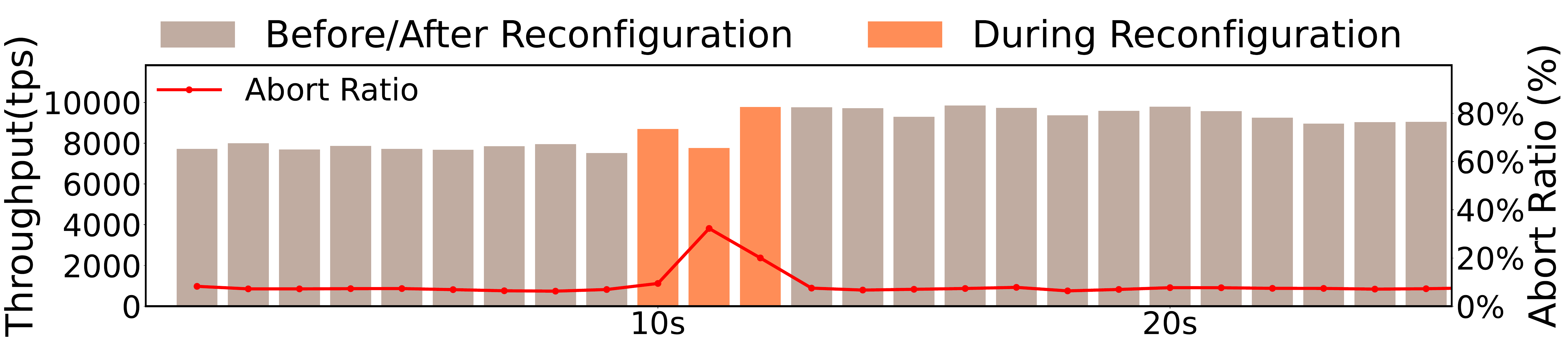}};
            \node[anchor=north east, font={\fontsize{\figEvOneTitlefontsize}{1.2\figEvOneTitlefontsize}\selectfont\bfseries}, 
                  draw, fill=white, inner sep=2pt, xshift=1.29\figEvOneTitlexshift, yshift=1.3\figEvOneTitleyshift] 
                at (image.north east) {(c) \lzk};
        \end{tikzpicture}
        \phantomcaption
        \label{fig:ev1-f4-c}
    \end{subfigure}
    
    \vspace{-.3in}
    \caption{\edit{Realtime Throughput of User Transactions {\small (TPC-C)}}}

    \label{fig:ev1-f4}
\end{figure}
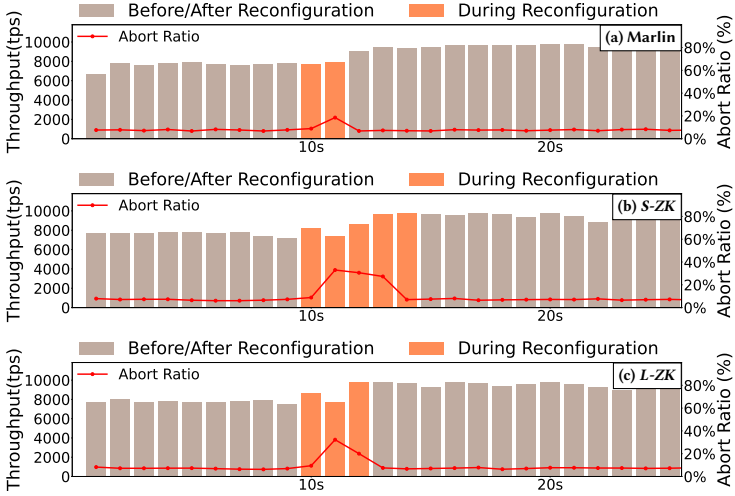



As shown in ~\figref{fig:ev1-f4}, \system completes the migration 2.5$\times$ and 1.5$\times$ faster than \szk and \lzk, respectively. The improvement comes from \system's distributed coordination design that partitions \gtbl across multiple nodes, avoiding the centralized bottleneck inherent to \zk-based systems. The duration of migration in TPC-C is significantly shorter than in the YCSB experiments because the number of granules to migrate is smaller (6.4K vs. 200K). The performance gap between \lzk and \szk is mainly due to the superior hardware configuration of \lzk. Furthermore, \system incurs less degradation of user transactions during migration, as reflected in higher user throughput and lower abort rate. This improvement comes from lower-latency and higher-throughput migration transactions, which reduce conflicts with ongoing user transactions.

\reviseEnd
\subsection{Cost vs. Migration Duration Tradeoff}
\label{ssec:exp_perf_tradeoff}
\begin{figure}[t]
        \vspace{-.10in}

    	\begin{subfigure}{0.7\columnwidth}%
            \centering
            \includegraphics[width=0.8\linewidth]{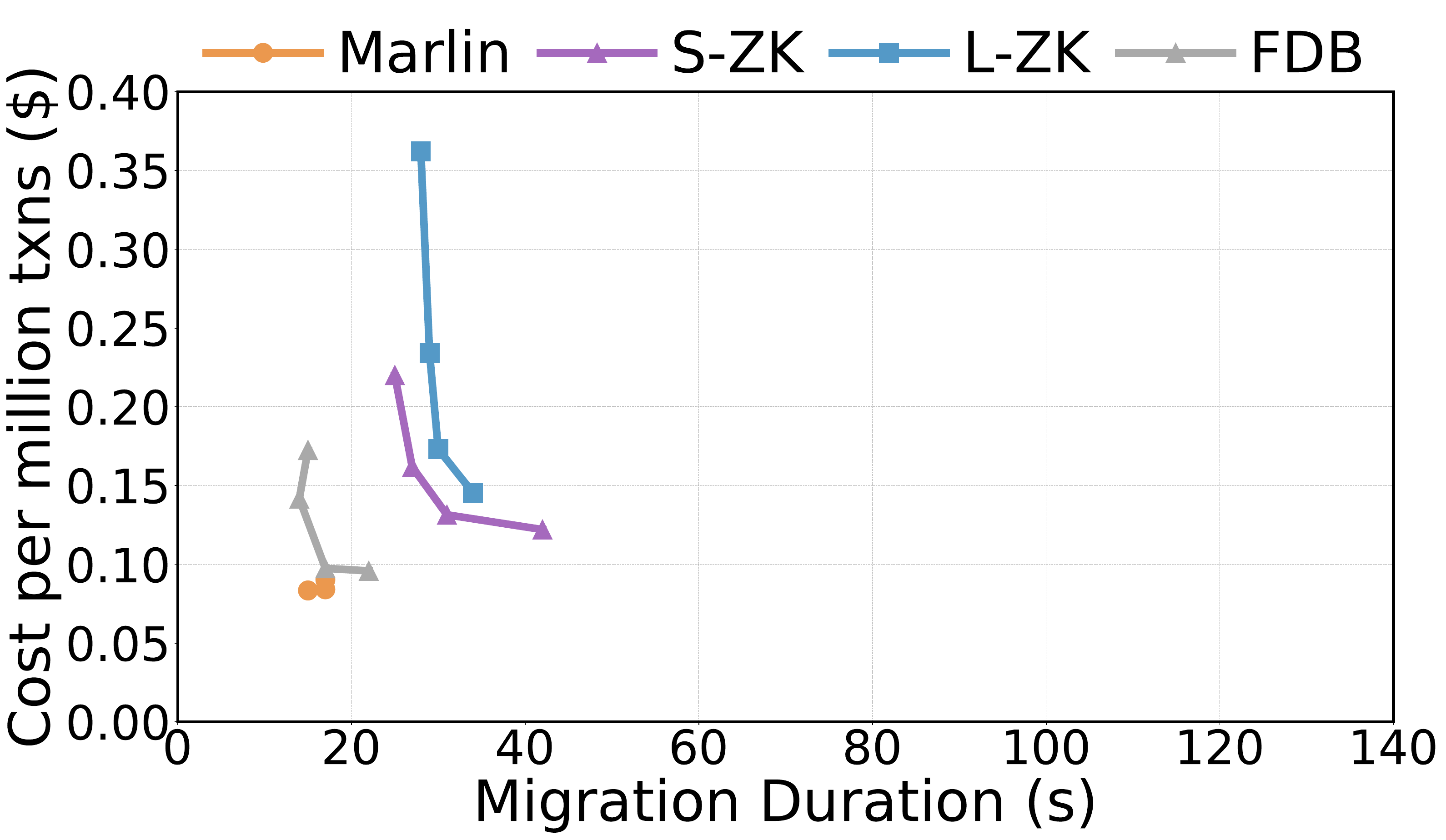}  
            \vspace{-.05in}
            \caption{Cost per Transaction vs. Migration Duration}
            \label{fig:ev2_f1_a}
        \end{subfigure}%
        \hfill
        \begin{subfigure}{0.45\columnwidth}%
            \centering
            \includegraphics[width=\linewidth]{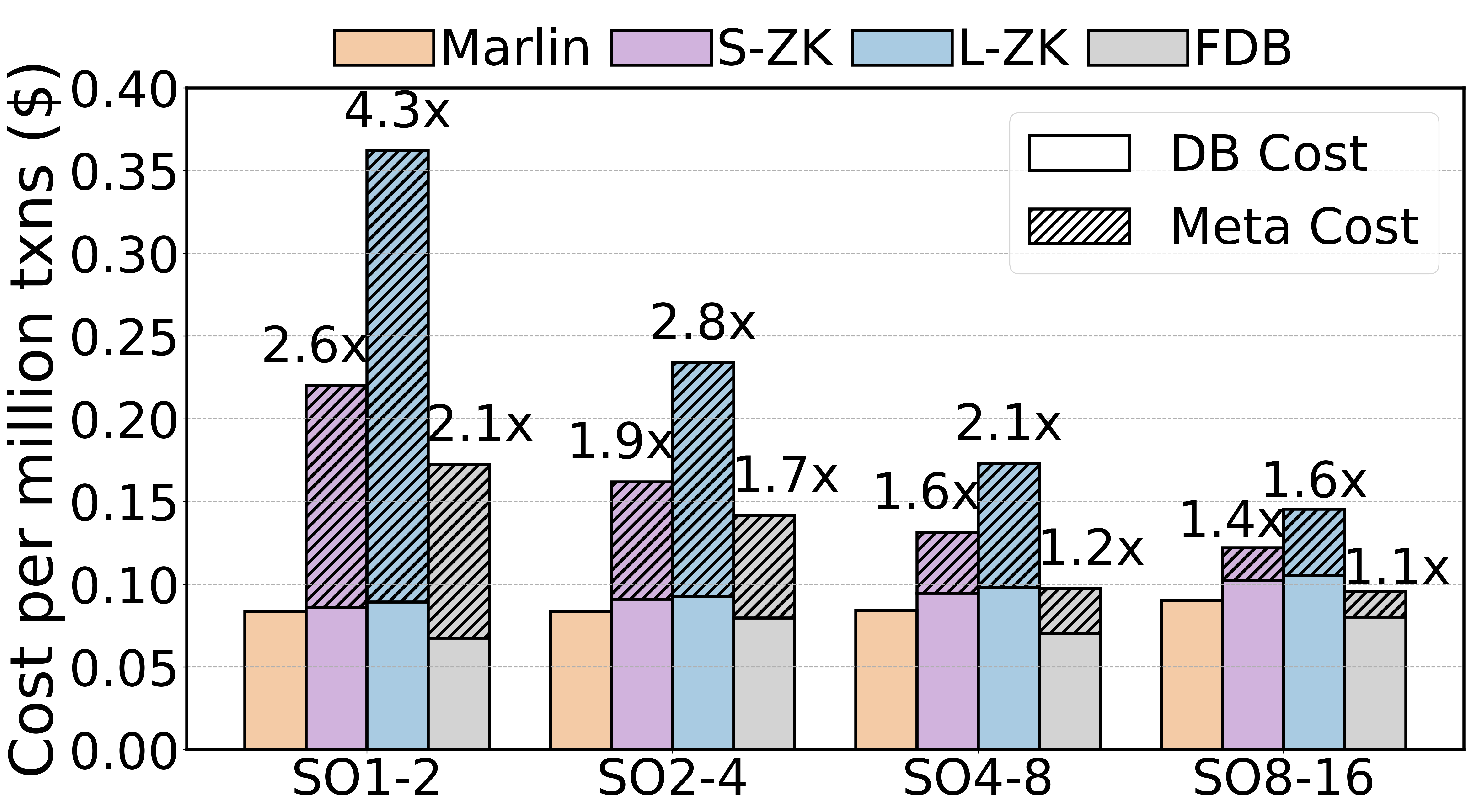}  
            \vspace{-.2in}
            \caption{Cost of UserTxn}
            \label{fig:ev2_f1_b}
        \end{subfigure}%
          \hspace{0.02\columnwidth}%
        \begin{subfigure}{0.45\columnwidth}%
            \centering
            \includegraphics[width=\linewidth]{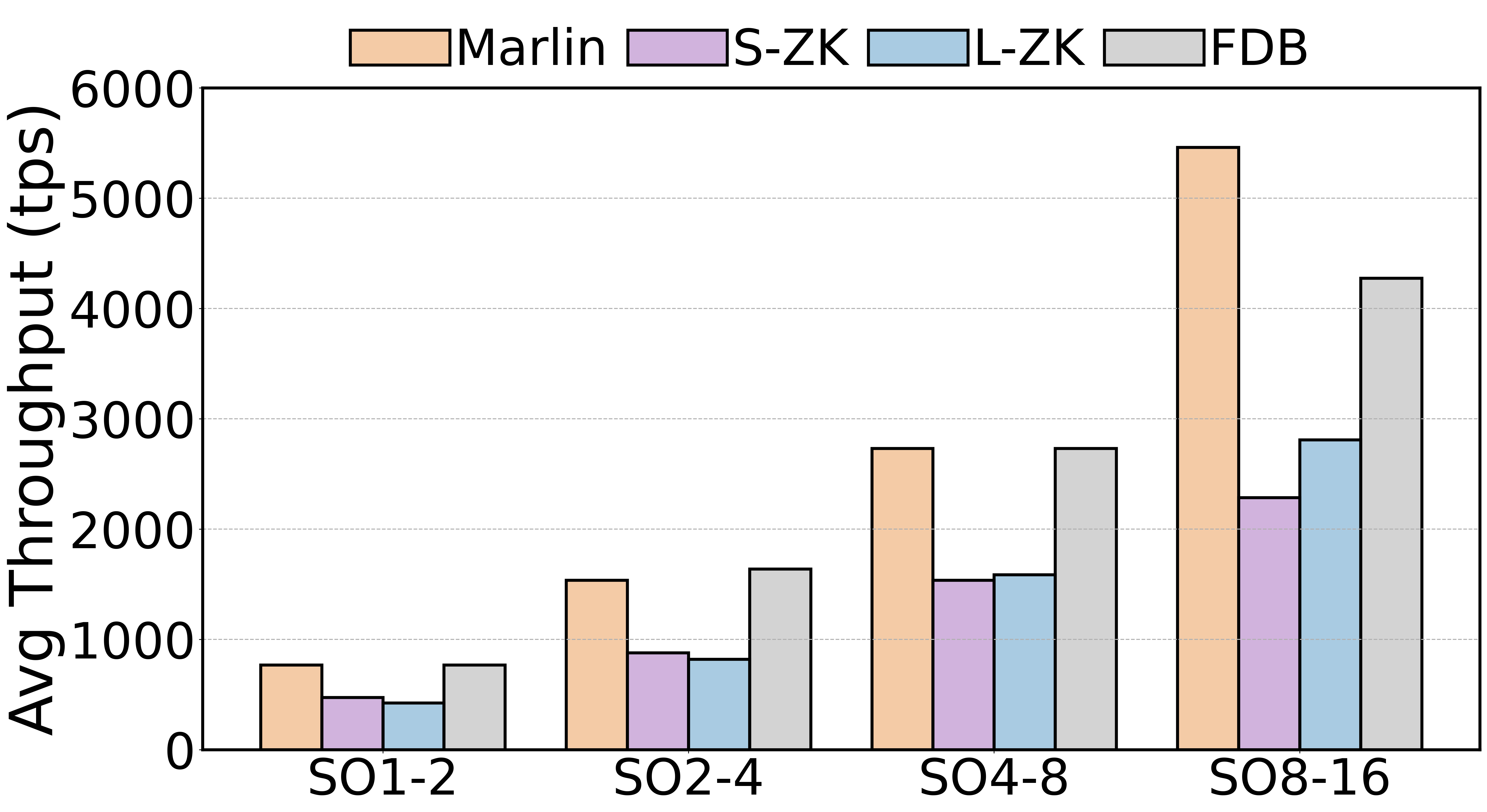}  
            \vspace{-.2in}
            \caption{Migration Throughput}
            \label{fig:ev2_f1_c}
        \end{subfigure}%
    \figcolor{}
    \vspace{-0.1in}
    \caption{\edit{Cost vs. Migration Performance {\small (Single-Region)}}}

    \label{fig:ev2_f1}
\end{figure}

We evaluate whether the benefits of \system persist across various scale-out scenarios with different data scales \edit{under YCSB workload}. The four settings—\textit{SO1-2}, \textit{SO2-4}, \textit{SO4-8}, and \textit{SO8-16}—represent scale-outs from 1 to 2, 2 to 4, 4 to 8, and 8 to 16 nodes. These scenarios correspond to workloads of 100, 200, 400, and 800 clients, with table sizes of 3, 6, 12, and 24 GB. As the scale increases, the number of migration transactions grow from $\sim$10K to $\sim$100K, with migration concurrency doubling at each step. In every data scale, the static workloads exceed the cluster's initial capacity, requiring a scale-out to double the cluster size to handle the workload effectively.

\figref{fig:ev2_f1_a} shows \system consistently achieves the best balance between cost and migration duration across all scales. It maintains the lowest cost per user transaction and shortest migration duration, with up to 4.4$\times$ lower cost than \lzk in \textit{SO1-2} and 2.5$\times$ faster migration than \szk in \textit{SO8-16}. The cost-benefit is because \system eliminates the upfront cost for an external coordination service. The migration duration benefit is due to \system's partitioned design for \gtbl, which avoids centralized bottlenecks and therefore achieves higher migration throughput than \zk-based approaches. While \szk is more cost-effective than \lzk due to its smaller hardware configuration, it causes longer migration durations compared to \lzk, especially at large-scale migration (e.g., \textit{SO8-16}) as it reaches its scalability limit. This is evident in \figref{fig:ev2_f1_c}, where the migration throughput in \system increases linearly with the intensity of migration workloads, while \szk shows exhibits diminishing gains, especially near the \textit{SO8-16} scale.

\edit{Compared to \szk and \lzk, \fdb demonstrates shorter migration durations, benefiting from FoundationDB’s internal partitioning that offers better scalability. However, \fdb's scalability remains constrained by its fixed resources, as it does not automatically rescale with the underlying database and requires manual reconfiguration. In contrast, \system’s scalability naturally grows with the scale of the coordinated database. This is evident in \figref{fig:ev2_f1_c}, where \system’s migration throughput increases linearly with database scale, while the throughput of \zk-based and \fdb approaches continues to grow but at a diminishing rate. \fdb incurs a higher cost per transaction (up to 2.1$\times$) compared to \system, due to the overhead of the external coordination cluster.}

\edit{For external coordination approaches}, \figref{fig:ev2_f1_b} shows that \textit{Meta Cost} constitutes a decreasing portion (e.g., from 75\% to 28\% in \lzk) of the total system cost when cluster size increases from \textit{SO1-2} to \textit{SO8-16}. As the cluster scales, \textit{Meta Cost} becomes less significant relative to \textit{DB Cost}. Therefore, \system provides greater cost benefits under smaller workloads. However, when the cluster scales, the scalability advantage of \system becomes more pronounced, leading to shorter migration duration and better elasticity. Therefore, \system either achieves cost efficiency or provides better elasticity on all database scales, striking the best balance on cost and migration performance.


\subsection{Performance in Geo-distributed Settings}
\label{ssec:exp_perf_geo}
\begin{figure}[t]
        \vspace{-.10in}

    	\begin{subfigure}{0.7\columnwidth}%
            \centering
            \includegraphics[width=0.8\linewidth]{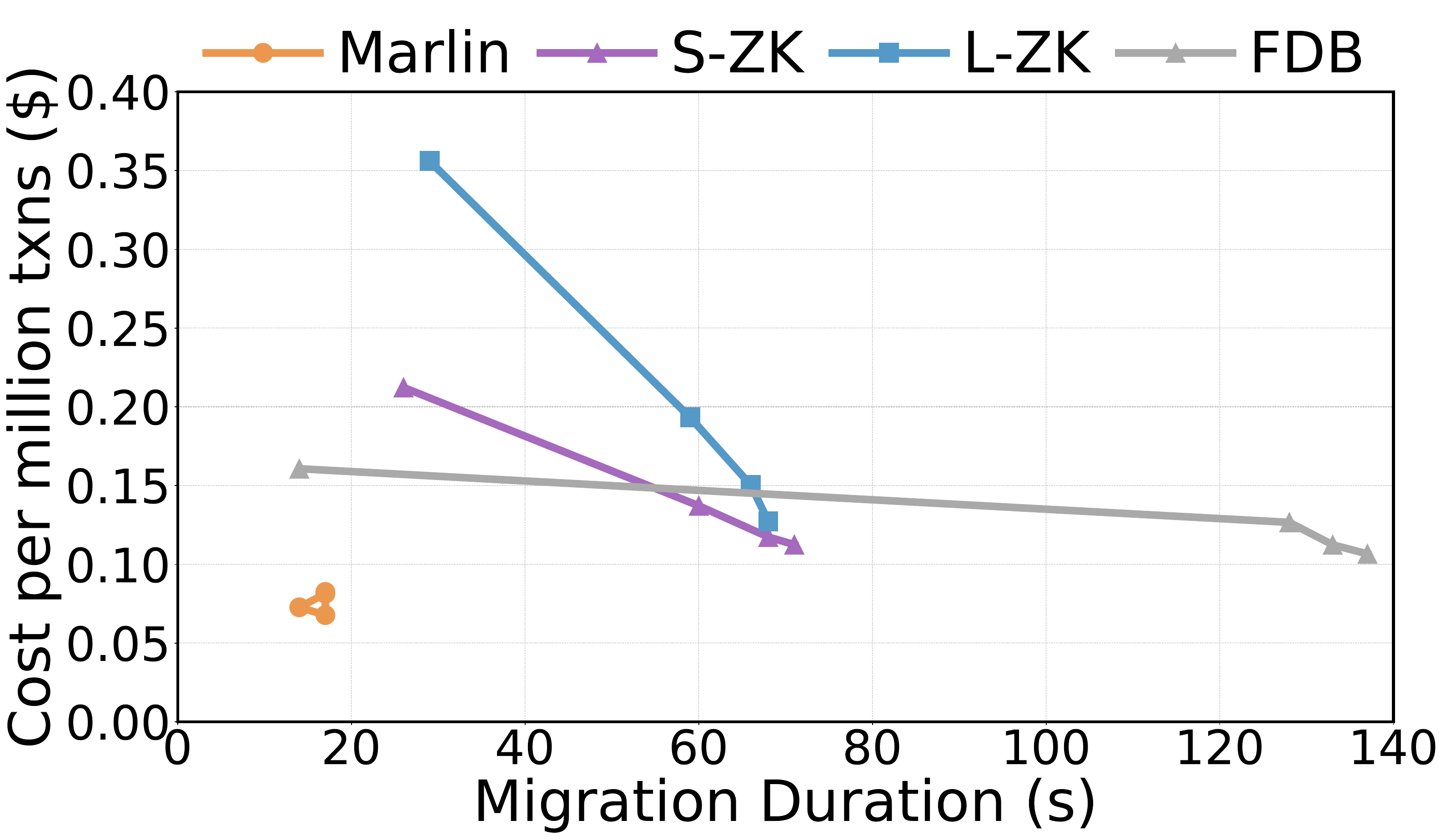}  
            \label{fig:ev3_f1_a}
        \end{subfigure}%

    \figcolor{}
        \vspace{-.1in}

    \caption{Cost vs. Migration Duration {\small (Geo-Distributed)}}

    \label{fig:ev3_f1}
\end{figure}

Geo-distributed databases are common in the cloud, driven by the need for improved performance and compliance with regional regulatory requirements. These systems strategically position compute servers closer to clients across different geographical locations. In this experiment, we deploy clients and compute nodes across up to four regions: US West, Asia East, UK South, and Australia East. The workloads are the same as those in the scale-out scenarios in ~\secref{ssec:exp_perf_tradeoff}. Clients and compute nodes were evenly distributed across regions, with each client accessing only local compute nodes. The disaggregated storage service is co-located with its respective compute nodes. \zk \edit{and \fdb are} deployed in US West. For each scale-out, compute nodes within each region are doubled.

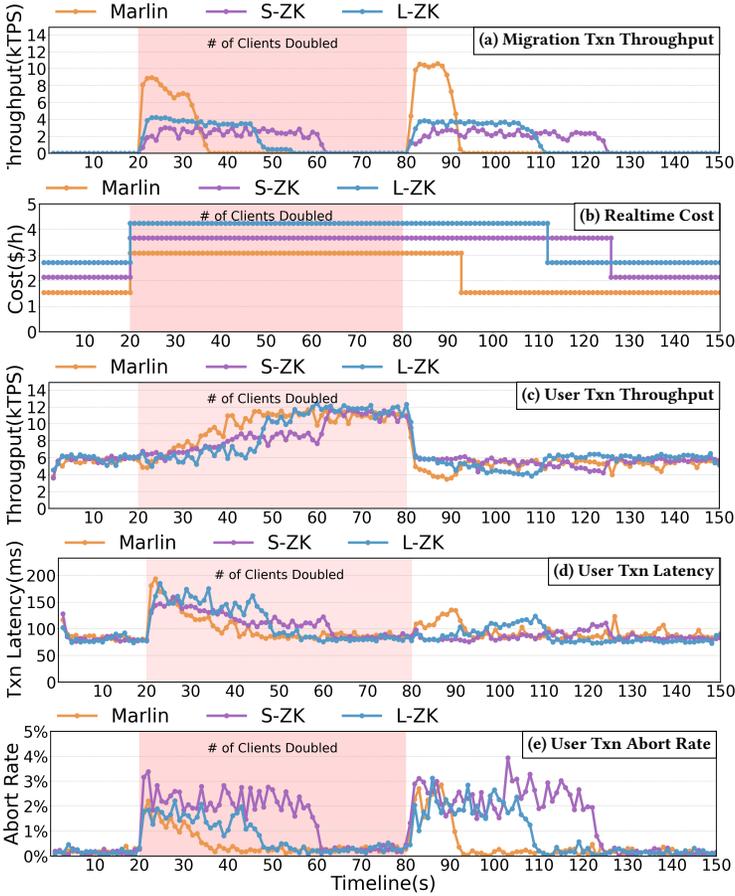
\begin{figure}[t]
        \vspace{-.10in}

    \centering
    \newcommand{\usrmarlin}{figures/plots/experiment/ev4-dynamic-throughput.png}
    \newcommand{\migrmarlin}{figures/plots/experiment/ev4-dynamic-latency.png}
    \newcommand{\usrszk}{figures/plots/experiment/ev4-dynamic-abort.png}
    \newcommand{\migrszk}{figures/plots/experiment/ev4-dynamic-migr-tps.png}
    \newcommand{\usrlzk}{figures/plots/experiment/ev4-dynamic-realtime-cost.png}

    \newlength{\subfigspace}
    \newlength{\titlexshift}
    \newlength{\titleyshift}
    \newlength{\titlefontsize}

    \setlength{\subfigspace}{-.15in}
    \setlength{\titlexshift}{-5pt}
    \setlength{\titleyshift}{-1pt}
    \setlength{\titlefontsize}{6.5pt}  


    \begin{subfigure}[t]{0.7\columnwidth}
        \centering
        \begin{tikzpicture}
            \node[inner sep=0pt] (image) {\includegraphics[width=\linewidth]{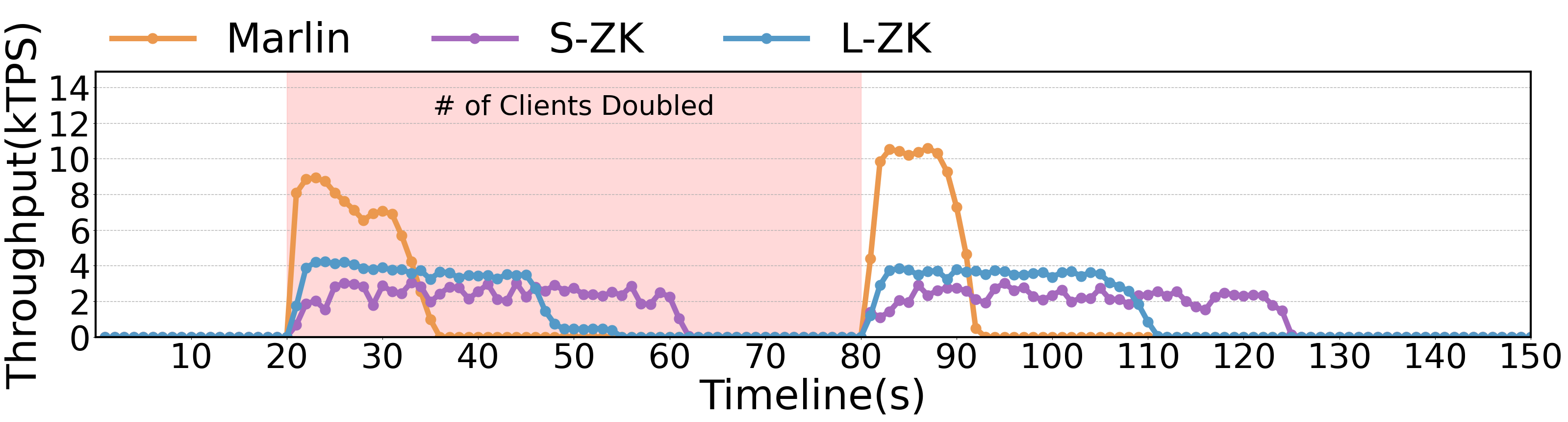}};
            \node[anchor=north east, font={\fontsize{\titlefontsize}{1.2\titlefontsize}\selectfont\bfseries}, draw, fill=white, inner sep=2pt, xshift=1.3\titlexshift, yshift=12.5\titleyshift] 
                at (image.north east) {(a) Migration Txn Throughput};
        \end{tikzpicture}
        \phantomcaption 
        \label{fig:ev4-f1-a}
    \end{subfigure}
    \par\vspace{\subfigspace}
            \vspace{-.15in}

    \begin{subfigure}[t]{0.7\columnwidth}
        \centering
        \begin{tikzpicture}
            \node[inner sep=0pt] (image) {\includegraphics[width=\linewidth]{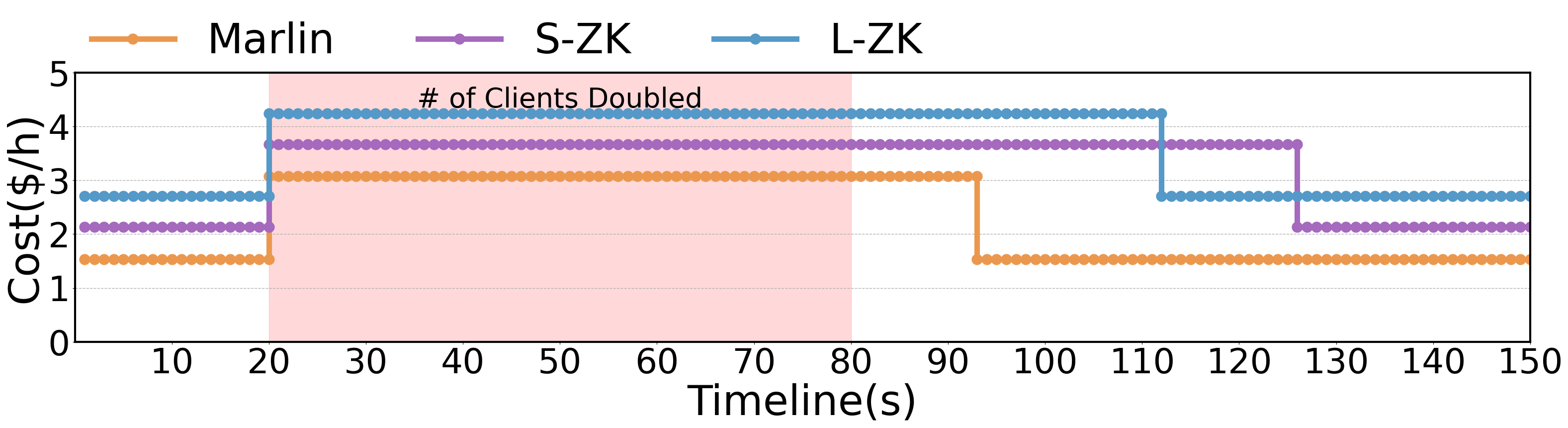}};
            \node[anchor=north east, font={\fontsize{\titlefontsize}{1.2\titlefontsize}\selectfont\bfseries}, draw, fill=white, inner sep=2pt, xshift=1.3\titlexshift, yshift=12.5\titleyshift] 
                at (image.north east) {(b) Realtime Cost};
        \end{tikzpicture}
        \phantomcaption 
        \label{fig:ev4-f1-b}
    \end{subfigure}
    \par\vspace{\subfigspace}
                \vspace{-.15in}

    \begin{subfigure}[t]{0.7\columnwidth}
        \centering
        \begin{tikzpicture}
            \node[inner sep=0pt] (image) {\includegraphics[width=\linewidth]{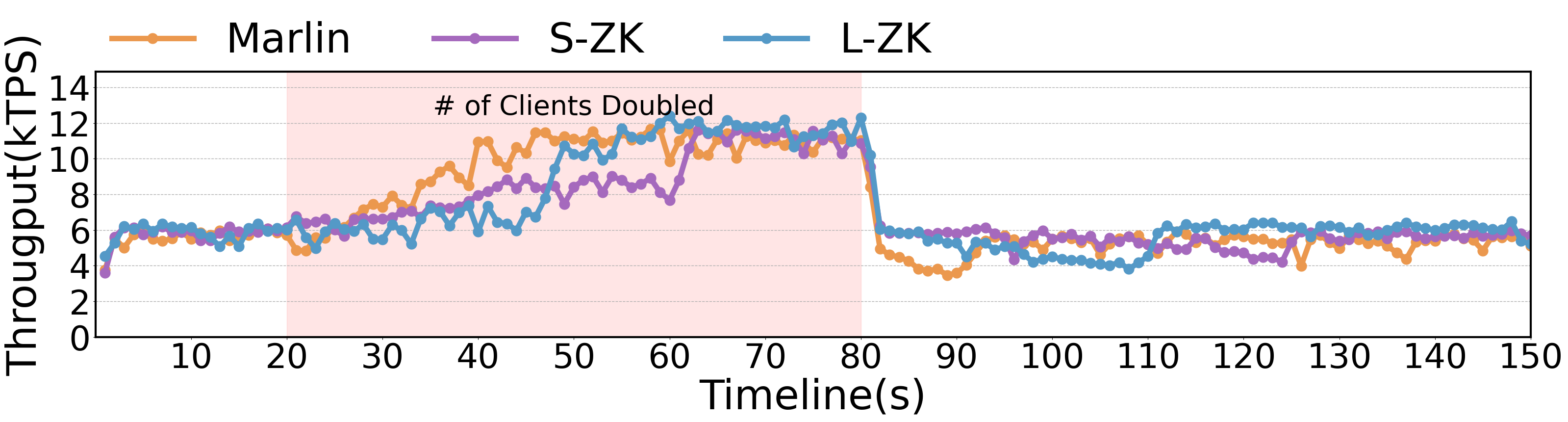}};
            \node[anchor=north east, font={\fontsize{\titlefontsize}{1.2\titlefontsize}\selectfont\bfseries}, draw, fill=white, inner sep=2pt, xshift=1.3\titlexshift, yshift=12.5\titleyshift] 
                at (image.north east) {(c) User Txn Throughput};
        \end{tikzpicture}
        \phantomcaption 
        \label{fig:ev4-f1-c}
    \end{subfigure}
    \par\vspace{\subfigspace}
                \vspace{-.15in}

    \begin{subfigure}[t]{0.7\columnwidth}
        \centering
        \begin{tikzpicture}
            \node[inner sep=0pt] (image) {\includegraphics[width=\linewidth]{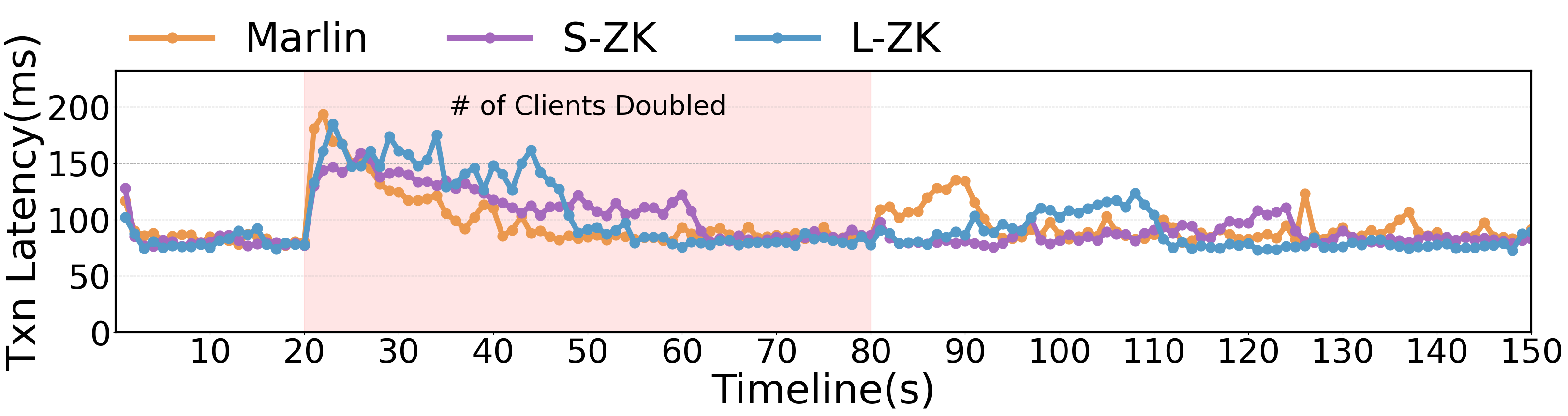}};
            \node[anchor=north east, font={\fontsize{\titlefontsize}{1.2\titlefontsize}\selectfont\bfseries}, draw, fill=white, inner sep=2pt, xshift=1.3\titlexshift, yshift=12.5\titleyshift] 
                at (image.north east) {(d) User Txn Latency};
        \end{tikzpicture}
        \phantomcaption      
        \label{fig:ev4-f1-d}
    \end{subfigure}
    \par\vspace{\subfigspace}
                \vspace{-.15in}

    \begin{subfigure}[t]{0.7\columnwidth}
        \centering
        \begin{tikzpicture}
            \node[inner sep=0pt] (image) {\includegraphics[width=\linewidth]{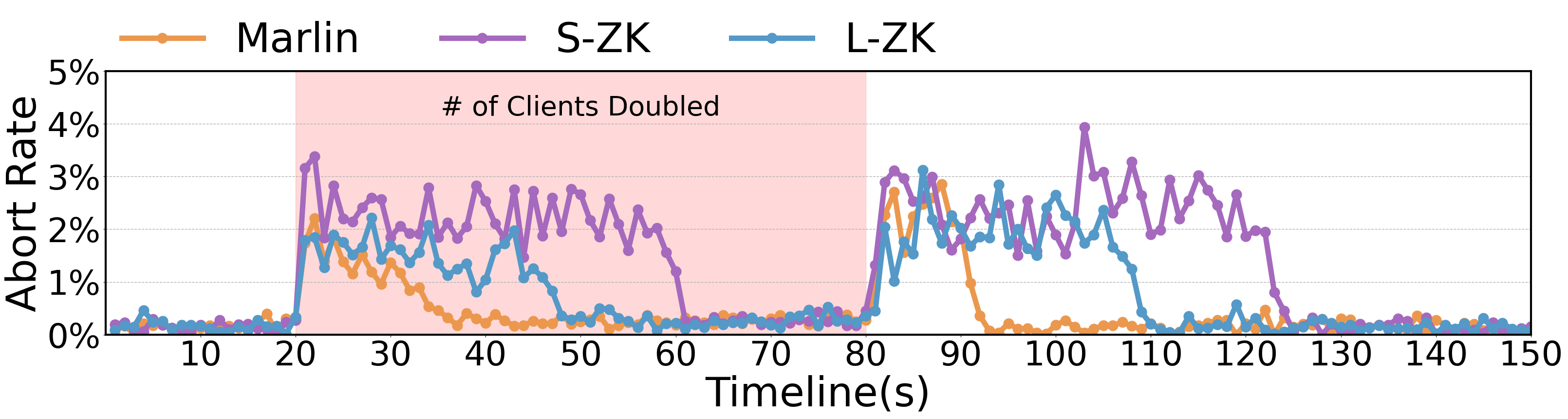}};
            \node[anchor=north east, font={\fontsize{\titlefontsize}{1.2\titlefontsize}\selectfont\bfseries}, draw, fill=white, inner sep=2pt, xshift=1.3\titlexshift, yshift=12.5\titleyshift] 
                at (image.north east) {(e) User Txn Abort Rate};
        \end{tikzpicture}
        \phantomcaption 
        \label{fig:ev4-f1-e}
    \end{subfigure}
    \vspace{-.3in}
    \caption{Realtime Performance of Dynamic Workloads}

    \label{fig:ev4-f1}
\end{figure}

As illustrated in \figref{fig:ev3_f1}, \system achieved up to 4.9$\times$ shorter migration duration than \zk-based methods \edit{and up to 9.5$\times$ shorter than \fdb} across all scales. This improvement is notably more pronounced in the geo-distributed setting, compared to the 2.5$\times$ gain observed in the single-region case in ~\secref{ssec:exp_perf_tradeoff}. The wider performance gap is due to the centralized nature of \zk-based approaches, which incur higher latency from cross-region communication. In contrast, \system’s distributed metadata management allows each region to manage its \gtbl partition independently, inherently co-locating coordination with compute and eliminating cross-region communication during migration. Interestingly, \lzk's migration duration was similar to \szk at all scales, as its hardware advantage was offset by cross-region latency. \edit{While \fdb scales better than \zk, it incurs significantly longer migration times because each migration triggers a metadata update in \fdb, requiring multiple cross-region round trips (e.g., \textit{GetReadVersion} and commit requests), whereas \zk-based methods require only one.} Regarding cost, \system remains the most cost-efficient by eliminating a dedicated coordination service. The extent of \system’s cost advantage remains consistent with single-region deployments.

\reviseBegin
While \zk-based approaches and \fdb can deploy one coordination cluster in each region to reduce cross-region latency, this increases the number of coordination clusters, leading to higher costs and operational complexity. For example, placing a \zk cluster in every region can raise the \textit{Meta Cost} by up to 4$\times$ compared to the current setup. In contrast, \system co-locates coordination with compute nodes by design, avoiding additional cost overhead. These benefits make \system particularly effective in geo-distributed settings.
\reviseEnd


\subsection{Dynamic Workloads}
\label{ssec:exp_perf_sens}
This experiment simulates a bursty workload scenario. The workload starts with 400 clients, scales to 800 at the 20th second, holds for 60 seconds, and drops back to 400 at the 80th second. The cluster begins with 8 compute nodes, scales out to 16, then returns to 8. An efficient coordination mechanism enables rapid scale-out and scale-in.


\noindent \textbf{Performance.} As shown in \figref{fig:ev4-f1-a}, \system achieves significantly higher migration throughput than \zk-based approaches, completing scale-out 2.6$\times$ and 2.3$\times$ faster than \szk and \lzk, respectively, and scale-in 3.8$\times$ and 2.6$\times$ faster. Rapid scaling enables the database to reach higher user transaction throughput ($\sim$12k TPS) more quickly, as seen in \figref{fig:ev4-f1-c}. Furthermore, \system minimizes the performance impact of reconfigurations on user transactions. This can be evident by \figref{fig:ev4-f1-c} and \figref{fig:ev4-f1-d}: the latency and abort rates of user transactions return to normal levels faster than \zk-based approaches.


\noindent \textbf{Cost efficiency.} As shown in ~\figref{fig:ev4-f1-b}, \system incurs lowest real-time cost under the dynamic workload. This benefit comes from two aspects. \circleb{1} \system avoids the upfront cost of external coordination cluster. From 0th to 20th second, \system reduces the cost by 1.4$\times$ and 1.8$\times$ compared to \szk and \lzk. \circleb{2} The rapid scale-in releases compute nodes more promptly, improving resource utilization and cost efficiency. \system reduces compute nodes 12 seconds later after the user workload drops, while \szk and \lzk reduce compute nodes after 45 seconds and 32 seconds.

\reviseBegin
\subsection{Membership Change Performance}\label{ssec:exp_mem_perf}
\begin{figure}[t]
        \vspace{-.1in}

    	\begin{subfigure}{\columnwidth}%
            \centering
            \includegraphics[height=3.3cm,keepaspectratio]
            {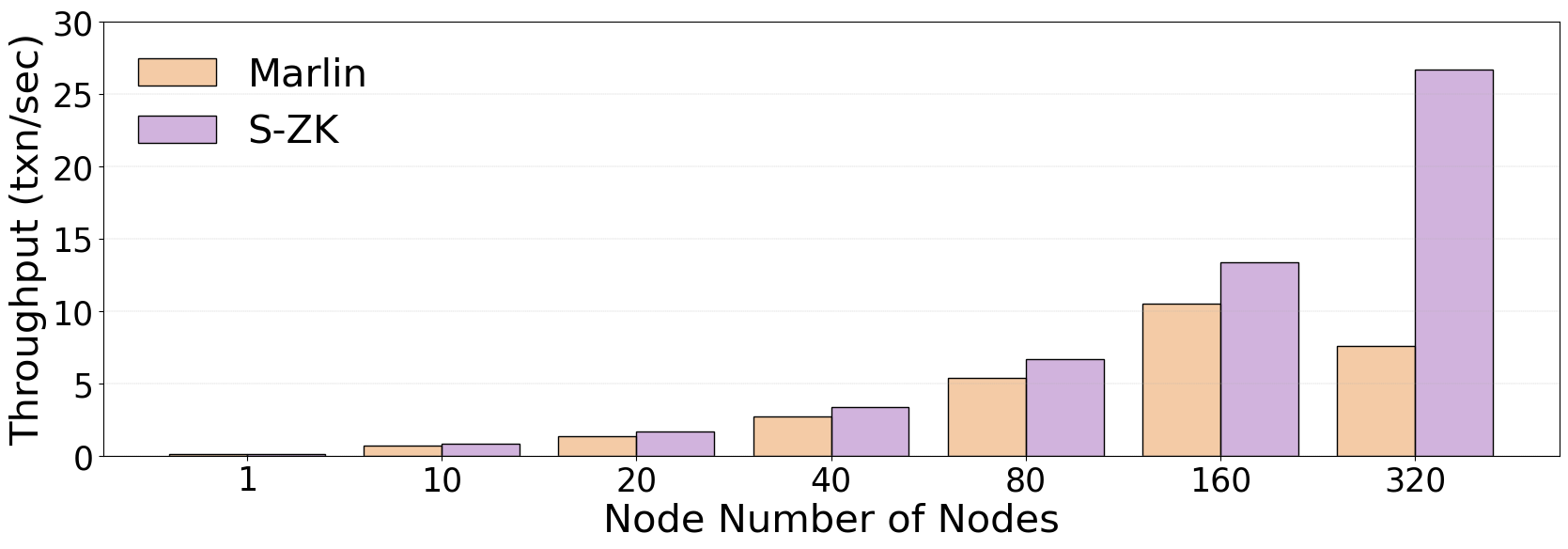}  
            \label{fig:ev5_f1_a}
        \end{subfigure}%

    \figcolor{}
    \vspace{-.15in}
    \caption{\edit{\mtbl Stress Test}}

    \label{fig:ev5_f1}
\end{figure}




Previous experiments primarily investigated migration performance, limited to updates on \gtbl. This experiment specifically evaluates the performance of group membership updates under \textit{intensive membership changes}, effectively stress-testing \mtbl. To reflect realistic loads, we simulate each compute node with one thread continuously issuing membership update requests, including node additions and removals. We scale the number of nodes by increasing threads number. Each thread issues a membership update every 15 seconds, similar to the monitoring interval utilized by autoscaling systems to determine the necessity for reconfiguration.

As shown in ~\figref{fig:ev5_f1}, \system performs comparably to \zk-based approaches up to 160 nodes. Beyond that point, performance degrades due to the overhead of optimistic concurrency control in the TryLog() API for \syslog, which incurs retries under high contention. This performance disparity is independent of our decision to integrate the coordination service into \system.

In practice, group membership changes are infrequent and initiated by a small subset of nodes rather than the entire cluster simultaneously, resulting in much lower intensity than our stress test. Thus, \system's membership change performance is likely sufficient in real-world scenarios. Moreover, established techniques for improving optimistic concurrency control under high contention (e.g., \cite{Yu2016, Harder1984, Mensace1982, Reiher1988, Carey1987}) could be incorporated into \system, which we leave for future work.



\reviseEnd

\section{Related Work}
\label{sec:related}


Existing database systems adopt one of the following three approaches to fault-tolerant metadata coordination. We discuss their relevance and difference with \system in this section.

\noindent \textbf{Single-master coordination services.} Prevalent converged coordination services take a single-master architecture, including lock services such as Chubby~\cite{burrows2006chubby} and general configuration stores such as Apache \zk~\cite{hunt2010zookeeper}, Doozer~\cite{doozer}, etcd~\cite{etcd}, and FireScroll~\cite{firescroll}. They are internally replicated for fault-tolerance and backed by an atomic broadcast or consensus protocol, e.g. ZAB~\cite{zab}, Paxos~\cite{paxos}, or Raft~\cite{raft}, for consistency of the ordering of operations. Such coordination services introduce cost overheads and deliver unideal performance in response to autoscaling workloads. 


\noindent \textbf{Sharded coordination services.} Some works divide coordination metadata into multiple shards and assign each shard to a separate consensus group, each with an independent master node. This approach is equivalent to deploying multiple coordination service clusters to balance the load of update requests. Consul~\cite{consul} is a primary example of this approach that offers little cross-shard consistency guarantees. ZooNet~\cite{zoonet} stitches the consistency of metadata operations between shards by fixing one coordination service cluster per datacenter and adding remote learners. Zelos~\cite{zelos} is a sharded \zk API atop Delos, a shared log system~\cite{delos}. Other examples include the rebuilt \zk in Clickhouse~\cite{clickhouse} and the transactional metadata key-value stores in Snowflake~\cite{snowflake} and ByConity~\cite{byconity}. Sharded services improve metadata scalability but have three significant drawbacks. First, they risk weakening metadata consistency for cross-shard operations, which are common in autoscaling migrations. Second, they dedicate even more resources to metadata coordination, enlarging the cost overhead. Third, they increase system architecture complexity. 


\noindent \textbf{Coordination without a dedicated service.} Some cloud systems have explored techniques to facilitate metadata coordination without a dedicated service. KRaft~\cite{kraft} proposes a simplified architecture for Kafka, a consistent messaging service~\cite{kafka}, by replacing \zk with embedded metadata quorums. FaasKeeper~\cite{FaaSKeeper} is a serverless variant of \zk, built entirely upon on-demand serverless functions and shared disaggregated storage. 


\mcomment{\wjh{discuss cockroachdb/neondb/polardb etc here as well}}

\vspace{-0.1in}
\section{Conclusion}
\label{sec:conclusion}

We propose \system, a cloud-native coordination mechanism for autoscaling databases with storage disaggregation. By integrating coordination within the database, \system achieves linear scalability, cost efficiency, and operational simplicity. It enables safe failover without external coordination services by permitting cross-node updates on coordination state, and introduces \mcommit, an optimized commit protocol, to ensure strong transactional guarantees under race conditions.
\everypar{\looseness=-1}




\clearpage

\bibliographystyle{ACM-Reference-Format}
\bibliography{base}


\begin{thebibliography}{102}


\ifx \showCODEN    \undefined \def \showCODEN     #1{\unskip}     \fi
\ifx \showDOI      \undefined \def \showDOI       #1{#1}\fi
\ifx \showISBNx    \undefined \def \showISBNx     #1{\unskip}     \fi
\ifx \showISBNxiii \undefined \def \showISBNxiii  #1{\unskip}     \fi
\ifx \showISSN     \undefined \def \showISSN      #1{\unskip}     \fi
\ifx \showLCCN     \undefined \def \showLCCN      #1{\unskip}     \fi
\ifx \shownote     \undefined \def \shownote      #1{#1}          \fi
\ifx \showarticletitle \undefined \def \showarticletitle #1{#1}   \fi
\ifx \showURL      \undefined \def \showURL       {\relax}        \fi
\providecommand\bibfield[2]{#2}
\providecommand\bibinfo[2]{#2}
\providecommand\natexlab[1]{#1}
\providecommand\showeprint[2][]{arXiv:#2}

\bibitem[\protect\citeauthoryear{??}{etc}{2024}]%
        {etcd}
 \bibinfo{year}{2024}\natexlab{}.
\newblock \bibinfo{title}{etcd}.
\newblock \bibinfo{howpublished}{\url{https://github.com/etcd-io/etcd}}.
\newblock
\newblock
\shownote{Accessed: 2024-01-01.}


\bibitem[\protect\citeauthoryear{??}{gRP}{2024}]%
        {gRPC}
 \bibinfo{year}{2024}\natexlab{}.
\newblock \bibinfo{title}{gRPC}.
\newblock \bibinfo{howpublished}{\url{https://grpc.io/}}.
\newblock
\newblock
\shownote{Accessed: 2024-01-01.}


\bibitem[\protect\citeauthoryear{{Ali Zaveri and Suyog Mapara and David Devecsery}}{{Ali Zaveri and Suyog Mapara and David Devecsery}}{2022}]%
        {zelos}
\bibfield{author}{\bibinfo{person}{{Ali Zaveri and Suyog Mapara and David Devecsery}}.} \bibinfo{year}{2022}\natexlab{}.
\newblock \bibinfo{title}{Introducing Zelos: A ZooKeeper API leveraging Delos}.
\newblock \bibinfo{howpublished}{\url{https://engineering.fb.com/2022/06/08/developer-tools/zelos/}}.
\newblock
\newblock
\shownote{Accessed: 2024-10-16.}


\bibitem[\protect\citeauthoryear{{Amazon Web Services}}{{Amazon Web Services}}{2023a}]%
        {redshift_spectrum}
\bibfield{author}{\bibinfo{person}{{Amazon Web Services}}.} \bibinfo{year}{2023}\natexlab{a}.
\newblock \bibinfo{booktitle}{\emph{{Amazon Redshift Spectrum}}}.
\newblock
\urldef\tempurl%
\url{https://aws.amazon.com/redshift/redshift-spectrum/}
\showURL{%
\tempurl}
\newblock
\shownote{Accessed: 2024-11-13.}


\bibitem[\protect\citeauthoryear{{Amazon Web Services}}{{Amazon Web Services}}{2023b}]%
        {aws_s3_express_one_zone}
\bibfield{author}{\bibinfo{person}{{Amazon Web Services}}.} \bibinfo{year}{2023}\natexlab{b}.
\newblock \bibinfo{booktitle}{\emph{Amazon S3 Express One Zone Storage Class}}.
\newblock
\urldef\tempurl%
\url{https://aws.amazon.com/s3/storage-classes/express-one-zone/}
\showURL{%
\tempurl}
\newblock
\shownote{Accessed: 2025-05-02.}


\bibitem[\protect\citeauthoryear{{Amazon Web Services}}{{Amazon Web Services}}{2023c}]%
        {aqua_redshift}
\bibfield{author}{\bibinfo{person}{{Amazon Web Services}}.} \bibinfo{year}{2023}\natexlab{c}.
\newblock \bibinfo{booktitle}{\emph{{AQUA for Amazon Redshift}}}.
\newblock
\urldef\tempurl%
\url{https://aws.amazon.com/blogs/aws/new-aqua-advanced-query-accelerator-for-amazon-redshift//}
\showURL{%
\tempurl}
\newblock
\shownote{Accessed: 2024-11-13.}


\bibitem[\protect\citeauthoryear{Antonopoulos, Budovski, Diaconu, Hernandez~Saenz, Hu, Kodavalla, Kossmann, Lingam, Minhas, Prakash, et~al\mbox{.}}{Antonopoulos et~al\mbox{.}}{2019}]%
        {socrates}
\bibfield{author}{\bibinfo{person}{Panagiotis Antonopoulos}, \bibinfo{person}{Alex Budovski}, \bibinfo{person}{Cristian Diaconu}, \bibinfo{person}{Alejandro Hernandez~Saenz}, \bibinfo{person}{Jack Hu}, \bibinfo{person}{Hanuma Kodavalla}, \bibinfo{person}{Donald Kossmann}, \bibinfo{person}{Sandeep Lingam}, \bibinfo{person}{Umar~Farooq Minhas}, \bibinfo{person}{Naveen Prakash}, {et~al\mbox{.}}} \bibinfo{year}{2019}\natexlab{}.
\newblock \showarticletitle{Socrates: The new sql server in the cloud}. In \bibinfo{booktitle}{\emph{Proceedings of the 2019 International Conference on Management of Data}}. \bibinfo{pages}{1743--1756}.
\newblock


\bibitem[\protect\citeauthoryear{Azure}{Azure}{2025}]%
        {azure_table_storage}
\bibfield{author}{\bibinfo{person}{Microsoft Azure}.} \bibinfo{year}{2025}\natexlab{}.
\newblock \bibinfo{title}{Azure Table Storage}.
\newblock \bibinfo{howpublished}{Online}.
\newblock
\urldef\tempurl%
\url{https://azure.microsoft.com/en-us/products/storage/tables}
\showURL{%
\tempurl}
\newblock
\shownote{Accessed: 2025-01-15.}


\bibitem[\protect\citeauthoryear{Babaoglu and Toueg}{Babaoglu and Toueg}{1993}]%
        {babaoglu1993nonblocking}
\bibfield{author}{\bibinfo{person}{Ozalp Babaoglu} {and} \bibinfo{person}{Sam Toueg}.} \bibinfo{year}{1993}\natexlab{}.
\newblock \showarticletitle{Understanding non-blocking atomic commitment}. In \bibinfo{booktitle}{\emph{Distributed systems}}.
\newblock


\bibitem[\protect\citeauthoryear{Balakrishnan, Flinn, Shen, Dharamshi, Jafri, Shi, Ghosh, Hassan, Sagar, Shi, et~al\mbox{.}}{Balakrishnan et~al\mbox{.}}{2020a}]%
        {balakrishnan2020virtual}
\bibfield{author}{\bibinfo{person}{Mahesh Balakrishnan}, \bibinfo{person}{Jason Flinn}, \bibinfo{person}{Chen Shen}, \bibinfo{person}{Mihir Dharamshi}, \bibinfo{person}{Ahmed Jafri}, \bibinfo{person}{Xiao Shi}, \bibinfo{person}{Santosh Ghosh}, \bibinfo{person}{Hazem Hassan}, \bibinfo{person}{Aaryaman Sagar}, \bibinfo{person}{Rhed Shi}, {et~al\mbox{.}}} \bibinfo{year}{2020}\natexlab{a}.
\newblock \showarticletitle{Virtual consensus in delos}. In \bibinfo{booktitle}{\emph{14th USENIX Symposium on Operating Systems Design and Implementation (OSDI 20)}}. \bibinfo{pages}{617--632}.
\newblock


\bibitem[\protect\citeauthoryear{Balakrishnan, Flinn, Shen, Dharamshi, Jafri, Shi, Ghosh, Hassan, Sagar, Shi, Liu, Gruszczynski, Zhang, Hoang, Yossef, Richard, and Song}{Balakrishnan et~al\mbox{.}}{2020b}]%
        {delos}
\bibfield{author}{\bibinfo{person}{Mahesh Balakrishnan}, \bibinfo{person}{Jason Flinn}, \bibinfo{person}{Chen Shen}, \bibinfo{person}{Mihir Dharamshi}, \bibinfo{person}{Ahmed Jafri}, \bibinfo{person}{Xiao Shi}, \bibinfo{person}{Santosh Ghosh}, \bibinfo{person}{Hazem Hassan}, \bibinfo{person}{Aaryaman Sagar}, \bibinfo{person}{Rhed Shi}, \bibinfo{person}{Jingming Liu}, \bibinfo{person}{Filip Gruszczynski}, \bibinfo{person}{Xianan Zhang}, \bibinfo{person}{Huy Hoang}, \bibinfo{person}{Ahmed Yossef}, \bibinfo{person}{Francois Richard}, {and} \bibinfo{person}{Yee~Jiun Song}.} \bibinfo{year}{2020}\natexlab{b}.
\newblock \showarticletitle{Virtual Consensus in Delos}. In \bibinfo{booktitle}{\emph{14th USENIX Symposium on Operating Systems Design and Implementation (OSDI 20)}}. \bibinfo{publisher}{USENIX Association}, \bibinfo{pages}{617--632}.
\newblock
\showISBNx{978-1-939133-19-9}
\urldef\tempurl%
\url{https://www.usenix.org/conference/osdi20/presentation/balakrishnan}
\showURL{%
\tempurl}


\bibitem[\protect\citeauthoryear{Bernstein, Bykov, Geller, Kliot, and Thelin}{Bernstein et~al\mbox{.}}{2014}]%
        {bernstein2014orleans}
\bibfield{author}{\bibinfo{person}{Phil Bernstein}, \bibinfo{person}{Sergey Bykov}, \bibinfo{person}{Alan Geller}, \bibinfo{person}{Gabriel Kliot}, {and} \bibinfo{person}{Jorgen Thelin}.} \bibinfo{year}{2014}\natexlab{}.
\newblock \showarticletitle{Orleans: Distributed virtual actors for programmability and scalability}. In \bibinfo{booktitle}{\emph{MSR-TR-2014--41}}.
\newblock


\bibitem[\protect\citeauthoryear{Bernstein}{Bernstein}{1987}]%
        {bernstein1987concurrency}
\bibfield{author}{\bibinfo{person}{Philip~A Bernstein}.} \bibinfo{year}{1987}\natexlab{}.
\newblock \bibinfo{booktitle}{\emph{Concurrency control and recovery in database systems}}. Vol.~\bibinfo{volume}{370}.
\newblock \bibinfo{publisher}{Addison-wesley New York}.
\newblock


\bibitem[\protect\citeauthoryear{Bernstein, Reid, and Das}{Bernstein et~al\mbox{.}}{2011}]%
        {bernstein2011hyder}
\bibfield{author}{\bibinfo{person}{Philip~A Bernstein}, \bibinfo{person}{Colin~W Reid}, {and} \bibinfo{person}{Sudipto Das}.} \bibinfo{year}{2011}\natexlab{}.
\newblock \showarticletitle{Hyder-A Transactional Record Manager for Shared Flash.}. In \bibinfo{booktitle}{\emph{CIDR}}, Vol.~\bibinfo{volume}{11}. \bibinfo{pages}{9--20}.
\newblock


\bibitem[\protect\citeauthoryear{Brantner, Florescu, Graf, Kossmann, and Kraska}{Brantner et~al\mbox{.}}{2008}]%
        {brantner2008building}
\bibfield{author}{\bibinfo{person}{Matthias Brantner}, \bibinfo{person}{Daniela Florescu}, \bibinfo{person}{David Graf}, \bibinfo{person}{Donald Kossmann}, {and} \bibinfo{person}{Tim Kraska}.} \bibinfo{year}{2008}\natexlab{}.
\newblock \showarticletitle{Building a database on S3}. In \bibinfo{booktitle}{\emph{Proceedings of the 2008 ACM SIGMOD international conference on Management of data}}. \bibinfo{pages}{251--264}.
\newblock


\bibitem[\protect\citeauthoryear{Burrows}{Burrows}{2006}]%
        {burrows2006chubby}
\bibfield{author}{\bibinfo{person}{Mike Burrows}.} \bibinfo{year}{2006}\natexlab{}.
\newblock \showarticletitle{The Chubby lock service for loosely-coupled distributed systems}. In \bibinfo{booktitle}{\emph{Proceedings of the 7th symposium on Operating systems design and implementation}}. \bibinfo{pages}{335--350}.
\newblock


\bibitem[\protect\citeauthoryear{{ByConity}}{{ByConity}}{2024}]%
        {byconity}
\bibfield{author}{\bibinfo{person}{{ByConity}}.} \bibinfo{year}{2024}\natexlab{}.
\newblock \bibinfo{title}{ByConity: an open source cloud data warehouse}.
\newblock \bibinfo{howpublished}{\url{https://github.com/ByConity/ByConity}}.
\newblock
\newblock
\shownote{Accessed: 2024-10-16.}


\bibitem[\protect\citeauthoryear{Cao, Liu, Cheng, Zheng, Li, Wu, Ouyang, Wang, Wang, Kuan, et~al\mbox{.}}{Cao et~al\mbox{.}}{2020}]%
        {cao2020polardb}
\bibfield{author}{\bibinfo{person}{Wei Cao}, \bibinfo{person}{Yang Liu}, \bibinfo{person}{Zhushi Cheng}, \bibinfo{person}{Ning Zheng}, \bibinfo{person}{Wei Li}, \bibinfo{person}{Wenjie Wu}, \bibinfo{person}{Linqiang Ouyang}, \bibinfo{person}{Peng Wang}, \bibinfo{person}{Yijing Wang}, \bibinfo{person}{Ray Kuan}, {et~al\mbox{.}}} \bibinfo{year}{2020}\natexlab{}.
\newblock \showarticletitle{$\{$POLARDB$\}$ meets computational storage: Efficiently support analytical workloads in $\{$Cloud-Native$\}$ relational database}. In \bibinfo{booktitle}{\emph{18th USENIX conference on file and storage technologies (FAST 20)}}. \bibinfo{pages}{29--41}.
\newblock


\bibitem[\protect\citeauthoryear{Carey}{Carey}{1987}]%
        {Carey1987}
\bibfield{author}{\bibinfo{person}{M.~J. Carey}.} \bibinfo{year}{1987}\natexlab{}.
\newblock \showarticletitle{Improving the performance of an optimistic concurrency control algorithm through timestamps and versions}.
\newblock \bibinfo{journal}{\emph{IEEE Transactions on Software Engineering}} \bibinfo{volume}{13}, \bibinfo{number}{6} (\bibinfo{year}{1987}), \bibinfo{pages}{746--751}.
\newblock


\bibitem[\protect\citeauthoryear{Chandra, Griesemer, and Redstone}{Chandra et~al\mbox{.}}{2007}]%
        {chandra2007paxos}
\bibfield{author}{\bibinfo{person}{Tushar~D Chandra}, \bibinfo{person}{Robert Griesemer}, {and} \bibinfo{person}{Joshua Redstone}.} \bibinfo{year}{2007}\natexlab{}.
\newblock \showarticletitle{Paxos made live: an engineering perspective}. In \bibinfo{booktitle}{\emph{Proceedings of the twenty-sixth annual ACM symposium on Principles of distributed computing}}. \bibinfo{pages}{398--407}.
\newblock


\bibitem[\protect\citeauthoryear{Chang, Dean, Ghemawat, Hsieh, Wallach, Burrows, Chandra, Fikes, and Gruber}{Chang et~al\mbox{.}}{2008}]%
        {chang2008bigtable}
\bibfield{author}{\bibinfo{person}{Fay Chang}, \bibinfo{person}{Jeffrey Dean}, \bibinfo{person}{Sanjay Ghemawat}, \bibinfo{person}{Wilson~C Hsieh}, \bibinfo{person}{Deborah~A Wallach}, \bibinfo{person}{Mike Burrows}, \bibinfo{person}{Tushar Chandra}, \bibinfo{person}{Andrew Fikes}, {and} \bibinfo{person}{Robert~E Gruber}.} \bibinfo{year}{2008}\natexlab{}.
\newblock \showarticletitle{Bigtable: A distributed storage system for structured data}.
\newblock \bibinfo{journal}{\emph{ACM Transactions on Computer Systems (TOCS)}} \bibinfo{volume}{26}, \bibinfo{number}{2} (\bibinfo{year}{2008}), \bibinfo{pages}{1--26}.
\newblock


\bibitem[\protect\citeauthoryear{{ClickHouse, Inc.}}{{ClickHouse, Inc.}}{2016}]%
        {clickhouse}
\bibfield{author}{\bibinfo{person}{{ClickHouse, Inc.}}} \bibinfo{year}{2016}\natexlab{}.
\newblock \bibinfo{title}{ClickHouse: Fast Open-Source OLAP DBMS}.
\newblock \bibinfo{howpublished}{\url{https://clickhouse.com/}}.
\newblock
\newblock
\shownote{Accessed: 2024-10-16.}


\bibitem[\protect\citeauthoryear{{Cockroach Labs}}{{Cockroach Labs}}{2024}]%
        {Cockroach}
\bibfield{author}{\bibinfo{person}{{Cockroach Labs}}.} \bibinfo{year}{2024}\natexlab{}.
\newblock \bibinfo{title}{CockroachDB}.
\newblock \bibinfo{howpublished}{\url{https://www.cockroachlabs.com/}}.
\newblock
\newblock
\shownote{Accessed: 2024-01-01.}


\bibitem[\protect\citeauthoryear{{Colin McCabe}}{{Colin McCabe}}{2020}]%
        {kraft}
\bibfield{author}{\bibinfo{person}{{Colin McCabe}}.} \bibinfo{year}{2020}\natexlab{}.
\newblock \bibinfo{title}{KIP-500: Replace ZooKeeper with a Self-Managed Metadata Quorum}.
\newblock \bibinfo{howpublished}{\url{https://cwiki.apache.org/confluence/display/KAFKA/KIP-500}}.
\newblock
\newblock
\shownote{Accessed: 2024-10-16.}


\bibitem[\protect\citeauthoryear{Cooper, Silberstein, Tam, Ramakrishnan, and Sears}{Cooper et~al\mbox{.}}{2010}]%
        {cooper2010benchmarking}
\bibfield{author}{\bibinfo{person}{Brian~F Cooper}, \bibinfo{person}{Adam Silberstein}, \bibinfo{person}{Erwin Tam}, \bibinfo{person}{Raghu Ramakrishnan}, {and} \bibinfo{person}{Russell Sears}.} \bibinfo{year}{2010}\natexlab{}.
\newblock \showarticletitle{Benchmarking cloud serving systems with YCSB}. In \bibinfo{booktitle}{\emph{Proceedings of the 1st ACM symposium on Cloud computing}}. \bibinfo{pages}{143--154}.
\newblock


\bibitem[\protect\citeauthoryear{Copik, Calotoiu, Zhou, Taranov, and Hoefler}{Copik et~al\mbox{.}}{2024}]%
        {FaaSKeeper}
\bibfield{author}{\bibinfo{person}{Marcin Copik}, \bibinfo{person}{Alexandru Calotoiu}, \bibinfo{person}{Pengyu Zhou}, \bibinfo{person}{Konstantin Taranov}, {and} \bibinfo{person}{Torsten Hoefler}.} \bibinfo{year}{2024}\natexlab{}.
\newblock \showarticletitle{FaaSKeeper: Learning from Building Serverless Services with ZooKeeper as an Example}. In \bibinfo{booktitle}{\emph{Proceedings of the 33rd International Symposium on High-Performance Parallel and Distributed Computing}} \emph{(\bibinfo{series}{HPDC ’24})}. \bibinfo{publisher}{ACM}, \bibinfo{pages}{94–108}.
\newblock
\urldef\tempurl%
\url{https://doi.org/10.1145/3625549.3658661}
\showDOI{\tempurl}


\bibitem[\protect\citeauthoryear{Corbett and et~al.}{Corbett and et~al.}{2012}]%
        {corbett12spanner}
\bibfield{author}{\bibinfo{person}{James~C. Corbett} {and} \bibinfo{person}{et al.}} \bibinfo{year}{2012}\natexlab{}.
\newblock \showarticletitle{{Spanner: Google's Globally-Distributed Database}}. In \bibinfo{booktitle}{\emph{OSDI}}. \bibinfo{pages}{251--264}.
\newblock


\bibitem[\protect\citeauthoryear{Corporation}{Corporation}{2024a}]%
        {IBMDB2DataSharing}
\bibfield{author}{\bibinfo{person}{IBM Corporation}.} \bibinfo{year}{2024}\natexlab{a}.
\newblock \bibinfo{booktitle}{\emph{IBM DB2 Data Sharing}}.
\newblock
\urldef\tempurl%
\url{https://www.ibm.com/docs/en/db2-for-zos/12?topic=db2-data-sharing}
\showURL{%
\tempurl}
\newblock
\shownote{IBM Documentation.}


\bibitem[\protect\citeauthoryear{Corporation}{Corporation}{2024b}]%
        {cosmosdb}
\bibfield{author}{\bibinfo{person}{Microsoft Corporation}.} \bibinfo{year}{2024}\natexlab{b}.
\newblock \bibinfo{title}{Azure Cosmos DB}.
\newblock
\newblock
\urldef\tempurl%
\url{https://azure.microsoft.com/en-us/products/cosmos-db}
\showURL{%
\tempurl}
\newblock
\shownote{Accessed: 2024-10-29.}


\bibitem[\protect\citeauthoryear{Cortez, Bonde, Muzio, Russinovich, Fontoura, and Bianchini}{Cortez et~al\mbox{.}}{2017}]%
        {cortez2017resource}
\bibfield{author}{\bibinfo{person}{Eli Cortez}, \bibinfo{person}{Anand Bonde}, \bibinfo{person}{Alexandre Muzio}, \bibinfo{person}{Mark Russinovich}, \bibinfo{person}{Marcus Fontoura}, {and} \bibinfo{person}{Ricardo Bianchini}.} \bibinfo{year}{2017}\natexlab{}.
\newblock \showarticletitle{Resource central: Understanding and predicting workloads for improved resource management in large cloud platforms}. In \bibinfo{booktitle}{\emph{Proceedings of the 26th Symposium on Operating Systems Principles}}. \bibinfo{pages}{153--167}.
\newblock


\bibitem[\protect\citeauthoryear{Cubukcu, Erdogan, Pathak, Sannakkayala, and Slot}{Cubukcu et~al\mbox{.}}{2021}]%
        {cubukcu2021citus}
\bibfield{author}{\bibinfo{person}{Umur Cubukcu}, \bibinfo{person}{Ozgun Erdogan}, \bibinfo{person}{Sumedh Pathak}, \bibinfo{person}{Sudhakar Sannakkayala}, {and} \bibinfo{person}{Marco Slot}.} \bibinfo{year}{2021}\natexlab{}.
\newblock \showarticletitle{Citus: Distributed postgresql for data-intensive applications}. In \bibinfo{booktitle}{\emph{Proceedings of the 2021 International Conference on Management of Data}}. \bibinfo{pages}{2490--2502}.
\newblock


\bibitem[\protect\citeauthoryear{Curino, Jones, Zhang, and Madden}{Curino et~al\mbox{.}}{2010}]%
        {curino2010schism}
\bibfield{author}{\bibinfo{person}{Carlo Curino}, \bibinfo{person}{Evan Philip~Charles Jones}, \bibinfo{person}{Yang Zhang}, {and} \bibinfo{person}{Samuel~R Madden}.} \bibinfo{year}{2010}\natexlab{}.
\newblock \showarticletitle{Schism: a workload-driven approach to database replication and partitioning}.
\newblock  (\bibinfo{year}{2010}).
\newblock


\bibitem[\protect\citeauthoryear{Dageville, Cruanes, Zukowski, Antonov, Avanes, Bock, Claybaugh, Engovatov, Hentschel, Huang, et~al\mbox{.}}{Dageville et~al\mbox{.}}{2016}]%
        {snowflake}
\bibfield{author}{\bibinfo{person}{Benoit Dageville}, \bibinfo{person}{Thierry Cruanes}, \bibinfo{person}{Marcin Zukowski}, \bibinfo{person}{Vadim Antonov}, \bibinfo{person}{Artin Avanes}, \bibinfo{person}{Jon Bock}, \bibinfo{person}{Jonathan Claybaugh}, \bibinfo{person}{Daniel Engovatov}, \bibinfo{person}{Martin Hentschel}, \bibinfo{person}{Jiansheng Huang}, {et~al\mbox{.}}} \bibinfo{year}{2016}\natexlab{}.
\newblock \showarticletitle{{The Snowflake Elastic Data Warehouse}}. In \bibinfo{booktitle}{\emph{SIGMOD}}.
\newblock


\bibitem[\protect\citeauthoryear{Das, Gupta, and Motivala}{Das et~al\mbox{.}}{2002}]%
        {das2002swim}
\bibfield{author}{\bibinfo{person}{Abhinandan Das}, \bibinfo{person}{Indranil Gupta}, {and} \bibinfo{person}{Ashish Motivala}.} \bibinfo{year}{2002}\natexlab{}.
\newblock \showarticletitle{Swim: Scalable weakly-consistent infection-style process group membership protocol}. In \bibinfo{booktitle}{\emph{Proceedings International Conference on Dependable Systems and Networks}}. IEEE, \bibinfo{pages}{303--312}.
\newblock


\bibitem[\protect\citeauthoryear{Das, Nishimura, Agrawal, and El~Abbadi}{Das et~al\mbox{.}}{2011}]%
        {das2011albatross}
\bibfield{author}{\bibinfo{person}{Sudipto Das}, \bibinfo{person}{Shoji Nishimura}, \bibinfo{person}{Divyakant Agrawal}, {and} \bibinfo{person}{Amr El~Abbadi}.} \bibinfo{year}{2011}\natexlab{}.
\newblock \showarticletitle{Albatross: Lightweight elasticity in shared storage databases for the cloud using live data migration}.
\newblock \bibinfo{journal}{\emph{Proceedings of the VLDB Endowment}} \bibinfo{volume}{4}, \bibinfo{number}{8} (\bibinfo{year}{2011}), \bibinfo{pages}{494--505}.
\newblock


\bibitem[\protect\citeauthoryear{Depoutovitch, Chen, Chen, Larson, Lin, Ng, Cui, Liu, Huang, Xiao, et~al\mbox{.}}{Depoutovitch et~al\mbox{.}}{2020}]%
        {depoutovitch2020taurus}
\bibfield{author}{\bibinfo{person}{Alex Depoutovitch}, \bibinfo{person}{Chong Chen}, \bibinfo{person}{Jin Chen}, \bibinfo{person}{Paul Larson}, \bibinfo{person}{Shu Lin}, \bibinfo{person}{Jack Ng}, \bibinfo{person}{Wenlin Cui}, \bibinfo{person}{Qiang Liu}, \bibinfo{person}{Wei Huang}, \bibinfo{person}{Yong Xiao}, {et~al\mbox{.}}} \bibinfo{year}{2020}\natexlab{}.
\newblock \showarticletitle{Taurus database: How to be fast, available, and frugal in the cloud}. In \bibinfo{booktitle}{\emph{Proceedings of the 2020 ACM SIGMOD International Conference on Management of Data}}. \bibinfo{pages}{1463--1478}.
\newblock


\bibitem[\protect\citeauthoryear{Documentation}{Documentation}{2025a}]%
        {s3_etag}
\bibfield{author}{\bibinfo{person}{Amazon Web~Services Documentation}.} \bibinfo{year}{2025}\natexlab{a}.
\newblock \bibinfo{booktitle}{\emph{Amazon S3 ETag Overview}}.
\newblock
\urldef\tempurl%
\url{https://docs.aws.amazon.com/AmazonS3/latest/API/API_Object.html}
\showURL{%
\tempurl}
\newblock
\shownote{Accessed: 2025-01-12.}


\bibitem[\protect\citeauthoryear{Documentation}{Documentation}{2025b}]%
        {gcs_etag}
\bibfield{author}{\bibinfo{person}{Google~Cloud Documentation}.} \bibinfo{year}{2025}\natexlab{b}.
\newblock \bibinfo{booktitle}{\emph{Google Cloud Storage ETag Usage}}.
\newblock
\urldef\tempurl%
\url{https://cloud.google.com/secret-manager/docs/etags}
\showURL{%
\tempurl}
\newblock
\shownote{Accessed: 2025-01-12.}


\bibitem[\protect\citeauthoryear{Documentation}{Documentation}{2025c}]%
        {azure_etag}
\bibfield{author}{\bibinfo{person}{Microsoft Documentation}.} \bibinfo{year}{2025}\natexlab{c}.
\newblock \bibinfo{booktitle}{\emph{Optimistic concurrency in Azure Storage with ETags}}.
\newblock
\urldef\tempurl%
\url{https://learn.microsoft.com/en-us/azure/storage/blobs/concurrency-manage}
\showURL{%
\tempurl}
\newblock
\shownote{Accessed: 2025-01-12.}


\bibitem[\protect\citeauthoryear{Elhemali, Gallagher, Tang, Gordon, Huang, Chen, Idziorek, Wang, Krog, Zhu, et~al\mbox{.}}{Elhemali et~al\mbox{.}}{2022}]%
        {elhemali2022dynamodb}
\bibfield{author}{\bibinfo{person}{Mostafa Elhemali}, \bibinfo{person}{Niall Gallagher}, \bibinfo{person}{Bin Tang}, \bibinfo{person}{Nick Gordon}, \bibinfo{person}{Hao Huang}, \bibinfo{person}{Haibo Chen}, \bibinfo{person}{Joseph Idziorek}, \bibinfo{person}{Mengtian Wang}, \bibinfo{person}{Richard Krog}, \bibinfo{person}{Zongpeng Zhu}, {et~al\mbox{.}}} \bibinfo{year}{2022}\natexlab{}.
\newblock \showarticletitle{Amazon $\{$DynamoDB$\}$: A Scalable, Predictably Performant, and Fully Managed $\{$NoSQL$\}$ Database Service}. In \bibinfo{booktitle}{\emph{2022 USENIX Annual Technical Conference (USENIX ATC 22)}}. \bibinfo{pages}{1037--1048}.
\newblock


\bibitem[\protect\citeauthoryear{Elmore, Arora, Taft, Pavlo, Agrawal, and El~Abbadi}{Elmore et~al\mbox{.}}{2015}]%
        {elmore2015squall}
\bibfield{author}{\bibinfo{person}{Aaron~J Elmore}, \bibinfo{person}{Vaibhav Arora}, \bibinfo{person}{Rebecca Taft}, \bibinfo{person}{Andrew Pavlo}, \bibinfo{person}{Divyakant Agrawal}, {and} \bibinfo{person}{Amr El~Abbadi}.} \bibinfo{year}{2015}\natexlab{}.
\newblock \showarticletitle{Squall: Fine-grained live reconfiguration for partitioned main memory databases}. In \bibinfo{booktitle}{\emph{Proceedings of the 2015 ACM SIGMOD International Conference on Management of Data}}. \bibinfo{pages}{299--313}.
\newblock


\bibitem[\protect\citeauthoryear{Eswaran, Gray, Lorie, and Traiger}{Eswaran et~al\mbox{.}}{1976}]%
        {eswaran1976notions}
\bibfield{author}{\bibinfo{person}{Kapali~P. Eswaran}, \bibinfo{person}{Jim~N Gray}, \bibinfo{person}{Raymond~A. Lorie}, {and} \bibinfo{person}{Irving~L. Traiger}.} \bibinfo{year}{1976}\natexlab{}.
\newblock \showarticletitle{The notions of consistency and predicate locks in a database system}.
\newblock \bibinfo{journal}{\emph{Commun. ACM}} \bibinfo{volume}{19}, \bibinfo{number}{11} (\bibinfo{year}{1976}), \bibinfo{pages}{624--633}.
\newblock


\bibitem[\protect\citeauthoryear{Ganesh, Kermarrec, and Massouli{\'e}}{Ganesh et~al\mbox{.}}{2003}]%
        {ganesh2003peer}
\bibfield{author}{\bibinfo{person}{Ayalvadi~J Ganesh}, \bibinfo{person}{A-M Kermarrec}, {and} \bibinfo{person}{Laurent Massouli{\'e}}.} \bibinfo{year}{2003}\natexlab{}.
\newblock \showarticletitle{Peer-to-peer membership management for gossip-based protocols}.
\newblock \bibinfo{journal}{\emph{IEEE transactions on computers}} \bibinfo{volume}{52}, \bibinfo{number}{2} (\bibinfo{year}{2003}), \bibinfo{pages}{139--149}.
\newblock


\bibitem[\protect\citeauthoryear{Gmach, Rolia, Cherkasova, and Kemper}{Gmach et~al\mbox{.}}{2007}]%
        {gmach2007workload}
\bibfield{author}{\bibinfo{person}{Daniel Gmach}, \bibinfo{person}{Jerry Rolia}, \bibinfo{person}{Ludmila Cherkasova}, {and} \bibinfo{person}{Alfons Kemper}.} \bibinfo{year}{2007}\natexlab{}.
\newblock \showarticletitle{Workload analysis and demand prediction of enterprise data center applications}. In \bibinfo{booktitle}{\emph{2007 IEEE 10th International Symposium on Workload Characterization}}. IEEE, \bibinfo{pages}{171--180}.
\newblock


\bibitem[\protect\citeauthoryear{{Google Cloud}}{{Google Cloud}}{2023}]%
        {alloydb}
\bibfield{author}{\bibinfo{person}{{Google Cloud}}.} \bibinfo{year}{2023}\natexlab{}.
\newblock \bibinfo{title}{{AlloyDB for PostgreSQL}}.
\newblock
\newblock
\urldef\tempurl%
\url{https://cloud.google.com/alloydb}
\showURL{%
\tempurl}
\newblock
\shownote{Accessed: 2024-11-13.}


\bibitem[\protect\citeauthoryear{Group}{Group}{2023}]%
        {postgresql}
\bibfield{author}{\bibinfo{person}{The PostgreSQL Global~Development Group}.} \bibinfo{year}{2023}\natexlab{}.
\newblock \bibinfo{title}{{PostgreSQL: The world's most advanced open source relational database}}.
\newblock
\newblock
\urldef\tempurl%
\url{https://www.postgresql.org/}
\showURL{%
\tempurl}
\newblock
\shownote{Accessed: 2024-10-29.}


\bibitem[\protect\citeauthoryear{Guo, Zeng, Wu, Hwang, Ren, Yu, Balakrishnan, and Bernstein}{Guo et~al\mbox{.}}{2022}]%
        {cornus}
\bibfield{author}{\bibinfo{person}{Zhihan Guo}, \bibinfo{person}{Xinyu Zeng}, \bibinfo{person}{Kan Wu}, \bibinfo{person}{Wuh{-}Chwen Hwang}, \bibinfo{person}{Ziwei Ren}, \bibinfo{person}{Xiangyao Yu}, \bibinfo{person}{Mahesh Balakrishnan}, {and} \bibinfo{person}{Philip~A. Bernstein}.} \bibinfo{year}{2022}\natexlab{}.
\newblock \showarticletitle{Cornus: Atomic Commit for a Cloud {DBMS} with Storage Disaggregation}.
\newblock \bibinfo{journal}{\emph{Proc. {VLDB} Endow.}} \bibinfo{volume}{16}, \bibinfo{number}{2} (\bibinfo{year}{2022}), \bibinfo{pages}{379--392}.
\newblock
\urldef\tempurl%
\url{https://doi.org/10.14778/3565816.3565837}
\showDOI{\tempurl}


\bibitem[\protect\citeauthoryear{{ha}}{{ha}}{2011}]%
        {doozer}
\bibfield{author}{\bibinfo{person}{{ha}}.} \bibinfo{year}{2011}\natexlab{}.
\newblock \bibinfo{title}{Doozer}.
\newblock \bibinfo{howpublished}{\url{https://github.com/ha/doozerd}}.
\newblock
\newblock
\shownote{Accessed: 2024-10-16.}


\bibitem[\protect\citeauthoryear{{HashiCorp}}{{HashiCorp}}{2020}]%
        {consul}
\bibfield{author}{\bibinfo{person}{{HashiCorp}}.} \bibinfo{year}{2020}\natexlab{}.
\newblock \bibinfo{title}{Consul}.
\newblock \bibinfo{howpublished}{\url{https://www.consul.io}}.
\newblock
\newblock
\shownote{Accessed: 2024-10-16.}


\bibitem[\protect\citeauthoryear{Huang, Liu, Cui, Fang, Ma, Xu, Shen, Tang, Zhou, Huang, et~al\mbox{.}}{Huang et~al\mbox{.}}{2020}]%
        {huang2020tidb}
\bibfield{author}{\bibinfo{person}{Dongxu Huang}, \bibinfo{person}{Qi Liu}, \bibinfo{person}{Qiu Cui}, \bibinfo{person}{Zhuhe Fang}, \bibinfo{person}{Xiaoyu Ma}, \bibinfo{person}{Fei Xu}, \bibinfo{person}{Li Shen}, \bibinfo{person}{Liu Tang}, \bibinfo{person}{Yuxing Zhou}, \bibinfo{person}{Menglong Huang}, {et~al\mbox{.}}} \bibinfo{year}{2020}\natexlab{}.
\newblock \showarticletitle{TiDB: a Raft-based HTAP database}.
\newblock \bibinfo{journal}{\emph{Proceedings of the VLDB Endowment}} \bibinfo{volume}{13}, \bibinfo{number}{12} (\bibinfo{year}{2020}), \bibinfo{pages}{3072--3084}.
\newblock


\bibitem[\protect\citeauthoryear{Hunt, Konar, Junqueira, and Reed}{Hunt et~al\mbox{.}}{2010}]%
        {hunt2010zookeeper}
\bibfield{author}{\bibinfo{person}{Patrick Hunt}, \bibinfo{person}{Mahadev Konar}, \bibinfo{person}{Flavio~P Junqueira}, {and} \bibinfo{person}{Benjamin Reed}.} \bibinfo{year}{2010}\natexlab{}.
\newblock \showarticletitle{$\{$ZooKeeper$\}$: Wait-free coordination for internet-scale systems}. In \bibinfo{booktitle}{\emph{2010 USENIX Annual Technical Conference (USENIX ATC 10)}}.
\newblock


\bibitem[\protect\citeauthoryear{Härder}{Härder}{1984}]%
        {Harder1984}
\bibfield{author}{\bibinfo{person}{Theo Härder}.} \bibinfo{year}{1984}\natexlab{}.
\newblock \showarticletitle{Observations on optimistic concurrency control schemes}.
\newblock \bibinfo{journal}{\emph{Information Systems}} \bibinfo{volume}{9}, \bibinfo{number}{2} (\bibinfo{year}{1984}), \bibinfo{pages}{111--120}.
\newblock


\bibitem[\protect\citeauthoryear{Junqueira, Reed, and Serafini}{Junqueira et~al\mbox{.}}{2011}]%
        {zab}
\bibfield{author}{\bibinfo{person}{Flavio~P. Junqueira}, \bibinfo{person}{Benjamin~C. Reed}, {and} \bibinfo{person}{Marco Serafini}.} \bibinfo{year}{2011}\natexlab{}.
\newblock \showarticletitle{Zab: High-performance broadcast for primary-backup systems}. In \bibinfo{booktitle}{\emph{2011 IEEE/IFIP 41st International Conference on Dependable Systems and Networks (DSN)}}. \bibinfo{pages}{245--256}.
\newblock
\urldef\tempurl%
\url{https://doi.org/10.1109/DSN.2011.5958223}
\showDOI{\tempurl}


\bibitem[\protect\citeauthoryear{Kreps}{Kreps}{2011}]%
        {kafka}
\bibfield{author}{\bibinfo{person}{Jay Kreps}.} \bibinfo{year}{2011}\natexlab{}.
\newblock \showarticletitle{Kafka : a Distributed Messaging System for Log Processing}.
\newblock
\urldef\tempurl%
\url{https://api.semanticscholar.org/CorpusID:18534081}
\showURL{%
\tempurl}


\bibitem[\protect\citeauthoryear{Kung and Robinson}{Kung and Robinson}{1981}]%
        {kung1981optimistic}
\bibfield{author}{\bibinfo{person}{Hsiang-Tsung Kung} {and} \bibinfo{person}{John~T Robinson}.} \bibinfo{year}{1981}\natexlab{}.
\newblock \showarticletitle{On optimistic methods for concurrency control}.
\newblock \bibinfo{journal}{\emph{ACM Transactions on Database Systems (TODS)}} \bibinfo{volume}{6}, \bibinfo{number}{2} (\bibinfo{year}{1981}), \bibinfo{pages}{213--226}.
\newblock


\bibitem[\protect\citeauthoryear{Lakshman and Malik}{Lakshman and Malik}{2010}]%
        {lakshman2010cassandra}
\bibfield{author}{\bibinfo{person}{Avinash Lakshman} {and} \bibinfo{person}{Prashant Malik}.} \bibinfo{year}{2010}\natexlab{}.
\newblock \showarticletitle{Cassandra: a decentralized structured storage system}.
\newblock \bibinfo{journal}{\emph{ACM SIGOPS operating systems review}} \bibinfo{volume}{44}, \bibinfo{number}{2} (\bibinfo{year}{2010}), \bibinfo{pages}{35--40}.
\newblock


\bibitem[\protect\citeauthoryear{Lamport}{Lamport}{2001}]%
        {paxos}
\bibfield{author}{\bibinfo{person}{Leslie Lamport}.} \bibinfo{year}{2001}\natexlab{}.
\newblock \showarticletitle{Paxos Made Simple}.
\newblock \bibinfo{journal}{\emph{ACM SIGACT News (Distributed Computing Column) 32, 4 (Whole Number 121, December 2001)}} (\bibinfo{date}{December} \bibinfo{year}{2001}), \bibinfo{pages}{51--58}.
\newblock
\urldef\tempurl%
\url{https://www.microsoft.com/en-us/research/publication/paxos-made-simple/}
\showURL{%
\tempurl}


\bibitem[\protect\citeauthoryear{Le, Fynn, Eslahi-Kelorazi, Soul{\'e}, and Pedone}{Le et~al\mbox{.}}{2019}]%
        {le2019dynastar}
\bibfield{author}{\bibinfo{person}{Long~Hoang Le}, \bibinfo{person}{Enrique Fynn}, \bibinfo{person}{Mojtaba Eslahi-Kelorazi}, \bibinfo{person}{Robert Soul{\'e}}, {and} \bibinfo{person}{Fernando Pedone}.} \bibinfo{year}{2019}\natexlab{}.
\newblock \showarticletitle{Dynastar: Optimized dynamic partitioning for scalable state machine replication}. In \bibinfo{booktitle}{\emph{2019 IEEE 39th International Conference on Distributed Computing Systems (ICDCS)}}. IEEE, \bibinfo{pages}{1453--1465}.
\newblock


\bibitem[\protect\citeauthoryear{Lev-Ari, Bortnikov, Keidar, and Shraer}{Lev-Ari et~al\mbox{.}}{2016}]%
        {zoonet}
\bibfield{author}{\bibinfo{person}{Kfir Lev-Ari}, \bibinfo{person}{Edward Bortnikov}, \bibinfo{person}{Idit Keidar}, {and} \bibinfo{person}{Alexander Shraer}.} \bibinfo{year}{2016}\natexlab{}.
\newblock \showarticletitle{Modular Composition of Coordination Services}. In \bibinfo{booktitle}{\emph{2016 USENIX Annual Technical Conference (USENIX ATC 16)}}. \bibinfo{publisher}{USENIX Association}, \bibinfo{address}{Denver, CO}, \bibinfo{pages}{251--264}.
\newblock
\showISBNx{978-1-931971-30-0}
\urldef\tempurl%
\url{https://www.usenix.org/conference/atc16/technical-sessions/presentation/lev-ari}
\showURL{%
\tempurl}


\bibitem[\protect\citeauthoryear{Liroz-Gistau, Akbarinia, Pacitti, Porto, and Valduriez}{Liroz-Gistau et~al\mbox{.}}{2013}]%
        {liroz2013dynamic}
\bibfield{author}{\bibinfo{person}{Miguel Liroz-Gistau}, \bibinfo{person}{Reza Akbarinia}, \bibinfo{person}{Esther Pacitti}, \bibinfo{person}{Fabio Porto}, {and} \bibinfo{person}{Patrick Valduriez}.} \bibinfo{year}{2013}\natexlab{}.
\newblock \showarticletitle{Dynamic workload-based partitioning algorithms for continuously growing databases}.
\newblock \bibinfo{journal}{\emph{Transactions on Large-Scale Data-and Knowledge-Centered Systems XII}} (\bibinfo{year}{2013}), \bibinfo{pages}{105--128}.
\newblock


\bibitem[\protect\citeauthoryear{Menascé and Nakanishi}{Menascé and Nakanishi}{1982}]%
        {Mensace1982}
\bibfield{author}{\bibinfo{person}{D.~A. Menascé} {and} \bibinfo{person}{T. Nakanishi}.} \bibinfo{year}{1982}\natexlab{}.
\newblock \showarticletitle{Optimistic versus pessimistic concurrency control mechanisms in database management systems}.
\newblock \bibinfo{journal}{\emph{Information Systems}} \bibinfo{volume}{7}, \bibinfo{number}{1} (\bibinfo{year}{1982}), \bibinfo{pages}{13--27}.
\newblock


\bibitem[\protect\citeauthoryear{{Microsoft}}{{Microsoft}}{nd}]%
        {orleans_cluster_management}
\bibfield{author}{\bibinfo{person}{{Microsoft}}.} \bibinfo{year}{n.d.}\natexlab{}.
\newblock \bibinfo{title}{Cluster Management - Orleans Documentation}.
\newblock \bibinfo{howpublished}{\url{https://learn.microsoft.com/en-us/dotnet/orleans/implementation/cluster-management}}.
\newblock
\newblock
\shownote{Accessed: 2025-01-13.}


\bibitem[\protect\citeauthoryear{{Microsoft Azure}}{{Microsoft Azure}}{2023}]%
        {azure_sql_hyperscale}
\bibfield{author}{\bibinfo{person}{{Microsoft Azure}}.} \bibinfo{year}{2023}\natexlab{}.
\newblock \bibinfo{booktitle}{\emph{{Azure SQL Database Hyperscale}}}.
\newblock
\urldef\tempurl%
\url{https://azure.microsoft.com/en-us/products/azure-sql/database/hyperscale/}
\showURL{%
\tempurl}
\newblock
\shownote{Accessed: 2024-11-13.}


\bibitem[\protect\citeauthoryear{{Microsoft Azure}}{{Microsoft Azure}}{2024a}]%
        {AzureAppendBlobs}
\bibfield{author}{\bibinfo{person}{{Microsoft Azure}}.} \bibinfo{year}{2024}\natexlab{a}.
\newblock \bibinfo{title}{Azure Append Blobs}.
\newblock \bibinfo{howpublished}{\url{https://docs.microsoft.com/en-us/azure/storage/blobs/storage-blobs-introduction\#append-blobs}}.
\newblock
\newblock
\shownote{Accessed: 2024-01-01.}


\bibitem[\protect\citeauthoryear{{Microsoft Azure}}{{Microsoft Azure}}{2024b}]%
        {AzureBlobStorage}
\bibfield{author}{\bibinfo{person}{{Microsoft Azure}}.} \bibinfo{year}{2024}\natexlab{b}.
\newblock \bibinfo{title}{Azure Blob Storage}.
\newblock \bibinfo{howpublished}{\url{https://azure.microsoft.com/en-us/services/storage/blobs/}}.
\newblock
\newblock
\shownote{Accessed: 2024-01-01.}


\bibitem[\protect\citeauthoryear{{Microsoft Azure}}{{Microsoft Azure}}{2024c}]%
        {AzureStorageGPv2}
\bibfield{author}{\bibinfo{person}{{Microsoft Azure}}.} \bibinfo{year}{2024}\natexlab{c}.
\newblock \bibinfo{title}{Azure Storage Account Standard General-Purpose v2}.
\newblock \bibinfo{howpublished}{\url{https://docs.microsoft.com/en-us/azure/storage/common/storage-account-overview\#general-purpose-v2-accounts}}.
\newblock
\newblock
\shownote{Accessed: 2024-01-01.}


\bibitem[\protect\citeauthoryear{{Microsoft Azure}}{{Microsoft Azure}}{2024d}]%
        {AzureCloudService}
\bibfield{author}{\bibinfo{person}{{Microsoft Azure}}.} \bibinfo{year}{2024}\natexlab{d}.
\newblock \bibinfo{title}{Microsoft Azure Cloud Service Platform}.
\newblock \bibinfo{howpublished}{\url{https://azure.microsoft.com/en-us/overview/}}.
\newblock
\newblock
\shownote{Accessed: 2024-01-01.}


\bibitem[\protect\citeauthoryear{{Microsoft Corporation}}{{Microsoft Corporation}}{[n.d.]}]%
        {azuretablestorage}
\bibfield{author}{\bibinfo{person}{{Microsoft Corporation}}.} \bibinfo{year}{[n.d.]}\natexlab{}.
\newblock \bibinfo{title}{{Azure Table Storage}}.
\newblock \bibinfo{howpublished}{\url{https://azure.microsoft.com/en-us/services/storage/tables/}}.
\newblock
\newblock
\shownote{Accessed: 2023-10-29.}


\bibitem[\protect\citeauthoryear{{Microsoft Corporation}}{{Microsoft Corporation}}{2024}]%
        {microsoft_sql_server}
\bibfield{author}{\bibinfo{person}{{Microsoft Corporation}}.} \bibinfo{year}{2024}\natexlab{}.
\newblock \bibinfo{title}{{Microsoft SQL Server}}.
\newblock
\newblock
\urldef\tempurl%
\url{https://www.microsoft.com/en-us/sql-server/}
\showURL{%
\tempurl}
\newblock
\shownote{Accessed: 2024-10-13.}


\bibitem[\protect\citeauthoryear{Mohan and Narang}{Mohan and Narang}{1991}]%
        {mohan1991recovery}
\bibfield{author}{\bibinfo{person}{C Mohan} {and} \bibinfo{person}{Inderpal Narang}.} \bibinfo{year}{1991}\natexlab{}.
\newblock \showarticletitle{Recovery and coherency-control protocols for fast intersystem page transfer and fine-granularity locking in a shared disks transaction environment}. In \bibinfo{booktitle}{\emph{Proceedings of the 17th International Conference on Very Large Data Bases}}. \bibinfo{pages}{193--207}.
\newblock


\bibitem[\protect\citeauthoryear{{Neo4j, Inc.}}{{Neo4j, Inc.}}{2025}]%
        {neo4j}
\bibfield{author}{\bibinfo{person}{{Neo4j, Inc.}}} \bibinfo{year}{2025}\natexlab{}.
\newblock \bibinfo{title}{Neo4j: The World's Leading Graph Database}.
\newblock
\newblock
\urldef\tempurl%
\url{https://neo4j.com}
\showURL{%
\tempurl}
\newblock
\shownote{Accessed: 2025-01-14.}


\bibitem[\protect\citeauthoryear{{Neon Technologies}}{{Neon Technologies}}{2023}]%
        {neondb}
\bibfield{author}{\bibinfo{person}{{Neon Technologies}}.} \bibinfo{year}{2023}\natexlab{}.
\newblock \bibinfo{booktitle}{\emph{{Neon: Serverless Postgres}}}.
\newblock
\urldef\tempurl%
\url{https://neon.tech/}
\showURL{%
\tempurl}
\newblock
\shownote{Accessed: 2024-11-13.}


\bibitem[\protect\citeauthoryear{Ongaro and Ousterhout}{Ongaro and Ousterhout}{2014a}]%
        {ongaro2014search}
\bibfield{author}{\bibinfo{person}{Diego Ongaro} {and} \bibinfo{person}{John Ousterhout}.} \bibinfo{year}{2014}\natexlab{a}.
\newblock \showarticletitle{In search of an understandable consensus algorithm}. In \bibinfo{booktitle}{\emph{2014 USENIX annual technical conference (USENIX ATC 14)}}. \bibinfo{pages}{305--319}.
\newblock


\bibitem[\protect\citeauthoryear{Ongaro and Ousterhout}{Ongaro and Ousterhout}{2014b}]%
        {raft}
\bibfield{author}{\bibinfo{person}{Diego Ongaro} {and} \bibinfo{person}{John Ousterhout}.} \bibinfo{year}{2014}\natexlab{b}.
\newblock \showarticletitle{In Search of an Understandable Consensus Algorithm}. In \bibinfo{booktitle}{\emph{Proceedings of the 2014 USENIX Conference on USENIX Annual Technical Conference}} (Philadelphia, PA) \emph{(\bibinfo{series}{USENIX ATC'14})}. \bibinfo{publisher}{USENIX Association}, \bibinfo{address}{USA}, \bibinfo{pages}{305–320}.
\newblock
\showISBNx{9781931971102}


\bibitem[\protect\citeauthoryear{Oracle}{Oracle}{2024}]%
        {OracleRAC2022}
\bibfield{author}{\bibinfo{person}{Oracle}.} \bibinfo{year}{2024}\natexlab{}.
\newblock \bibinfo{title}{Oracle RAC}.
\newblock \bibinfo{howpublished}{\url{https://www.oracle.com/database/real-application-clusters/}}.
\newblock


\bibitem[\protect\citeauthoryear{Pang and Wang}{Pang and Wang}{2024}]%
        {pang2024understanding}
\bibfield{author}{\bibinfo{person}{Xi Pang} {and} \bibinfo{person}{Jianguo Wang}.} \bibinfo{year}{2024}\natexlab{}.
\newblock \showarticletitle{Understanding the performance implications of the design principles in storage-disaggregated databases}.
\newblock \bibinfo{journal}{\emph{Proceedings of the ACM on Management of Data}} \bibinfo{volume}{2}, \bibinfo{number}{3} (\bibinfo{year}{2024}), \bibinfo{pages}{1--26}.
\newblock


\bibitem[\protect\citeauthoryear{Prasaad, Cheung, and Suciu}{Prasaad et~al\mbox{.}}{2020}]%
        {prasaad2020handling}
\bibfield{author}{\bibinfo{person}{Guna Prasaad}, \bibinfo{person}{Alvin Cheung}, {and} \bibinfo{person}{Dan Suciu}.} \bibinfo{year}{2020}\natexlab{}.
\newblock \showarticletitle{Handling highly contended OLTP workloads using fast dynamic partitioning}. In \bibinfo{booktitle}{\emph{Proceedings of the 2020 ACM SIGMOD International Conference on Management of Data}}. \bibinfo{pages}{527--542}.
\newblock


\bibitem[\protect\citeauthoryear{Rahm}{Rahm}{1989}]%
        {rahm1989recovery}
\bibfield{author}{\bibinfo{person}{Erhard Rahm}.} \bibinfo{year}{1989}\natexlab{}.
\newblock \bibinfo{booktitle}{\emph{Recovery concepts for data sharing systems}}.
\newblock \bibinfo{publisher}{Citeseer}.
\newblock


\bibitem[\protect\citeauthoryear{{Redis}}{{Redis}}{2024}]%
        {Redis2009}
\bibfield{author}{\bibinfo{person}{{Redis}}.} \bibinfo{year}{2024}\natexlab{}.
\newblock \bibinfo{title}{Redis}.
\newblock \bibinfo{howpublished}{\url{http://redis.io/}}.
\newblock
\newblock
\shownote{Accessed: 2024-01-01.}


\bibitem[\protect\citeauthoryear{{Redpanda Data Inc.}}{{Redpanda Data Inc.}}{2023}]%
        {firescroll}
\bibfield{author}{\bibinfo{person}{{Redpanda Data Inc.}}} \bibinfo{year}{2023}\natexlab{}.
\newblock \bibinfo{title}{FireScroll}.
\newblock \bibinfo{howpublished}{\url{https://github.com/FireScroll/FireScroll}}.
\newblock
\newblock
\shownote{Accessed: 2024-10-16.}


\bibitem[\protect\citeauthoryear{Reiher and Popek}{Reiher and Popek}{1988}]%
        {Reiher1988}
\bibfield{author}{\bibinfo{person}{P.~L. Reiher} {and} \bibinfo{person}{G.~J. Popek}.} \bibinfo{year}{1988}\natexlab{}.
\newblock \showarticletitle{Optimistic concurrency control by predictive optimistic concurrency control}. In \bibinfo{booktitle}{\emph{Proceedings of the 1988 ACM SIGMOD International Conference on Management of Data (SIGMOD'88)}}. \bibinfo{publisher}{ACM}, \bibinfo{pages}{81--88}.
\newblock


\bibitem[\protect\citeauthoryear{Serafini, Taft, Elmore, Pavlo, Aboulnaga, and Stonebraker}{Serafini et~al\mbox{.}}{2016}]%
        {serafini2016clay}
\bibfield{author}{\bibinfo{person}{Marco Serafini}, \bibinfo{person}{Rebecca Taft}, \bibinfo{person}{Aaron~J Elmore}, \bibinfo{person}{Andrew Pavlo}, \bibinfo{person}{Ashraf Aboulnaga}, {and} \bibinfo{person}{Michael Stonebraker}.} \bibinfo{year}{2016}\natexlab{}.
\newblock \showarticletitle{Clay: Fine-grained adaptive partitioning for general database schemas}.
\newblock \bibinfo{journal}{\emph{Proceedings of the VLDB Endowment}} \bibinfo{volume}{10}, \bibinfo{number}{4} (\bibinfo{year}{2016}), \bibinfo{pages}{445--456}.
\newblock


\bibitem[\protect\citeauthoryear{Shen, Jia, Sela, Rainero, Song, van Renesse, and Weatherspoon}{Shen et~al\mbox{.}}{2016}]%
        {shen2016follow}
\bibfield{author}{\bibinfo{person}{Zhiming Shen}, \bibinfo{person}{Qin Jia}, \bibinfo{person}{Gur-Eyal Sela}, \bibinfo{person}{Ben Rainero}, \bibinfo{person}{Weijia Song}, \bibinfo{person}{Robbert van Renesse}, {and} \bibinfo{person}{Hakim Weatherspoon}.} \bibinfo{year}{2016}\natexlab{}.
\newblock \showarticletitle{Follow the sun through the clouds: Application migration for geographically shifting workloads}. In \bibinfo{booktitle}{\emph{Proceedings of the Seventh ACM Symposium on Cloud Computing}}. \bibinfo{pages}{141--154}.
\newblock


\bibitem[\protect\citeauthoryear{Skeen}{Skeen}{1981}]%
        {skeen1981nonblocking}
\bibfield{author}{\bibinfo{person}{Dale Skeen}.} \bibinfo{year}{1981}\natexlab{}.
\newblock \showarticletitle{Nonblocking commit protocols}. In \bibinfo{booktitle}{\emph{Proceedings of the 1981 ACM SIGMOD International Conference on Management of Data}}. \bibinfo{pages}{133--142}.
\newblock


\bibitem[\protect\citeauthoryear{Sloss, Dahlin, Rau, and Beyer}{Sloss et~al\mbox{.}}{2017}]%
        {sloss2017calculus}
\bibfield{author}{\bibinfo{person}{Benjamin~Treynor Sloss}, \bibinfo{person}{Mike Dahlin}, \bibinfo{person}{Vivek Rau}, {and} \bibinfo{person}{Betsy Beyer}.} \bibinfo{year}{2017}\natexlab{}.
\newblock \showarticletitle{The Calculus of Service Availability: You’re only as available as the sum of your dependencies.}
\newblock \bibinfo{journal}{\emph{Queue}} \bibinfo{volume}{15}, \bibinfo{number}{2} (\bibinfo{year}{2017}), \bibinfo{pages}{49--67}.
\newblock


\bibitem[\protect\citeauthoryear{Stoica, Morris, Liben-Nowell, Karger, Kaashoek, Dabek, and Balakrishnan}{Stoica et~al\mbox{.}}{2003}]%
        {stoica2003chord}
\bibfield{author}{\bibinfo{person}{Ion Stoica}, \bibinfo{person}{Robert Morris}, \bibinfo{person}{David Liben-Nowell}, \bibinfo{person}{David~R Karger}, \bibinfo{person}{M~Frans Kaashoek}, \bibinfo{person}{Frank Dabek}, {and} \bibinfo{person}{Hari Balakrishnan}.} \bibinfo{year}{2003}\natexlab{}.
\newblock \showarticletitle{Chord: a scalable peer-to-peer lookup protocol for internet applications}.
\newblock \bibinfo{journal}{\emph{IEEE/ACM Transactions on networking}} \bibinfo{volume}{11}, \bibinfo{number}{1} (\bibinfo{year}{2003}), \bibinfo{pages}{17--32}.
\newblock


\bibitem[\protect\citeauthoryear{Taft, Mansour, Serafini, Duggan, Elmore, Aboulnaga, Pavlo, and Stonebraker}{Taft et~al\mbox{.}}{2014}]%
        {taft2014store}
\bibfield{author}{\bibinfo{person}{Rebecca Taft}, \bibinfo{person}{Essam Mansour}, \bibinfo{person}{Marco Serafini}, \bibinfo{person}{Jennie Duggan}, \bibinfo{person}{Aaron~J Elmore}, \bibinfo{person}{Ashraf Aboulnaga}, \bibinfo{person}{Andrew Pavlo}, {and} \bibinfo{person}{Michael Stonebraker}.} \bibinfo{year}{2014}\natexlab{}.
\newblock \showarticletitle{E-store: Fine-grained elastic partitioning for distributed transaction processing systems}.
\newblock \bibinfo{journal}{\emph{Proceedings of the VLDB Endowment}} \bibinfo{volume}{8}, \bibinfo{number}{3} (\bibinfo{year}{2014}), \bibinfo{pages}{245--256}.
\newblock


\bibitem[\protect\citeauthoryear{Taft, Sharif, Matei, VanBenschoten, Lewis, Grieger, Niemi, Woods, Birzin, Poss, et~al\mbox{.}}{Taft et~al\mbox{.}}{2020}]%
        {taft2020cockroachdb}
\bibfield{author}{\bibinfo{person}{Rebecca Taft}, \bibinfo{person}{Irfan Sharif}, \bibinfo{person}{Andrei Matei}, \bibinfo{person}{Nathan VanBenschoten}, \bibinfo{person}{Jordan Lewis}, \bibinfo{person}{Tobias Grieger}, \bibinfo{person}{Kai Niemi}, \bibinfo{person}{Andy Woods}, \bibinfo{person}{Anne Birzin}, \bibinfo{person}{Raphael Poss}, {et~al\mbox{.}}} \bibinfo{year}{2020}\natexlab{}.
\newblock \showarticletitle{Cockroachdb: The resilient geo-distributed sql database}. In \bibinfo{booktitle}{\emph{Proceedings of the 2020 ACM SIGMOD international conference on management of data}}. \bibinfo{pages}{1493--1509}.
\newblock


\bibitem[\protect\citeauthoryear{Team}{Team}{2024}]%
        {ScyllaDB}
\bibfield{author}{\bibinfo{person}{ScyllaDB Team}.} \bibinfo{year}{2024}\natexlab{}.
\newblock \bibinfo{title}{ScyllaDB}.
\newblock \bibinfo{howpublished}{\url{https://www.scylladb.com/}}.
\newblock
\newblock
\shownote{Accessed: 2024-01-01.}


\bibitem[\protect\citeauthoryear{(TPC)}{(TPC)}{1992}]%
        {tpcc}
\bibfield{author}{\bibinfo{person}{Transaction Processing Performance~Council (TPC)}.} \bibinfo{year}{1992}\natexlab{}.
\newblock \bibinfo{title}{{TPC Benchmark C (TPC-C): Standard Specification}}.
\newblock \bibinfo{howpublished}{\url{http://www.tpc.org/tpcc/}}.
\newblock
\newblock
\shownote{Accessed: 2025-03-07.}


\bibitem[\protect\citeauthoryear{Verbitski, Gupta, Saha, Brahmadesam, Gupta, Mittal, Krishnamurthy, Maurice, Kharatishvili, and Bao}{Verbitski et~al\mbox{.}}{2017}]%
        {verbitski2017amazon}
\bibfield{author}{\bibinfo{person}{Alexandre Verbitski}, \bibinfo{person}{Anurag Gupta}, \bibinfo{person}{Debanjan Saha}, \bibinfo{person}{Murali Brahmadesam}, \bibinfo{person}{Kamal Gupta}, \bibinfo{person}{Raman Mittal}, \bibinfo{person}{Sailesh Krishnamurthy}, \bibinfo{person}{Sandor Maurice}, \bibinfo{person}{Tengiz Kharatishvili}, {and} \bibinfo{person}{Xiaofeng Bao}.} \bibinfo{year}{2017}\natexlab{}.
\newblock \showarticletitle{Amazon aurora: Design considerations for high throughput cloud-native relational databases}. In \bibinfo{booktitle}{\emph{Proceedings of the 2017 ACM International Conference on Management of Data}}. \bibinfo{pages}{1041--1052}.
\newblock


\bibitem[\protect\citeauthoryear{Verma, Dasgupta, Nayak, De, and Kothari}{Verma et~al\mbox{.}}{2009}]%
        {verma2009server}
\bibfield{author}{\bibinfo{person}{Akshat Verma}, \bibinfo{person}{Gargi Dasgupta}, \bibinfo{person}{Tapan~Kumar Nayak}, \bibinfo{person}{Pradipta De}, {and} \bibinfo{person}{Ravi Kothari}.} \bibinfo{year}{2009}\natexlab{}.
\newblock \showarticletitle{Server workload analysis for power minimization using consolidation}. In \bibinfo{booktitle}{\emph{Proceedings of the 2009 conference on USENIX Annual technical conference}}. \bibinfo{pages}{28--28}.
\newblock


\bibitem[\protect\citeauthoryear{Yang, Tschetter, L{\'e}aut{\'e}, Ray, Merlino, and Ganguli}{Yang et~al\mbox{.}}{2014}]%
        {yang2014druid}
\bibfield{author}{\bibinfo{person}{Fangjin Yang}, \bibinfo{person}{Eric Tschetter}, \bibinfo{person}{Xavier L{\'e}aut{\'e}}, \bibinfo{person}{Nelson Ray}, \bibinfo{person}{Gian Merlino}, {and} \bibinfo{person}{Deep Ganguli}.} \bibinfo{year}{2014}\natexlab{}.
\newblock \showarticletitle{Druid: A real-time analytical data store}. In \bibinfo{booktitle}{\emph{Proceedings of the 2014 ACM SIGMOD international conference on Management of data}}. \bibinfo{pages}{157--168}.
\newblock


\bibitem[\protect\citeauthoryear{Yang, Youill, Woicik, Liu, Yu, Serafini, Aboulnaga, and Stonebraker}{Yang et~al\mbox{.}}{2021}]%
        {fpdb}
\bibfield{author}{\bibinfo{person}{Yifei Yang}, \bibinfo{person}{Matt Youill}, \bibinfo{person}{Matthew Woicik}, \bibinfo{person}{Yizhou Liu}, \bibinfo{person}{Xiangyao Yu}, \bibinfo{person}{Marco Serafini}, \bibinfo{person}{Ashraf Aboulnaga}, {and} \bibinfo{person}{Michael Stonebraker}.} \bibinfo{year}{2021}\natexlab{}.
\newblock \showarticletitle{FlexPushdownDB: Hybrid Pushdown and Caching in a CloudDBMS}. In \bibinfo{booktitle}{\emph{VLDB}}.
\newblock


\bibitem[\protect\citeauthoryear{Yang, Yang, Han, Zhuang, Yang, Yang, Cheng, Zhao, Shi, Xi, et~al\mbox{.}}{Yang et~al\mbox{.}}{2022}]%
        {yang2022oceanbase}
\bibfield{author}{\bibinfo{person}{Zhenkun Yang}, \bibinfo{person}{Chuanhui Yang}, \bibinfo{person}{Fusheng Han}, \bibinfo{person}{Mingqiang Zhuang}, \bibinfo{person}{Bing Yang}, \bibinfo{person}{Zhifeng Yang}, \bibinfo{person}{Xiaojun Cheng}, \bibinfo{person}{Yuzhong Zhao}, \bibinfo{person}{Wenhui Shi}, \bibinfo{person}{Huafeng Xi}, {et~al\mbox{.}}} \bibinfo{year}{2022}\natexlab{}.
\newblock \showarticletitle{OceanBase: a 707 million tpmC distributed relational database system}.
\newblock \bibinfo{journal}{\emph{Proceedings of the VLDB Endowment}} \bibinfo{volume}{15}, \bibinfo{number}{12} (\bibinfo{year}{2022}), \bibinfo{pages}{3385--3397}.
\newblock


\bibitem[\protect\citeauthoryear{Yu, Pavlo, and Devadas}{Yu et~al\mbox{.}}{2016}]%
        {Yu2016}
\bibfield{author}{\bibinfo{person}{Xiangyao Yu}, \bibinfo{person}{Andrew Pavlo}, {and} \bibinfo{person}{Srinivas Devadas}.} \bibinfo{year}{2016}\natexlab{}.
\newblock \showarticletitle{TicToc: Time Traveling Optimistic Concurrency Control}. In \bibinfo{booktitle}{\emph{Proceedings of the 2016 International Conference on Management of Data (SIGMOD)}}. \bibinfo{publisher}{ACM}, \bibinfo{pages}{1629--1642}.
\newblock


\bibitem[\protect\citeauthoryear{Yu, Xia, Pavlo, Sanchez, Rudolph, and Devadas}{Yu et~al\mbox{.}}{2018}]%
        {yu2018sundial}
\bibfield{author}{\bibinfo{person}{Xiangyao Yu}, \bibinfo{person}{Yu Xia}, \bibinfo{person}{Andrew Pavlo}, \bibinfo{person}{Daniel Sanchez}, \bibinfo{person}{Larry Rudolph}, {and} \bibinfo{person}{Srinivas Devadas}.} \bibinfo{year}{2018}\natexlab{}.
\newblock \showarticletitle{Sundial: harmonizing concurrency control and caching in a distributed OLTP database management system}.
\newblock \bibinfo{journal}{\emph{Proceedings of the VLDB Endowment}} \bibinfo{volume}{11}, \bibinfo{number}{10} (\bibinfo{year}{2018}), \bibinfo{pages}{1289--1302}.
\newblock


\bibitem[\protect\citeauthoryear{{Yugabyte, Inc.}}{{Yugabyte, Inc.}}{2023}]%
        {yugabytedb}
\bibfield{author}{\bibinfo{person}{{Yugabyte, Inc.}}} \bibinfo{year}{2023}\natexlab{}.
\newblock \bibinfo{title}{YugabyteDB: A High-Performance Distributed SQL Database for Global, Internet-Scale Applications}.
\newblock \bibinfo{howpublished}{\url{https://www.yugabyte.com}}.
\newblock
\newblock
\shownote{Accessed: 2024-11-13.}


\bibitem[\protect\citeauthoryear{Zamanian, Binnig, Kraska, and Harris}{Zamanian et~al\mbox{.}}{2016}]%
        {zamanian2016end}
\bibfield{author}{\bibinfo{person}{Erfan Zamanian}, \bibinfo{person}{Carsten Binnig}, \bibinfo{person}{Tim Kraska}, {and} \bibinfo{person}{Tim Harris}.} \bibinfo{year}{2016}\natexlab{}.
\newblock \showarticletitle{The end of a myth: Distributed transactions can scale}.
\newblock \bibinfo{journal}{\emph{arXiv preprint arXiv:1607.00655}} (\bibinfo{year}{2016}).
\newblock


\bibitem[\protect\citeauthoryear{Zhou, Xu, Shraer, Namasivayam, Miller, Tschannen, Atherton, Beamon, Sears, Leach, et~al\mbox{.}}{Zhou et~al\mbox{.}}{2021}]%
        {zhou2021foundationdb}
\bibfield{author}{\bibinfo{person}{Jingyu Zhou}, \bibinfo{person}{Meng Xu}, \bibinfo{person}{Alexander Shraer}, \bibinfo{person}{Bala Namasivayam}, \bibinfo{person}{Alex Miller}, \bibinfo{person}{Evan Tschannen}, \bibinfo{person}{Steve Atherton}, \bibinfo{person}{Andrew~J Beamon}, \bibinfo{person}{Rusty Sears}, \bibinfo{person}{John Leach}, {et~al\mbox{.}}} \bibinfo{year}{2021}\natexlab{}.
\newblock \showarticletitle{Foundationdb: A distributed unbundled transactional key value store}. In \bibinfo{booktitle}{\emph{Proceedings of the 2021 International Conference on Management of Data}}. \bibinfo{pages}{2653--2666}.
\newblock


\bibitem[\protect\citeauthoryear{Ziegler, Bernstein, Leis, and Binnig}{Ziegler et~al\mbox{.}}{2023}]%
        {ziegler2023scalable}
\bibfield{author}{\bibinfo{person}{Tobias Ziegler}, \bibinfo{person}{Philip~A Bernstein}, \bibinfo{person}{Viktor Leis}, {and} \bibinfo{person}{Carsten Binnig}.} \bibinfo{year}{2023}\natexlab{}.
\newblock \showarticletitle{Is Scalable OLTP in the Cloud a Solved Problem?}. In \bibinfo{booktitle}{\emph{CIDR}}.
\newblock


\bibitem[\protect\citeauthoryear{Ziegler, Binnig, and Leis}{Ziegler et~al\mbox{.}}{2022}]%
        {ziegler2022scalestore}
\bibfield{author}{\bibinfo{person}{Tobias Ziegler}, \bibinfo{person}{Carsten Binnig}, {and} \bibinfo{person}{Viktor Leis}.} \bibinfo{year}{2022}\natexlab{}.
\newblock \showarticletitle{ScaleStore: A fast and cost-efficient storage engine using DRAM, NVMe, and RDMA}. In \bibinfo{booktitle}{\emph{Proceedings of the 2022 International Conference on Management of Data}}. \bibinfo{pages}{685--699}.
\newblock


\end{thebibliography}

\clearpage
\appendix
\includepdf[pages=-]{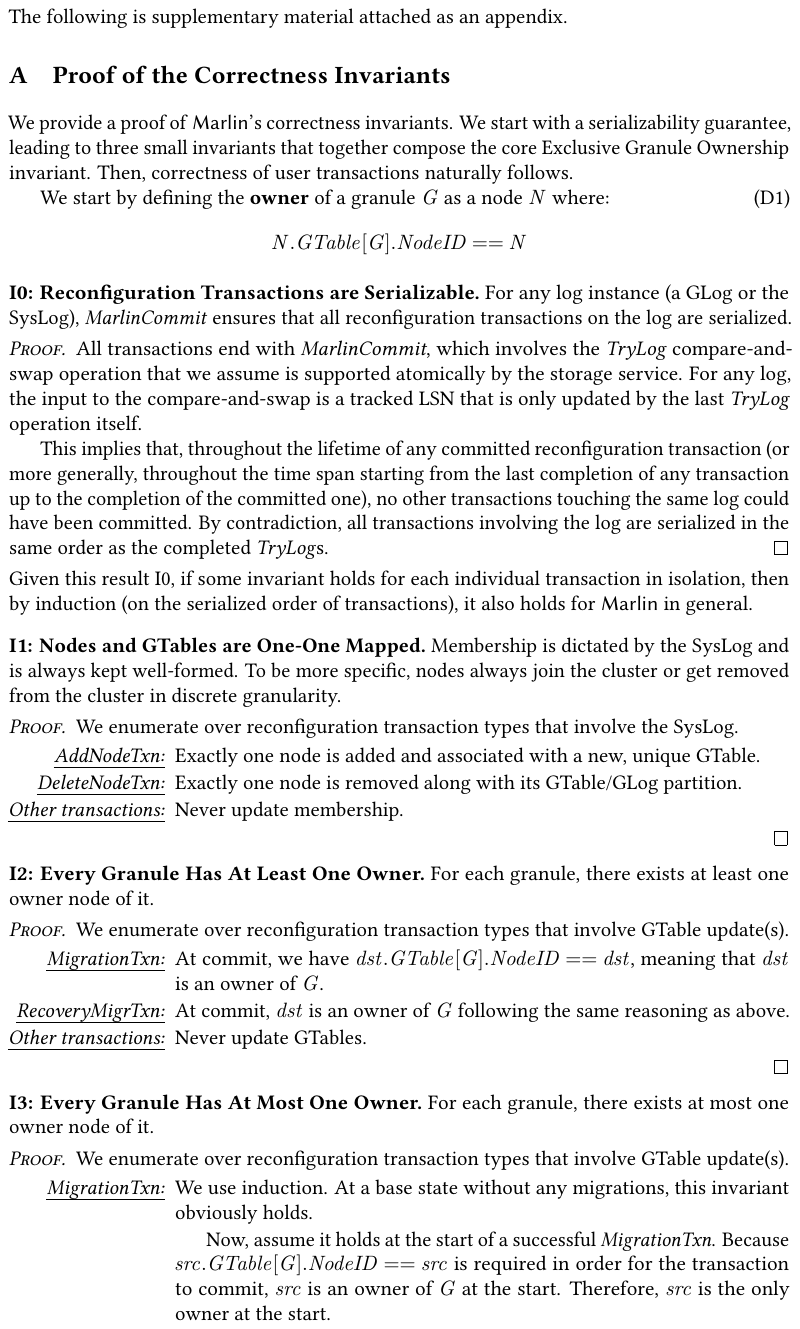}


\end{document}


\tlatex

\pagenumbering{gobble}

The following is supplementary material attached as an appendix.

\setlength{\parindent}{15pt}

\section{\edit{Proof of the Correctness Invariants}}
\edit{We provide a proof of \system's correctness invariants. We start with a serializability guarantee, leading to three small invariants that together compose the core Exclusive Granule Ownership invariant. Then, correctness of user transactions naturally follows.}

\edit{We start by defining the \textbf{owner} of a granule $G$ as a node $N$ where: \hfill (D1)}
\[
    \color{black}
    N.GTable[G].NodeID == N
\]

\vspace{-10pt}
\noindent\edit{\textbf{I0: Reconfiguration Transactions are Serializable.} For any log instance (a GLog or the \syslog), \textit{\mcommit} ensures that all reconfiguration transactions on the log are serialized.}

\vspace{-5pt}
\begin{proof}[\textsc{Proof}]
\edit{All transactions end with \textit{\mcommit}, which involves the \textit{TryLog} compare-and-swap operation that we assume is supported atomically by the storage service. For any log, the input to the compare-and-swap is a tracked LSN that is only updated by the last \textit{TryLog} operation itself.}

\edit{This implies that, throughout the lifetime of any committed reconfiguration transaction (or more generally, throughout the time span starting from the last completion of any transaction up to the completion of the committed one), no other transactions touching the same log could have been committed. By contradiction, all transactions involving the log are serialized in the same order as the completed \textit{TryLog}s.}
\end{proof}

\vspace{-5pt}
\noindent\edit{Given this result I0, if some invariant holds for each individual transaction in isolation, then by induction (on the serialized order of transactions), it also holds for \system in general.}

\vspace{8pt}
\noindent\edit{\textbf{I1: Nodes and GTables are One-One Mapped.} Membership is dictated by the \syslog and is always kept well-formed. To be more specific, nodes always join the cluster or get removed from the cluster in discrete granularity.}

\vspace{-5pt}
\begin{proof}[\textsc{Proof}]
\edit{We enumerate over reconfiguration transaction types that involve the \syslog.}

\begin{proofitems}
    \item[\prooflabel{AddNodeTxn}] \edit{Exactly one node is added and associated with a new, unique GTable.}
    \item[\prooflabel{DeleteNodeTxn}] \edit{Exactly one node is removed along with its GTable/GLog partition.}
    \item[\prooflabel{Other transactions}] \edit{Never update membership.}
\end{proofitems}
\end{proof}

\vspace{-3pt}
\noindent\edit{\textbf{I2: Every Granule Has At Least One Owner.} For each granule, there exists at least one owner node of it.}

\vspace{-5pt}
\begin{proof}[\textsc{Proof}]
\edit{We enumerate over reconfiguration transaction types that involve GTable update(s).}

\begin{proofitems}
    \item[\prooflabel{MigrationTxn}] \edit{At commit, we have $dst.GTable[G].NodeID == dst$, meaning that $dst$ is an owner of $G$.}
    \item[\prooflabel{RecoveryMigrTxn}] \edit{At commit, $dst$ is an owner of $G$ following the same reasoning as above.}
    \item[\prooflabel{Other transactions}] \edit{Never update GTables.}
\end{proofitems}
\end{proof}

\vspace{-3pt}
\noindent\edit{\textbf{I3: Every Granule Has At Most One Owner.} For each granule, there exists at most one owner node of it.}

\vspace{-5pt}
\begin{proof}[\textsc{Proof}]
\edit{We enumerate over reconfiguration transaction types that involve GTable update(s).}

\begin{proofitems}
    \item[\prooflabel{MigrationTxn}] \edit{We use induction. At a base state without any migrations, this invariant obviously holds.}
    
        \edit{Now, assume it holds at the start of a successful \textit{MigrationTxn}. Because $src.GTable[G].NodeID == src$ is required in order for the transaction to commit, $src$ is an owner of $G$ at the start. Therefore, $src$ is the only owner at the start.}
        
        \edit{After the transaction, $src$ is no longer an owner (GTable entry changed) while $dst$ is an owner (as previously shown). Therefore, $dst$ is the only owner at the end.}
        
    \item[\prooflabel{RecoveryMigrTxn}] \edit{We use induction similarly. Base state obviously holds, and that every transaction exactly swaps the owner of $G$ from $src$ to $dst$. Therefore, $dst$ is the only owner at the end.}
    
    \item[\prooflabel{Other transactions}] \edit{Never update GTables.}
\end{proofitems}
\end{proof}

\vspace{-3pt}
\noindent\edit{The conjunction of the invariants I1 $\wedge$ I2 $\wedge$ I3 is equivalent to the following concise invariant:}

\noindent\edit{\textbf{I4: Exclusive Granule Ownership.} For each granule $G$, there is always one and only one owner node $N$, where ownership is defined as D1.}

\vspace{-5pt}
\begin{proof}[\textsc{Proof}]
\edit{Obvious.}
\end{proof}

\noindent\edit{Finally, according to \system's patched user transaction logic, the following correctness invariant ultimately holds:}

\noindent\edit{\textbf{I5: Exclusive UserTxn Service.} For any granule $G$ at any time, exactly one node is capable of executing the committing path of any user transaction involving $G$.}

\vspace{-5pt}
\begin{proof}[\textsc{Proof}]
\edit{Let $N$ be the only owner of $G$ according to definition D1; the existence and uniqueness of $N$ is guaranteed by I4. $N$ is the only node that can ever execute and successfully commit a user request on $G$, because the additional guard at the start of a \system user transaction requires exactly D1 to enter the commit path.}
\end{proof}

\setlength{\parindent}{0pt}

\section{TLA$^+$ Specification of \system Migration}

We include the TLA$^+$ specification of \system's migration protocol below, written as a PlusCal algorithm that can be auto-translated into TLA$^+$. It has been model-checked for granule ownership correctness invariants on symbolic inputs of 3 nodes, 6 granules, and 6 migration transactions. Please refer to the inlined comments for details of the specification.

\vspace*{10pt}

\@x{}\moduleLeftDash\@xx{ {\MODULE} Marlin}\moduleRightDash\@xx{}%
\@x{ {\EXTENDS} FiniteSets ,\, Sequences ,\, Integers ,\, TLC}%
\@pvspace{8.0pt}%
\begin{lcom}{0}%
\begin{cpar}{0}{F}{F}{0}{0}{}%
Model inputs \& assumptions.
\end{cpar}%
\end{lcom}%
\@x{ {\CONSTANT} Nodes ,\,\@s{39.79}}%
\@y{\@s{0}%
 set of all compute nodes
}%
\@xx{}%
\@x{\@s{50.17} Granules ,\,\@s{26.84}}%
\@y{\@s{0}%
 set of all data granules
}%
\@xx{}%
\@x{\@s{50.17} NumMigrations\@s{4.09}}%
\@y{\@s{0}%
 number of migrations to run
}%
\@xx{}%
\@pvspace{8.0pt}%
\@x{ NodesAssumption \.{\defeq} \.{\land} IsFiniteSet ( Nodes )}%
\@x{\@s{96.87} \.{\land} Cardinality ( Nodes ) \.{\geq} 1}%
\@pvspace{8.0pt}%
\@x{ GranulesAssumption \.{\defeq} \.{\land} IsFiniteSet ( Granules )}%
 \@x{\@s{109.83} \.{\land} Cardinality ( Granules ) \.{\geq} Cardinality (
 Nodes )}%
\@pvspace{8.0pt}%
\@x{ NumMigrationsAssumption \.{\defeq} \.{\land} NumMigrations \.{\in} Nat}%
\@x{\@s{138.68} \.{\land} NumMigrations \.{\geq} 2}%
\@pvspace{8.0pt}%
\@x{ {\ASSUME} \.{\land} NodesAssumption}%
\@x{\@s{38.24} \.{\land} GranulesAssumption}%
\@x{\@s{38.24} \.{\land} NumMigrationsAssumption}%
\@pvspace{8.0pt}%
\@x{}\midbar\@xx{}%
\@pvspace{8.0pt}%
\begin{lcom}{0}%
\begin{cpar}{0}{F}{F}{0}{0}{}%
Useful constants \& typedefs.
\end{cpar}%
\end{lcom}%
\@x{ Range ( f ) \.{\defeq} \{ f [ x ] \.{:} x \.{\in} {\DOMAIN} f \}}%
\@pvspace{8.0pt}%
\@x{}%
\@y{\@s{0}%
a valid \ensuremath{GTable} is a map from granules to nodes
}%
\@xx{}%
\@x{ GTables \.{\defeq} [ Granules \.{\rightarrow} Nodes ]}%
\@pvspace{8.0pt}%
 \@x{ InitGTable \.{\defeq} {\CHOOSE} gtab \.{\in} GTables \.{:} Range ( gtab
 ) \.{=} Nodes}%
\@pvspace{8.0pt}%
\@x{}%
\@y{\@s{0}%
a valid \ensuremath{GLog} is a \ensuremath{log} of \ensuremath{GTable} update actions
}%
\@xx{}%
\@x{ Updates \.{\defeq} \{ update \.{\in} [ id \.{:}\@s{8.2} Nat ,\,}%
\@x{\@s{103.94} gran \.{:} Granules ,\,}%
\@x{\@s{103.94} old\@s{4.13} \.{:}\@s{4.1} Nodes ,\,}%
\@x{\@s{103.94} new \.{:}\@s{4.10} Nodes ] \.{:}}%
\@x{\@s{58.97} update . old \.{\neq} update . new \}}%
\@pvspace{8.0pt}%
 \@x{ Update ( i ,\, g ,\, o ,\, n ) \.{\defeq} [ id \.{\mapsto} i ,\, gran
 \.{\mapsto} g ,\, old \.{\mapsto} o ,\, new \.{\mapsto} n ]}%
\@pvspace{8.0pt}%
 \@x{ GLogs \.{\defeq} \{ log \.{\in} Seq ( Updates ) \.{:} Len ( log )
 \.{\leq} NumMigrations \}}%
\@pvspace{8.0pt}%
\@x{}%
\@y{\@s{0}%
disaggregated storage state: each node maintains its own \ensuremath{Glog},
}%
\@xx{}%
\@x{}%
\@y{\@s{0}%
and materializes it into its own view of \ensuremath{GTable
}}%
\@xx{}%
 \@x{ Storages \.{\defeq} [ glogs\@s{1.27} \.{:} [ Nodes \.{\rightarrow} GLogs
 ] ,\,}%
\@x{\@s{58.64} gtabs \.{:} [ Nodes \.{\rightarrow} GTables ] ]}%
\@pvspace{8.0pt}%
 \@x{ InitStorage \.{\defeq} [ glogs\@s{1.27} \.{\mapsto} [ n \.{\in} Nodes
 \.{\mapsto} {\langle} {\rangle} ] ,\,}%
 \@x{\@s{70.35} gtabs \.{\mapsto} [ n \.{\in} Nodes \.{\mapsto} InitGTable ]
 ]}%
\@pvspace{8.0pt}%
\@x{}\midbar\@xx{}%
\@pvspace{8.0pt}%
\begin{lcom}{0}%
\begin{cpar}{0}{F}{F}{0}{0}{}%
Main algorithm in \ensuremath{PlusCal}.
\end{cpar}%
\end{lcom}%
\pcalsymbolstrue
\csyntaxfalse
\@x{ {\p@algorithm} Marlin}%
\@pvspace{8.0pt}%
\@x{ {\p@variable} storage \.{=} InitStorage ,\,\@s{4.1}}%
\@y{\@s{0}%
 disaggregated storage state
}%
\@xx{}%
\@x{\@s{43.22} nextUpdateId \.{=} 0 ,\,\@s{20.59}}%
\@y{\@s{0}%
 randomly unique \ensuremath{updateIds
}}%
\@xx{}%
\@x{\@s{43.22} numDone \.{=} 0\@s{42.38}}%
\@y{\@s{0}%
 number of migrations done
}%
\@xx{}%
\@pvspace{8.0pt}%
\@x{ {\p@define}}%
 \@x{\@s{16.4} IsNewUpdate ( n ,\, u ) \.{\defeq} u . id \.{\notin} \{ v . id
 \.{:} v \.{\in} Range ( storage . glogs [ n ] ) \}}%
\@pvspace{8.0pt}%
\@x{\@s{16.4}}%
\@y{\@s{0}%
consider execution terminated when all migrations have been done and
}%
\@xx{}%
\@x{\@s{16.4}}%
\@y{\@s{0}%
 all nodes\mbox{'} \ensuremath{GTables} are the same
}%
\@xx{}%
\@x{\@s{16.4} terminated \.{\defeq}}%
\@x{\@s{32.8} \.{\land} numDone \.{=} NumMigrations}%
\@x{\@s{32.8} \.{\land} Cardinality ( Range ( storage . gtabs ) ) \.{=} 1}%
\@x{ {\p@end} {\p@define} {\p@semicolon}}%
\@pvspace{8.0pt}%
\@x{}%
\@y{\@s{0}%
Node picks a granule and a peer to do a migration push.
}%
\@xx{}%
\@x{}%
\@y{\@s{0}%
This is the ``\ensuremath{MigrationTxn}'' that happens proactively between
 src and dst
}%
\@xx{}%
\@x{}%
\@y{\@s{0}%
where src is the old owner and dst is the new owner.
}%
\@xx{}%
\@x{ {\p@macro} DoMigrate ( n ) {\p@begin}}%
\@x{\@s{16.4} {\p@await} numDone \.{<} NumMigrations {\p@semicolon}}%
\@x{\@s{16.4} {\p@with} g\@s{0.25} \.{\in} Granules ,\,}%
\@x{\@s{42.59} p \.{\in} \{ p \.{\in} Nodes \.{:} p \.{\neq} n \}}%
\@x{\@s{16.4} {\p@do}}%
\@x{\@s{32.8}}%
\@y{\@s{0}%
a migration txn first checks that on both source and target, their
}%
\@xx{}%
\@x{\@s{32.8}}%
\@y{\@s{0}%
 \ensuremath{GTables} both say that k is assigned to the source right now
}%
\@xx{}%
\@x{\@s{32.8} {\p@await} \.{\land} storage . gtabs [ n ] [ g ] \.{=} n}%
 \@x{\@s{63.14} \.{\land} storage . gtabs [ p ] [ g ]\@s{0.64} \.{=} n
 {\p@semicolon}}%
\@x{\@s{32.8}}%
\@y{\@s{0}%
append to both source and target \ensuremath{GLogs} to assign \ensuremath{g}
 to the target
}%
\@xx{}%
 \@x{\@s{32.8} storage . glogs [ n ] \.{:=} Append ( @ ,\, Update (
 nextUpdateId ,\, g ,\, n ,\, p ) ) \.{\p@barbar}}%
 \@x{\@s{32.8} storage . glogs [ p ]\@s{0.64} \.{:=} Append ( @ ,\, Update (
 nextUpdateId ,\, g ,\, n ,\, p ) ) \.{\p@barbar}}%
\@x{\@s{32.8}}%
\@y{\@s{0}%
materialize into their \ensuremath{GTables
}}%
\@xx{}%
\@x{\@s{32.8} storage . gtabs [ n ] [ g ] \.{:=} p \.{\p@barbar}}%
\@x{\@s{32.8} storage . gtabs [ p ] [ g ]\@s{0.64} \.{:=} p {\p@semicolon}}%
\@x{\@s{16.4} {\p@end} {\p@with} {\p@semicolon}}%
 \@x{\@s{16.4} nextUpdateId \.{:=} nextUpdateId \.{+} 1
 {\p@semicolon}\@s{4.1}}%
\@y{\@s{0}%
in practice, this is say a random hash
}%
\@xx{}%
\@x{\@s{16.4} numDone \.{:=} numDone \.{+} 1 {\p@semicolon}}%
\@x{ {\p@end} {\p@macro} {\p@semicolon}}%
\@pvspace{8.0pt}%
\@x{}%
\@y{\@s{0}%
Node receives gossiping of an update action from some peer (or probes the
 \ensuremath{log
}}%
\@xx{}%
\@x{}%
\@y{\@s{0}%
 of some peer and finds a difference; both ways are mathematically
 equivalent).
}%
\@xx{}%
\@x{}%
\@y{\@s{0}%
This is part of the ``\ensuremath{MetaRefresh}'' that non-owners do to keep
 state up-to-date.
}%
\@xx{}%
\@x{ {\p@macro} DoRefresh ( n ) {\p@begin}}%
 \@x{\@s{16.4} {\p@with} p\@s{0.39} \.{\in} \{ p \.{\in} Nodes \.{:} p
 \.{\neq} n \} ,\,}%
\@x{\@s{42.59} u \.{\in} Range ( storage . glogs [ p ] )}%
\@x{\@s{16.4} {\p@do}}%
\@x{\@s{32.8}}%
\@y{\@s{0}%
this update action is not in my \ensuremath{GLog} yet, and that the old
 holder
}%
\@xx{}%
\@x{\@s{32.8}}%
\@y{\@s{0}%
of the updated granule matches what\mbox{'}s in my \ensuremath{GTable
}}%
\@xx{}%
\@x{\@s{32.8} {\p@await} \.{\land} IsNewUpdate ( n ,\, u )}%
 \@x{\@s{63.14} \.{\land} storage . gtabs [ n ] [ u . gran ] \.{=} u . old
 {\p@semicolon}}%
\@x{\@s{32.8}}%
\@y{\@s{0}%
append it to my \ensuremath{GLog} and update my \ensuremath{GTable} view
}%
\@xx{}%
\@x{\@s{32.8} storage . glogs [ n ] \.{:=} Append ( @ ,\, u ) \.{\p@barbar}}%
 \@x{\@s{32.8} storage . gtabs [ n ] [ u . gran ] \.{:=} u . new
 {\p@semicolon}}%
\@x{\@s{16.4} {\p@end} {\p@with} {\p@semicolon}}%
\@x{ {\p@end} {\p@macro} {\p@semicolon}}%
\@pvspace{8.0pt}%
\@x{}%
\@y{\@s{0}%
Compute node main loop.
}%
\@xx{}%
\@x{ {\p@process} Node \.{\in} Nodes}%
\@x{ {\p@begin}}%
 \@x{\@s{16.4} nloop\@s{.5}\textrm{:}\@s{3} {\p@while} {\lnot} terminated
 {\p@do}}%
\@x{\@s{32.8} {\p@either}}%
\@x{\@s{49.19} DoMigrate ( self ) {\p@semicolon}}%
\@x{\@s{32.8} {\p@or}}%
\@x{\@s{49.19} DoRefresh ( self ) {\p@semicolon}}%
\@x{\@s{32.8} {\p@end} {\p@either} {\p@semicolon}}%
\@x{\@s{16.4} {\p@end} {\p@while} {\p@semicolon}}%
\@x{ {\p@end} {\p@process} {\p@semicolon}}%
\@pvspace{8.0pt}%
\@x{ {\p@end} {\p@algorithm} {\p@semicolon}}%
\@xx{}%
\pcalshadingfalse \pcalsymbolsfalse
\@pvspace{8.0pt}%
\@x{}\bottombar\@xx{}%

\@pvspace{8.0pt}%
\@x{}\moduleLeftDash\@xx{ {\MODULE} Marlin\_MC}\moduleRightDash\@xx{}%
\@x{ {\EXTENDS} Marlin}%
\@pvspace{8.0pt}%
\begin{lcom}{0}%
\begin{cpar}{0}{F}{F}{0}{0}{}%
\ensuremath{TLC} config-related defs.
\end{cpar}%
\end{lcom}%
\@x{ ConditionalPerm ( set ) \.{\defeq} {\IF} Cardinality ( set ) \.{>} 1}%
\@x{\@s{123.22} \.{\THEN} Permutations ( set )}%
\@x{\@s{123.22} \.{\ELSE} \{ \}}%
\@pvspace{8.0pt}%
\@x{ SymmetricPerms \.{\defeq}\@s{20.5} ConditionalPerm ( Nodes )}%
\@x{\@s{87.62} \.{\cup}\@s{15.97} ConditionalPerm ( Granules )}%
\@pvspace{8.0pt}%
\@x{ ConstNumMigrations \.{\defeq} 6}%
\@pvspace{8.0pt}%
\@x{}\midbar\@xx{}%
\@pvspace{8.0pt}%
\begin{lcom}{0}%
\begin{cpar}{0}{F}{F}{0}{0}{}%
Type check invariant.
\end{cpar}%
\end{lcom}%
\@x{ TypeOK \.{\defeq} \.{\land} storage\@s{11.51} \.{\in} Storages}%
\@x{\@s{56.14} \.{\land} numDone \.{\geq} 0}%
\@x{\@s{56.14} \.{\land} numDone \.{\leq} ConstNumMigrations}%
\@pvspace{8.0pt}%
\@x{ {\THEOREM} Spec \.{\implies} {\Box} TypeOK}%
\@pvspace{8.0pt}%
\@x{}\midbar\@xx{}%
\@pvspace{8.0pt}%
\begin{lcom}{0}%
\begin{cpar}{0}{F}{F}{0}{0}{}%
\ensuremath{NoDualOwnership} invariant.
\end{cpar}%
\end{lcom}%
 \@x{ PairsOfNodes \.{\defeq} \{ pn \.{\in} {\SUBSET} Nodes \.{:} Cardinality
 ( pn ) \.{=} 2 \}}%
\@pvspace{8.0pt}%
\@x{ NoDualOwnership \.{\defeq}}%
\@x{\@s{16.4} \A\, g \.{\in} Granules \.{:}}%
\@x{\@s{27.72} \A\, pn \.{\in} PairsOfNodes \.{:}}%
 \@x{\@s{39.04} \.{\LET}\@s{16.4} n1\@s{8.09} \.{\defeq} {\CHOOSE} n \.{\in}
 pn \.{:} {\TRUE}}%
 \@x{\@s{39.04} \.{\IN}\@s{4.1} \.{\LET} n2 \.{\defeq} {\CHOOSE} n \.{\in} pn
 \.{:} n \.{\neq} n1}%
 \@x{\@s{63.54} \.{\IN} {\lnot} ( \.{\land} storage . gtabs [ n1 ] [ g
 ]\@s{0.21} \.{=} n1}%
\@x{\@s{94.71} \.{\land} storage . gtabs [ n2 ] [ g ] \.{=} n2 )}%
\@pvspace{8.0pt}%
\@x{ {\THEOREM} Spec \.{\implies} {\Box} NoDualOwnership}%
\@pvspace{8.0pt}%
\@x{}\midbar\@xx{}%
\@pvspace{8.0pt}%
\begin{lcom}{0}%
\begin{cpar}{0}{F}{F}{0}{0}{}%
\ensuremath{HasOneOwnership} invariant.
\end{cpar}%
\end{lcom}%
\@x{ HasOneOwnership \.{\defeq}}%
\@x{\@s{16.4} \A\, g \.{\in} Granules \.{:}}%
\@x{\@s{27.72} \E\, n \.{\in} Nodes \.{:}}%
\@x{\@s{39.04} storage . gtabs [ n ] [ g ] \.{=} n}%
\@pvspace{8.0pt}%
\@x{ {\THEOREM} Spec \.{\implies} {\Box} HasOneOwnership}%
\@pvspace{8.0pt}%
\@x{}\bottombar\@xx{}%

\@pvspace{8.0pt}%
\@x{}\moduleLeftDash\@xx{\texttt{Marlin\_MC.cfg}}\moduleRightDash\@xx{}%
\begin{verbatim}
    SPECIFICATION Spec

    CONSTANTS 
        Nodes = {n1, n2, n3}
        Granules = {g1, g2, g3, g4, g5, g6}
        NumMigrations <- ConstNumMigrations
    
    SYMMETRY SymmetricPerms
    
    INVARIANTS 
        TypeOK
        NoDualOwnership
        HasOneOwnership
    
    CHECK_DEADLOCK TRUE
\end{verbatim}
\@x{}\bottombar\@xx{}%